\documentclass{aa}
\usepackage{natbib}
\usepackage{pdflscape} 
\usepackage{wasysym}
\usepackage[varg]{txfonts}
\usepackage{graphicx}
\usepackage{hyperref}
\bibpunct{(}{)}{;}{a}{}{,} 
\usepackage{longtable}

\usepackage{afterpage}

\def\ms{\hbox{\,m\,s$^{-1}$}}         
\def\cms{\hbox{\,cm\,s$^{-1}$}}       
\def\m2s2{\hbox{\,m$^{2}$\,s$^{-2}$}} 

\def\logrhk{$\log$(R$^{\prime}_{HK}$)}

\usepackage{color}

\begin{document}

\title{Radial Velocity Fitting Challenge
\thanks{Based on observations collected at the La Silla Parana Observatory,
ESO (Chile), with the HARPS spectrograph at the 3.6-m telescope. The data of the RV fitting challenge are only available in electronic form
at the CDS via anonymous ftp to cdsarc.u-strasbg.fr (130.79.128.5) or via \url{http://cdsweb.u-strasbg.fr/cgi-bin/qcat?J/A+A/}.}}

\subtitle{I. Simulating the data set including realistic stellar radial-velocity signals}

\author{X. Dumusque\inst{1,2}
            \thanks{Society in Science -- Branco Weiss Fellow (url: \url{http://www.society-in-science.org)}}}

\institute{Observatoire de Gen\`eve, Universit\'e de Gen\`eve, 51 ch. des Maillettes, CH-1290 Versoix, Switzerland \email{xavier.dumusque@unige.ch} 
              \and Harvard-Smithsonian Center for Astrophysics, 60 Garden Street, Cambridge, Massachusetts 02138, USA
              }

\date{Received XXX; accepted XXX}

\abstract
{Stellar signals are the
main limitation for precise radial-velocity (RV) measurements.
These signals arise from the photosphere of the stars. The \ms\,perturbation created by these signals prevents the detection and mass characterization of small-mass planetary candidates such as Earth-twins. Several methods have been proposed to mitigate stellar signals in RV measurements. However, without
precisely knowing the stellar and planetary signals in real observations, it is extremely difficult to test the efficiency of these methods.}
{The goal of the RV fitting challenge is to generate simulated RV data including stellar and planetary signals and to perform a blind test within the community to test the efficiency of the different methods proposed to recover planetary signals despite stellar signals.}
{In this first paper, we describe the simulation used to model the measurements of the RV fitting challenge. Each simulated
planetary system includes the signals from instrumental noise, stellar oscillations, granulation, supergranulation, stellar activity, and observed and simulated planetary systems. In addition to RV variations, this simulation also models the effects of instrumental noise and stellar signals on activity observables obtained by HARPS-type high-resolution spectrographs, that is, the calcium activity index \logrhk\,and the bisector span and full width at half maximum of the cross-correlation function.}
{We publish the 15 systems used for the RV fitting challenge including the details about the planetary systems that were injected into each of them (data available at CDS and here: \url{https://rv-challenge.wikispaces.com}).}
{}

\keywords{techniques: radial velocities -- stars: oscillations --  stars: activity -- Sun: activity -- Sun: starspots -- Sun: faculae, plages
                 }

\maketitle
\titlerunning{The Radial Velocity Fitting Challenge}
\authorrunning{X. Dumusque}

\section{Introduction} \label{sect:1}

The radial-velocity (RV) technique is an indirect method that measures the stellar wobble induced by a planet orbiting its host star with Doppler spectrostrcopy. It is fundamental for transit surveys, where it is used to confirm the exoplanet candidates and to understand their composition through measuring the planet mass. Based on this, the planet density is derived with the help of the radius measured using the transit light curve. Important results have been achieved in the \emph{Kepler} era, such as the confirmation of Kepler-78 with the HIRES and the HARPS-N spectrographs \citep{Sanchis-Ojeda-2013,Howard-2013b,Pepe-2013}, the characterization of the exoplanet Kepler 10-c, which has 17 Earth masses \citep[][]{Dumusque-2014}, and the indication that all planets below $\sim 6\,M_{\oplus}$ have a similar rocky composition \citep[][]{Dressing-2015}. However, the RV technique is strongly limited by the faintness of \emph{Kepler} targets. This will not be the case for upcoming photometric space missions such as TESS \citep[][]{Ricker-2014} and PLATO \citep[][]{Rauer-2014}, which will deliver hundreds of good candidates for RV follow-up. Obtaining the planet density is also important to estimate the scale height of the atmosphere that might exist on a planet, and thus form an idea of the detectability of chemical species in this atmosphere. This will be crucial in the \emph{James Webb Space Telescope} era, when this scale height needs to be known before dozens of hours of telescope time are spent trying to detect chemical species in the atmosphere of rocky planets orbiting M
dwarfs.  In addition to complementing transit detections that are restricted to a very limited orbital configuration in terms of inclinations, precise RVs may also be the only technique for detecting low-mass planets orbiting nearby bright stars, for which the atmospheres will be characterized with direct-imaging future space missions.

The RV technique is sensitive not only to possible companions, but also to signals arising from the photosphere of the host star, called stellar signals here. Now that the \ms\,precision level has been reached by the best spectrographs, it is obvious that stars introduce signals at a similar level, which strongly complicates the detection and measurement of small-mass planets.  There are several examples in the literature that show this, and we present here only a few of them. Gl581 is assumed to have between three and six planets \citep[][]{Hatzes-2016,Anglada-Escude-2015, Robertson-2014, Baluev-2013, Vogt-2012, Gregory-2011, Vogt-2010b, Mayor-2009b}, HD40307 between four and six planets \citep[][]{Diaz-2016,Tuomi-2013a}, and GJ667C between three and seven planets \citep[][]{Feroz-2014,Anglada-Escude-2012a,Gregory-2012}. In all these systems, the number of detected planets strongly depends on the model used to fit the data, and this is a sign that stellar signals are not properly modeled. Different models exists to mitigate the effect of stellar signals, but without precisely knowing   the contribution of stellar and planetary signals in real observations, it is extremely difficult to determine
how efficient the different methods are in correcting for stellar signals. We note that the HARPS-N solar telescope \citep[][]{Dumusque-2015b} will deliver the perfect real data set on which the RV effect of stellar signals can be better determined. Likewise, these
future data sets will help in determining the efficiency with
which different techniques can recover tiny planetary signals such as those from Venus despite stellar signals. Other facilities are also observing the Sun-as-a-star using high-resolution spectroscopy: the SONG spectrograph \citep[][]{Palle-2013}, the PEPSI spectrograph \citep[][]{Strassmeier-2015}, and the G\"ottingen FTS spectrograph \citep[][]{Lemke-2016}. The data provided by these facilities, often allowing higher spectral resolution than HARPS-N, are complementary to the data obtained by the HARPS-N solar telescope and therefore will be extremely useful for characterizing the RV effect of the stellar signal in more detail through analyzing spectral line variations (spectral line bisector and full width at half maximum, FWHM).

To be able to test the efficiency of the different techniques in recovering tiny planetary signals despite stellar signals, we started the RV fitting challenge initiative. The idea of the RV fitting challenge is
\begin{itemize}
\item to generate simulated RV time series that include realistic planetary and stellar signals, and 
\item to share these times series with the community so that different teams using different techniques to account for stellar signal can search for the injected planetary signal. 
\end{itemize}
When the planetary signal injected in the data is
known exactly, it is \emph{\textup{a posteriori}} possible to test the efficiency of the different methods used to search for planetary signals despite stellar signals. This RV fitting challenge exercise was performed between October 2014 and June 2015, and an analysis of the results of the different teams will be presented in a forthcoming paper \citep[][]{Dumusque-2016b}. We note that a similar challenge was previously posed in the exoplanet field to test the efficiency of different algorithms to recover planetary transit in photometric light curves \citep[][]{Moutou-2005b}. 

The goal of this first paper is to describe the simulation we
used to generate the different RV time series of the RV fitting challenge and to present the planets that were artificially injected into each of them. Although complex models can be used to estimate the effect of the different stellar signals (see Sect. \ref{sect:10}), we adopt here a simple approach based on real observations that therefore already includes the RV effect of stellar oscillations and granulation. We then model the RV variation induced by activity using a simplistic simulation of active region appearance on the solar surface, based on empirical properties derived from solar observations. This approach allows us to easily and rapidly generate RV times series that are similar to real RV measurements. We can therefore use these time series to further test the efficiency of the different methods used to search for planetary signals despite stellar signals.

A \emph{wiki} was created for the purpose of the RV fitting challenge to invite participants to collaborate in designing the optimal data set, and to share it. This can be accessed at \url{https://rv-challenge.wikispaces.com} to follow the preliminary discussions before the data set of
the RV fitting challenge was created, to download the data set (data also available at CDS), and to see preliminary results.

Section \ref{sect:10}  introduces the different stellar signals that affect RV measurements. In Sect. \ref{sect:2} we present the simulation that we used to model RV stellar signals and instrumental noise, followed in Sect. \ref{sect:3} by a description of the model used to generate RV variations induced by stellar activity. In Sect. \ref{sect:4} we perform a comparison between real and simulated data to check that our simulation of stellar signals gives realistic results, and finally, we conclude in Sect. \ref{sect:5}.

\section{Stellar signals and their effects on precise RV measurements} \label{sect:10}

Stellar signals affecting RV measurements can be separated
into four categories: 
\begin{itemize}
\item stellar oscillations on timescales of a few minutes for solar-like stars,
\item stellar granulation and supergranulation on timescales of a few minutes to 48 hours,
\item short-term stellar activity on the timescale of the stellar rotation period, which is induced by rotation in the presence of evolving surface magnetic inhomogeneities (mainly spots and plages), and
\item long-term stellar activity on timescales of several years, which is induced by stellar magnetic cycles.
\end{itemize}

\emph{Stellar oscillations} are produced by pressure waves (p-modes) propagating at the surface of solar-type stars and inducing a dilatation and contraction of external envelopes over timescales of a few minutes \citep[5  minutes for the Sun,][]{Kjeldsen-1995,Ulrich-1970,Leighton-1962,Evans-1962}. The RV signature of these modes typically varies between 10 and 400 \cms, depending on the stellar type and evolutionary stage. The amplitude and period of oscillation modes increases with mass along the main sequence. Theory predicts that the frequencies of p-modes increase with the square root of the stellar mean density and that their amplitudes are proportional to the luminosity over mass ratio (Christensen-Dalsgaard 2004). The most precise spectrographs today can reach sub-\ms\,RV precision and are therefore capable of directly resolving stellar oscillations on solar-type stars other than the Sun \citep[][]{Arentoft-2008,Bouchy-2001,Bedding-2001,Martic-1999}.

\emph{Stellar granulation and supergranulation} that are due to the convective nature of solar-type stars also affect RV measurements at the order of the \ms level \citep[][]{Dumusque-2011}. These convective phenomena can be found throughout the stellar surface, except in active regions where convection is significantly lower \citep[e.g.,][]{Brandt-1990,Livingston-1982,Dravins-1982}. The surface convection has been studied in detail on the Sun and also by using 3D simulations \citep[e.g.,][]{Beeck-2013a,Beeck-2013b,Holzreuter-2013,de-la-Cruz-Rodriguez-2011,Asplund-2000,Dravins-1981}. While most of these simulations often modeled a few spectral lines and in a very small box compared to the solar surface, some other models allow reproducing the disk-integrated visible spectrum, which can then be used to measure full-disk RV variations as is done for exoplanet observations \citep[][]{Meunier-2015,Cegla-2013,Allende-Prieto-2013}. From these simulations and from high-spatial resolution observations of the Sun, it is very well known that photospheric lines that emerge from the granules and intergranular lanes (which surround granules) are not only shifted with respect to each other as
a result of the opposite velocity field, but are highly asymmetric and are characterized by bisectors of opposite slope. The strong asymmetry of spectral lines originating from granules is largely cancelled out by the opposite asymmetry of the intergranular lanes. The line shifts and asymmetries observed in stellar spectra are therefore much smaller than the asymmetries and shifts of the spectral lines that originate from either the granules or the intergranular lanes.

\emph{Short-term stellar activity} is induced by stellar rotation in the presence of active regions, dark spots, and bright plages. When the star rotates, these active regions induce an RV variation by two different physical processes. Because the temperatures in these regions differ from the average surface temperature, their emerging flux is different. Sunspots are $\sim$700 K cooler than the effective temperature of the Sun \citep[][]{Meunier-2010a}
and therefore have a much lower flux than quiet solar photosphere regions. A spot will therefore break the flux balance between the blueshifted approaching limb and the redshifted receding limb of a rotating star and will induce an RV variation as it passes across the visible stellar disk. A plage at the disk center is only slightly hotter than the average effective temperature and will induce a weak flux effect. A plage on the limb will be brighter because of the center-to-limb brightness dependence \citep[][]{Unruh-1999, Frazier-1971}, but at this location, the star emits less light as a result of limb darkening. Independently of its location, a plage will therefore induce a weaker flux effect than a spot, even if plages tend to be an order of magnitude more extended than spots \citep[][]{Chapman-2001}. The second effect is induced because strong local magnetic fields inhibit convection inside active regions \citep[][]{Brandt-1990,Cavallini-1985a,Livingston-1982,Dravins-1982}. This inhibition of convection suppresses the convective blueshift effect, $\sim$300 \ms\,for the Sun, inside active regions. These regions therefore appear redshifted compared to the quiet photosphere \citep[see Fig. 3 in][]{Cavallini-1985a}, which implies an RV variation as active regions appear and disappear from the visible part of the stellar disk through rotation.

\emph{Long-term stellar activity}  due to a solar-like magnetic cycle affects RV measurements on a timescale of several years. The idea that magnetic cycles perturb RVs was first reported
by \citet{Campbell-1988} and \citet{Dravins-1985}. The level of magnetic activity in solar-type stars has been studied for more than 50 years using the Mount Wilson S-index and the \logrhk\,index \citep[][]{Wilson-1963}, and it has been shown that this index and the magnetic field strength are
closely related \citep[e.g.,][]{Schrijver-1989}. Variation in the activity level of thousands of FGK dwarfs has been studied \citep[][]{Hall-2007,Baliunas-1995,Wilson-1978}, but it was realized
only eventually that RVs can be influenced by magnetic cycles. A few attempts of measuring such an effect in the Sun yielded somewhat contradictory results: \citet{Deming-1994} found a peak-to-peak RV variation of 28 \ms  throughout the solar cycle, while \citet{McMillan-1993} obtained constant RVs within $\sim$4 \ms. Measurements of similar RV effects in other stars have remained inconclusive \citep[][]{Santos-2010a,Isaacson-2010,Wright-2005,Paulson-2002} until the work of \citet{Lovis-2011b} and \citet{Dumusque-2011c}. Solar-like magnetic cycles are characterized by an increasing filling factor of active regions when the activity level rises. Because convection is strongly reduced in active regions as a
result of the magnetic field (see previous paragraph), the star will appear redder (positive velocity) during its high-activity phase. A positive correlation between the RVs and the activity level is therefore observed \citep[][]{Meunier-2010a,Lindegren-2003}. This is the most commonly given explanation, although some other physical processes might be responsible for this long-term RV-activity correlation, for example, the variation in surface flows proposed by \citet{Makarov-2010}.

\section{Simulating instrumental noise and stellar signals} \label{sect:2}

The goal of the RV fitting challenge is to measure the efficiency of the different techniques that are used to search for planetary signals embedded in stellar signals. This measurement can only be made on simulated RV measurements, for which we precisely
know the planetary signals that should be recovered. These simulated RV time series need to be realistic, however, if we wish to be able to use the conclusions of the RV fitting challenge for real data sets. In this section, we describe the model that we used to simulate in a simple way realistic RV, BIS SPAN, FWHM, and \logrhk\,time series.

\subsection{Instrumental noise and stellar signal induced by oscillations, granulation, and supergranulation} \label{sect:2-0}

Instrumental noise and stellar signals induced by oscillations, granulation, and supergranulation are known to have timescales shorter than a few days.
These types of signals can be studied using continuous high-cadence observations of solar-type stars on a timescale of a week \citep[][]{Dumusque-2011a}. These observations are crucial for asteroseismology studies and have been collected by the HARPS spectrograph \citep[][]{Teixeira-2009,Bazot-2007,Bouchy-2005b}. In particular, the G8 dwarf $\tau$\,Ceti (\object{HD10700}) was observed with a cadence of 54 to 71 seconds over six consecutive nights \citep[][]{Teixeira-2009}. We used these real measurements to estimate the RV effect of instrumental noise and stellar signals induced by oscillations, granulation, and supergranulation, and to generate realistic RV data that included all these signals.

The strategy we employed to generate realistic data that include instrumental noise and stellar signals is to first transform the time-domain RVs of $\tau$\,Ceti into the frequency domain, fit the power of the signal as a function of frequency, and finally return to the time domain while changing the phase of all frequencies randomly \citep[][]{Dumusque-2011a}. Changing the phases allows us to reconstruct RV measurements that are differently sampled than the original data, but still include similar instrumental and stellar signals.

\begin{figure*}[t]
\begin{center}
\includegraphics[width=16.4cm]{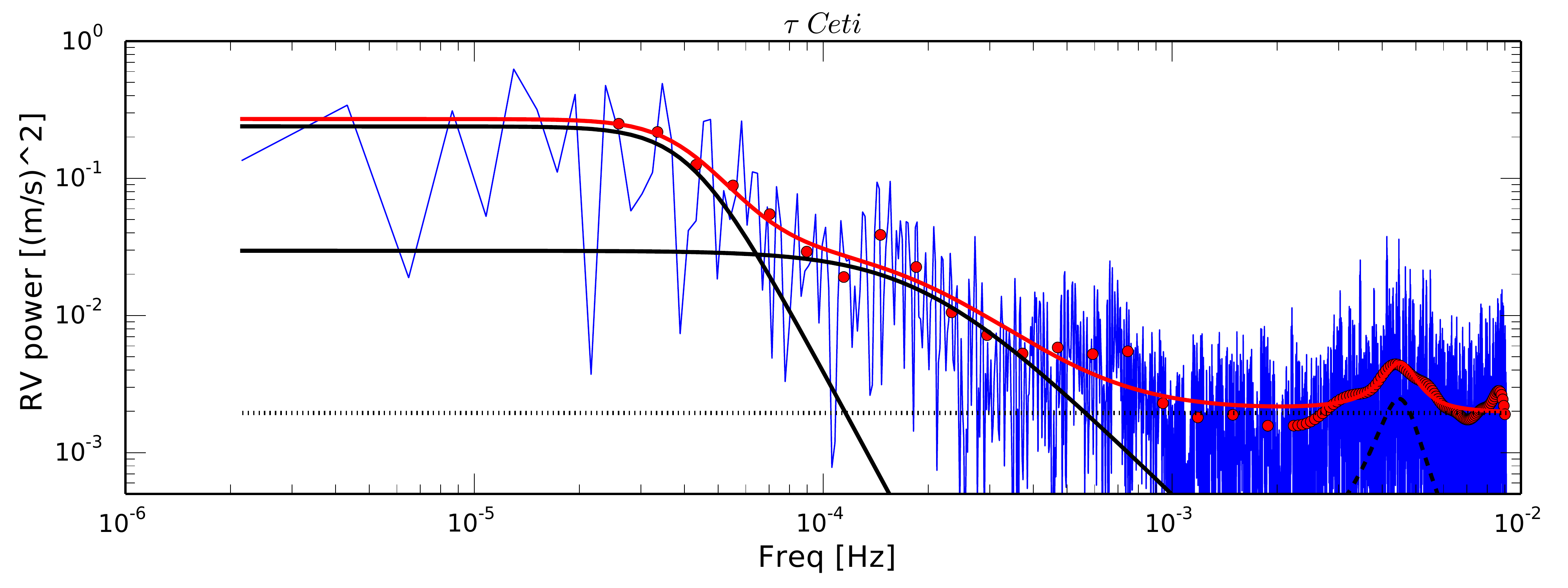}
\caption{Raw RVs of $\tau$\,Ceti in the frequency domain (blue line) and binned in frequency (red dots). The best fit to the data (red line), using the model presented in Eq. \ref{eq:2-0-1}, is obtained considering two power laws to account for supergranulation and granulation (black continuous lines), a Lorentzian profile to account for stellar oscillations (dashed line), and a constant to include white instrumental noise (dotted line).}
\label{fig:2-0-0}
\end{center}
\end{figure*}
To transform the time-domain RV measurements of $\tau$\,Ceti into the frequency domain, we used the generalized Lomb-Scargle periodogram \citep[GLS,][]{Zechmeister-2009,Scargle-1982,Lomb-1976a}. Figure \ref{fig:2-0-0} shows the result of this transformation, which highlights four different components:
\begin{itemize} 
\item Supergranulation and granulation at low frequency, which we can fit using an empirical power law as initially proposed by \citet{Harvey-1984} and reviewed by \citet{Andersen-1994} and \citet{Palle-1995}.\item Stellar oscillations that create a bump in the power spectrum seen at high frequency. We can account for this feature using a Lorentzian profile \citep[][]{Lefebvre-2008},
\item Instrumental noise at very high frequency. Within the assumption that this noise is white, we can fit it using a constant value spanning the entire frequency range. The value of this constant is only seen at very high frequency because of the strong oscillation, granulation, and supergranulation signals at lower frequencies.
\end{itemize} 
Considering these different sources of signal, we can fit the frequency-domain RV measurements using the following model:
\begin{equation} \label{eq:2-0-1}
P(\nu) = \sum_{i=1}^2{\frac{A_i}{1+(B_i\nu)^{C_i}}}+A_L\frac{\Gamma^2}{(\nu-\nu_0)^2+\Gamma^2}+C_{inst},
\end{equation}
where $A_i$, $B_i$ , and $C_i$ are parameters that account for granulation and supergranulation, and $i$ differentiates one source of the signal from the other. $A_L$, $\Gamma,$ and $\nu_0$ are the amplitude, the FWHM and the center of the bump created by stellar oscillations, and $C_{inst}$ is the constant amplitude of the white instrumental noise. More details about this fit can be found in \citet{Dumusque-2011a}.


         


\begin{figure*}[t]
\begin{center}
\includegraphics[width=16.4cm]{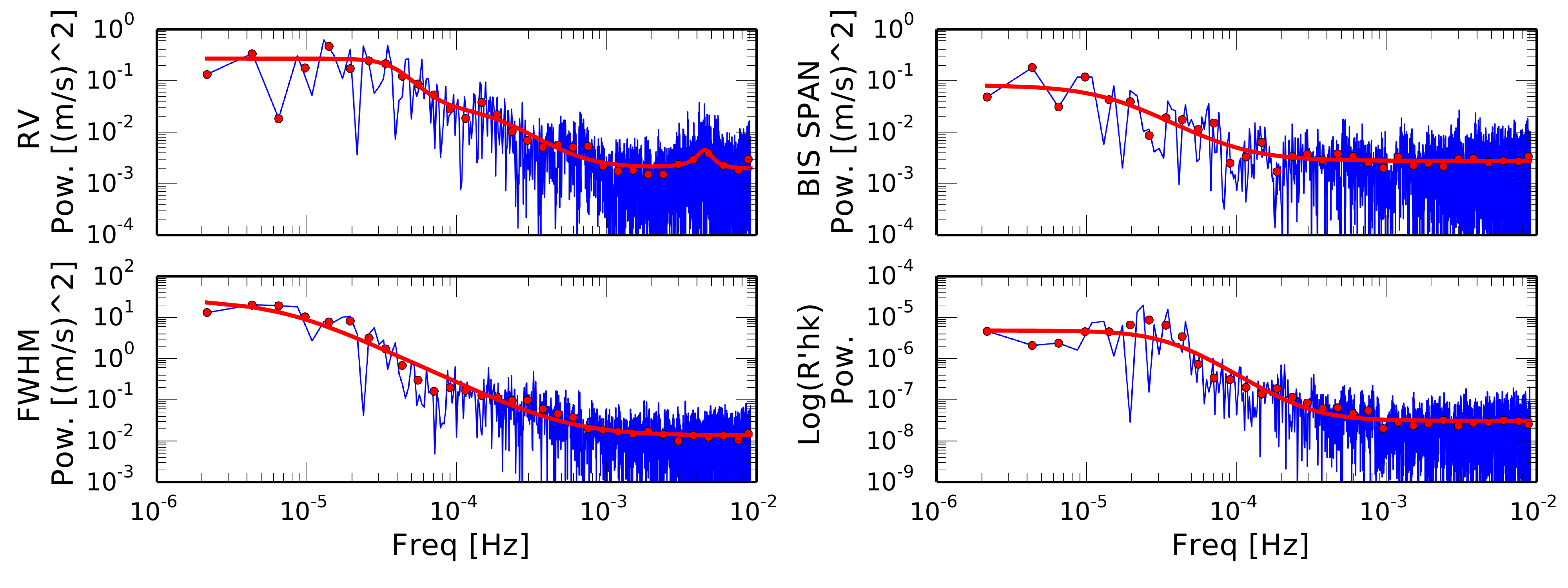}
\caption{Raw RV, BIS SPAN, FWHM and \logrhk\,of $\tau$\,Ceti in the frequency domain (blue line) and binned in frequency (red dots). The best fit of each observable (red line) is obtained using the model presented in Eq. \ref{eq:2-0-1}
for the RVs and the model presented in Eq. \ref{eq:2-0-2} for the other observables.}
\label{fig:2-0-1}
\end{center}
\end{figure*}

The RV is not the only observable to be affected by stellar signals. The bisector span \citep[BIS SPAN,][]{Queloz-2001} and the FWHM of the cross-correlation function \citep[CCF,][]{Pepe-2002a,Baranne-1996} are also affected. The power spectrum of these two observables in Fig. \ref{fig:2-0-1} shows the supergranulation and granulation signals at low frequency and the signal of instrumental noise at very high frequency. However, we do not detect a bump caused
by stellar oscillations. We decided to fit the power spectrum of the BIS SPAN and the FWHM using a model similar to Eq. \ref{eq:2-0-1}, but only considering one component to fit granulation and supergranulation and one constant to account for instrumental noise. In this case, we have
\begin{equation} \label{eq:2-0-2}
P(\nu) = \frac{A}{1+(B\nu)^{C}}+C_{inst}.
\end{equation}
The calcium activity index \logrhk\,in the frequency domain shows a similar behavior as for the BIS SPAN and the FWHM. We therefore fit this observable using the same model.

The best fit of each observable in the frequency domain contains the information on the amplitude of the stellar and instrumental signals as a function of frequency. Returning to the time domain using the inverse transformation of the GLS applied to the best fit allows creating simulated RV, BIS SPAN, FWHM, and \logrhk\,measurements. These simulated measurements include stellar and instrumental signals of similar amplitude as observed in the original measurements of $\tau$\,\,Ceti. During the inversion process, the phase of each point in frequency must be changed to generate the data for any observational date. We did not elect to use the phases calculated by the GLS periodogram because it will reproduce the real data after inversion, setting the RVs to zero during the observation gaps. We chose a uniformly random phase between $0$ and $2\pi$ for each point in frequency before applying the inverse GLS periodogram transformation.

\begin{figure*}
\begin{center}
\includegraphics[width=8.2cm]{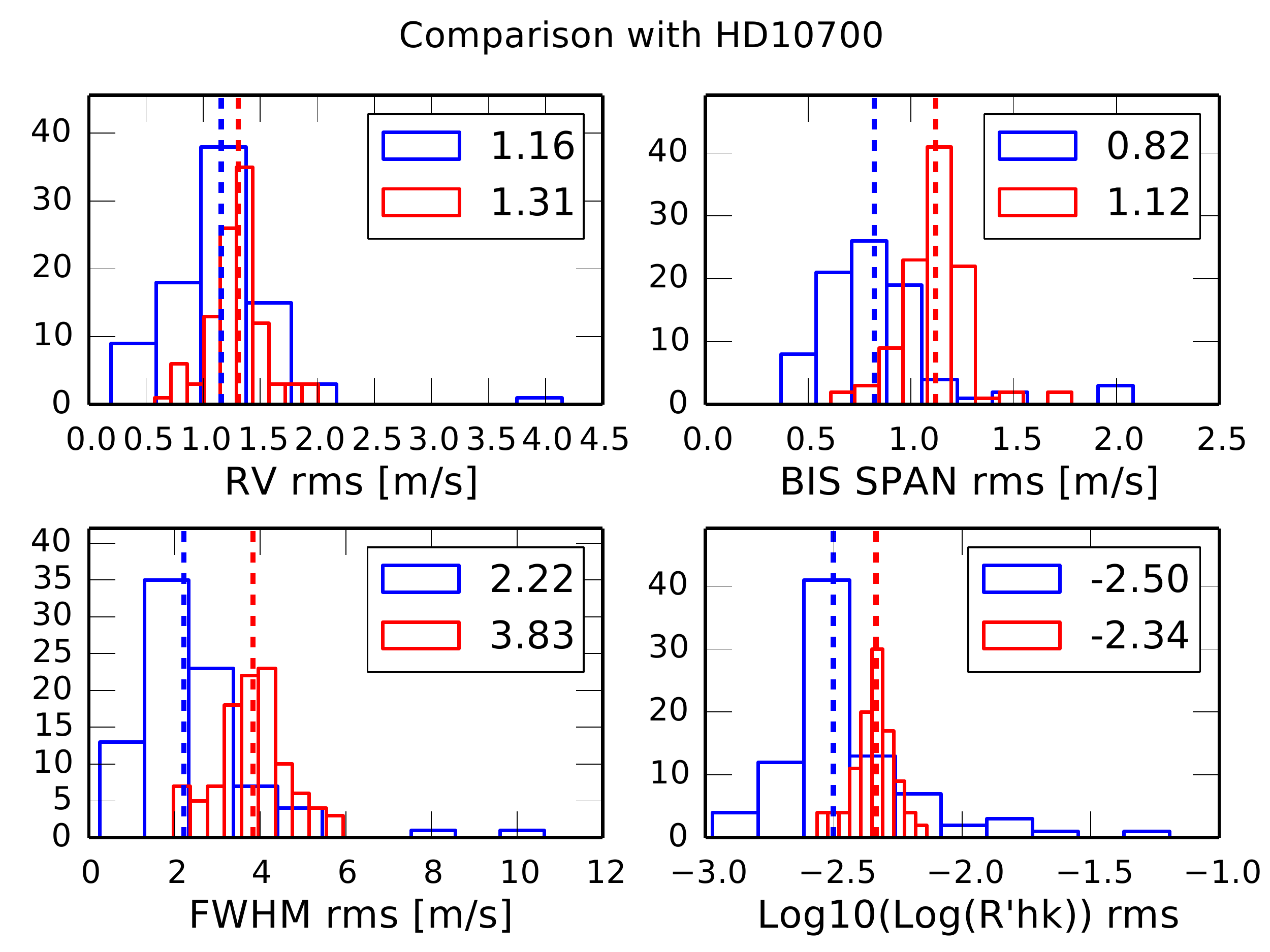}
\includegraphics[width=8.2cm]{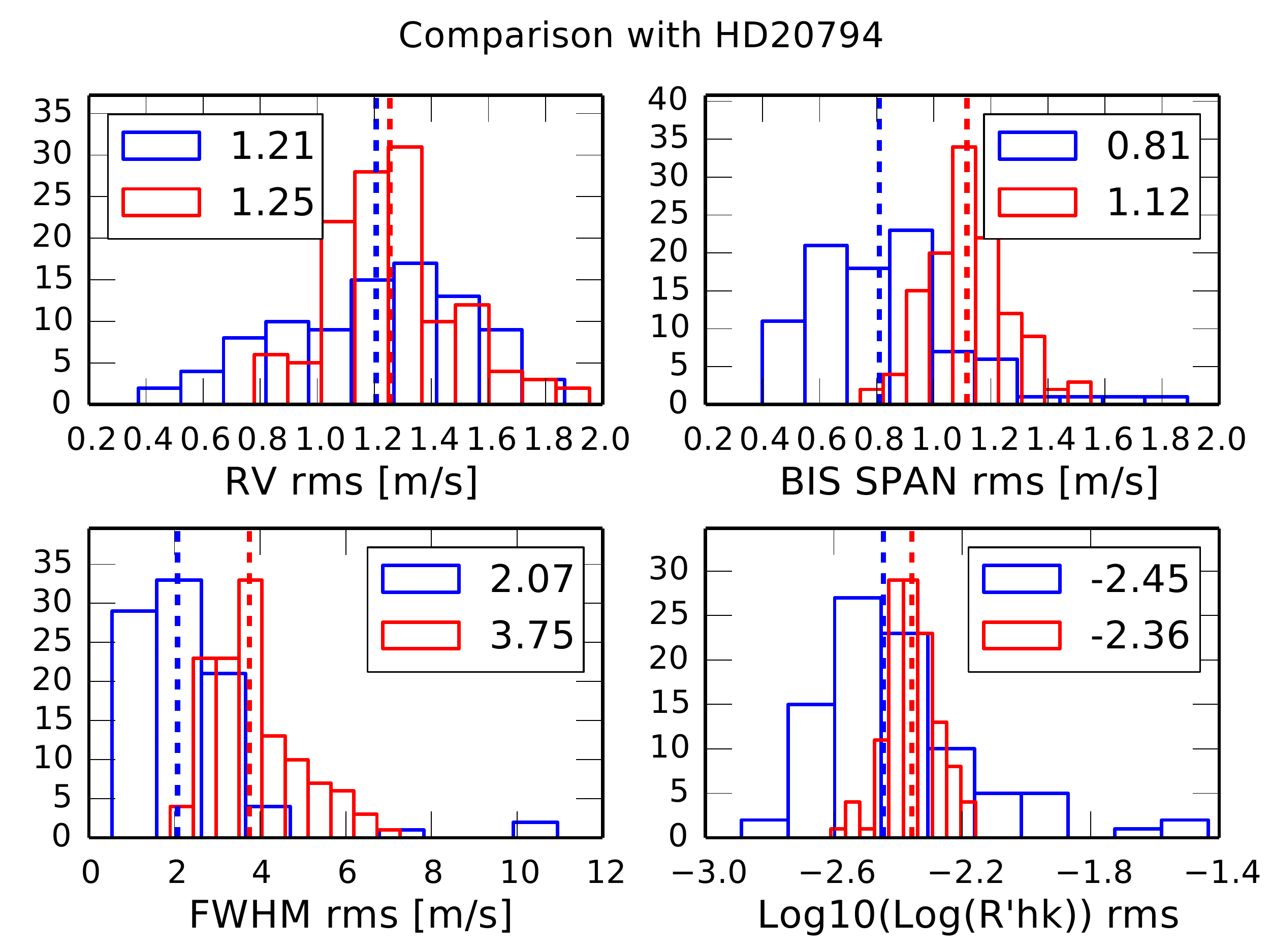}
\includegraphics[width=8.2cm]{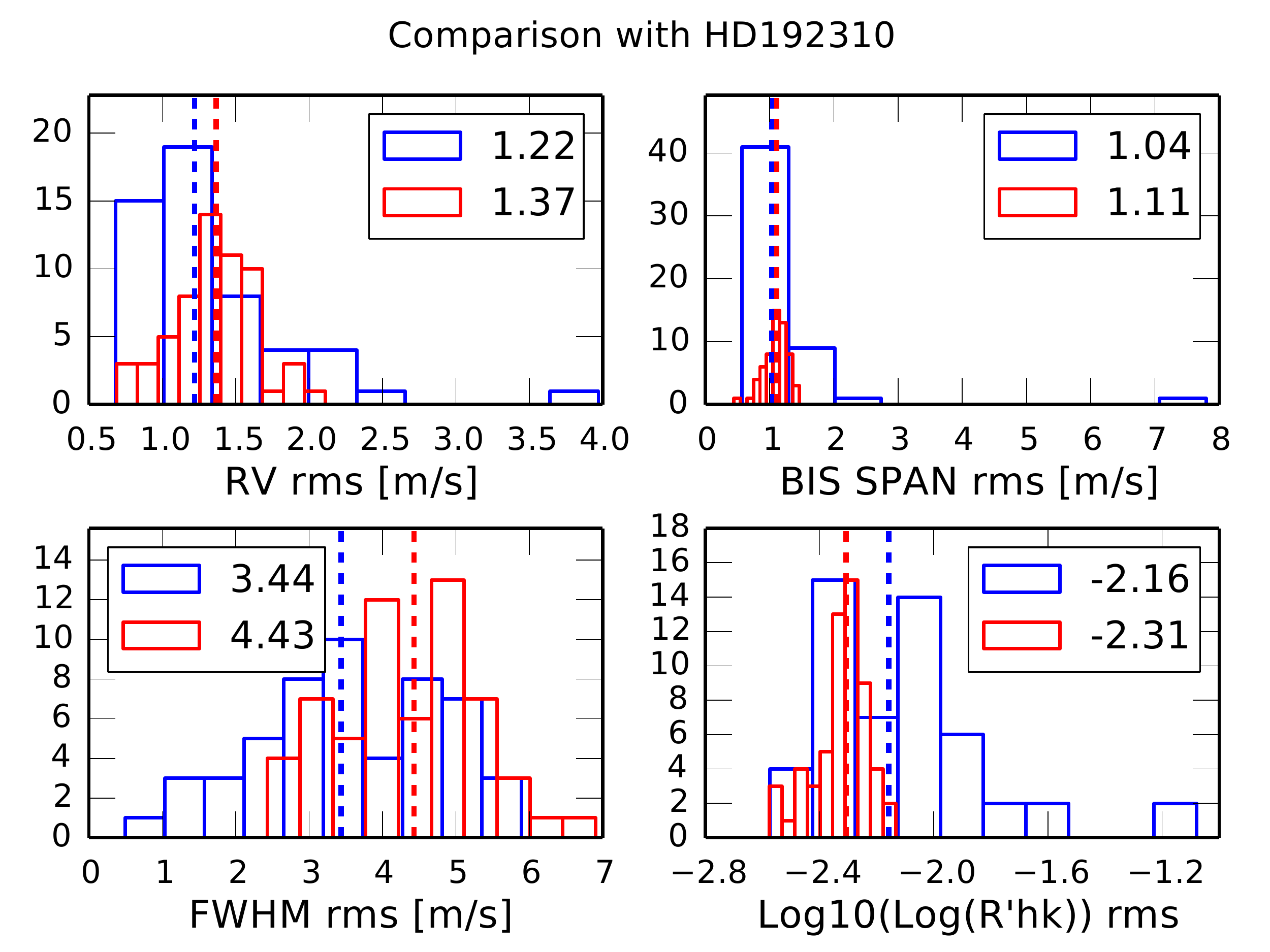}
\includegraphics[width=8.2cm]{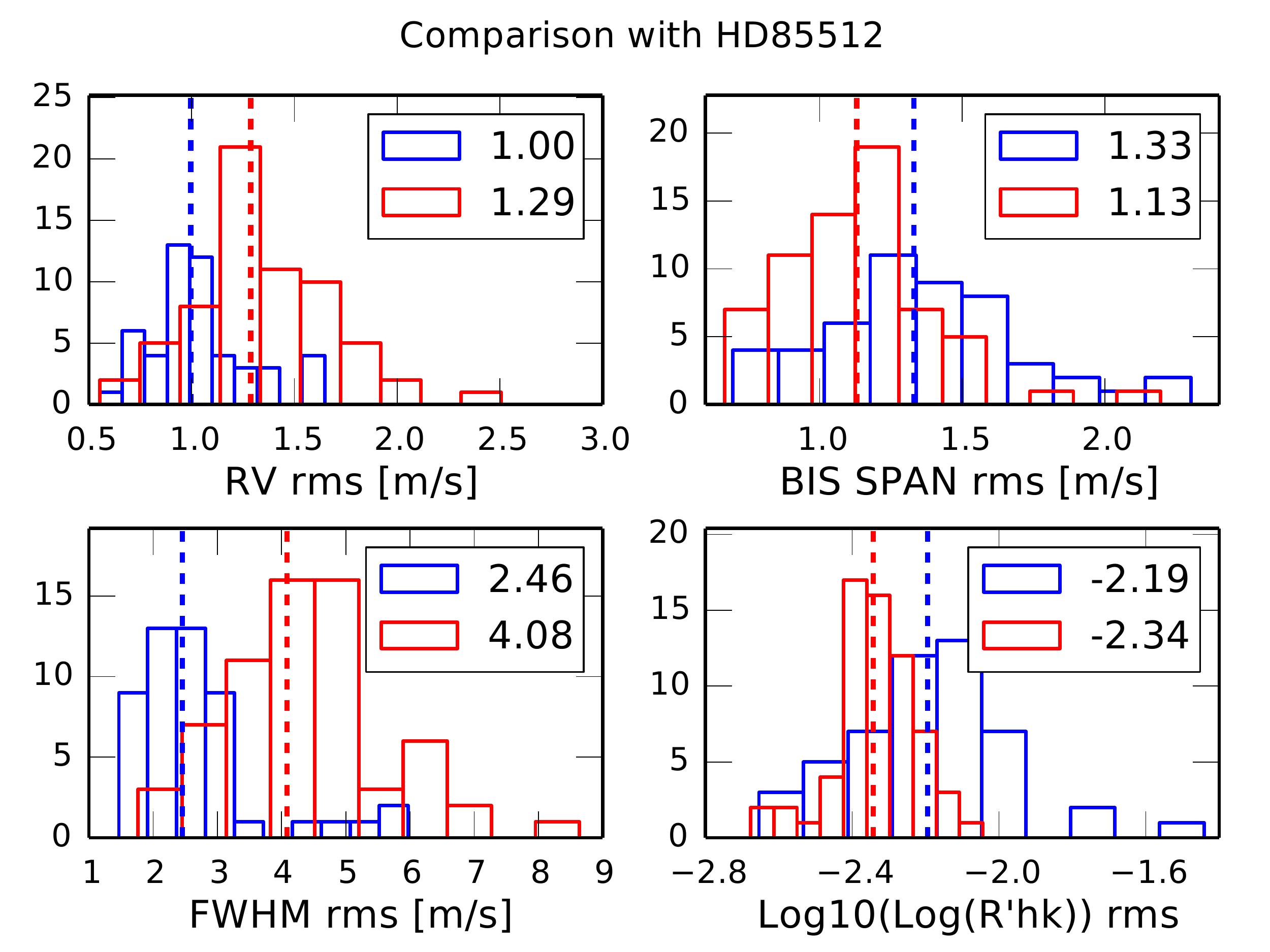}
\caption{Comparison between the simulated and real data for $\tau$\,\,Ceti (HD10700), HD20794, HD192310, and HD85512. For each star, the simulated data are generated with the same calendar as the observed measurements. We then compute the rms of the RVs, the BIS SPAN, the FWHM, and the \logrhk\,for chunks of ten consecutive days each, including at least seven observations. Each subplot shows the histogram of the rms computed based on the observed (blue) and simulated (red) data, with the corresponding median highlighted by the dashed vertical lines and its value reported in the legend. The agreement between the histograms and the median values demonstrates that the simulated data that include granulation signal, oscillation signal, and instrumental noise are similar to the real observations performed with HARPS.}
\label{fig:2-0-2}
\end{center}
\end{figure*}

To check that simulated data that include stellar and instrumental signals are realistic when compared to observations, we selected four quiet solar-type stars that have thoroughly been observed with HARPS:  $\tau$\,Ceti (G8V), HD20794 (G8V), HD192310 (K3V), and HD85512 (K5V). For each star, we generated simulated data using the same calendar as the real measurements. We then computed the rms of the RVs, the BIS SPAN, the FWHM, and the \logrhk\,for all chunks of ten consecutive days, including at least seven observations. In Fig. \ref{fig:2-0-2} we show the results of this comparison. The agreement between the rms measured on the observed and simulated data demonstrates that we are able to create simulated data that include stellar and instrumental signals similar to real observations performed with HARPS.

\subsection{Signal induced by stellar activity} \label{sect:2-1}

The vast majority of solar-type stars presents a solar-like magnetic cycle \citep[][]{Lovis-2011b,Dumusque-2011}. During one cycle, the activity of a star changes from a quiet phase where only a few active regions appear on its surface to an active phase where a significant fraction of its surface is covered by active regions. These active regions, consisting of dark spots or bright plages, induce a deformation of spectral lines because their difference in contrast with respect to the photosphere breaks the flux balance between the blueshifted approaching and the
redshifted receding limb of a rotating star. In addition, inhibited convection by strong local magnetic fields in these active regions also modifies the shape of spectral lines. Both of these effects induce a change in the barycenter of each spectral line, therefore implying a spurious RV shift. More details about the effect of spots and plages on stellar spectra and RV measurements are reported
in \citet{Dumusque-2014b}, \citet{Meunier-2010a}, \citet{Desort-2007}, \citet{Saar-1997b}, and references therein.

Stellar activity can be separated into two components: one on short timescales, induced by stellar rotation in the presence of evolving active regions, and  another on long timescales, induced by the increasing or decreasing number of active regions that are a result of solar-like magnetic cycles. We discuss the effect of short-term activity  in
the next section. Long-term activity is addressed in Sect. \ref{sect:2-1-2}.

For an active region of a given size, contrast, latitude, and longitude, and a given rotation period, we can use the SOAP 2.0 code to estimate the effects of this active region on photometry, RV, BIS SPAN, and FWHM as the star rotates\footnote{\url{http://astro.up.pt/resources/soap2/}} \citep[][]{Dumusque-2014b}. To estimate the photometric, RV, BIS SPAN, and FWHM variations induced by an active region, SOAP 2.0 divides the stellar surface into thousands of cells, each of which contains the average line profile of a solar active region or of the solar quiet photosphere, depending on whether the cells cover active regions or not. These line profiles are derived by computing the CCF\footnote{The CCFs are obtained by cross correlation of the spectral atlases with the binary mask commonly used on HARPS measurements to derive the CCF of G dwarfs. This binary mask exhibits holes at the position of each spectral line, each hole being weighted by the depth of the corresponding spectral line. This allows giving more weight to the deep lines because they contain more RV information \citep[][]{Pepe-2002a}. The RV precision on a spectral line is inversely proportional to the derivative of the spectrum \citep[see Eq. 3 in][]{Bouchy-2001b}. Thus deep spectral lines have wings with stronger slopes, thus higher derivative values, which in the end implies a better RV precision.} of the NSO solar spectral atlases for a quiet-photosphere region \citep[][]{Wallace-1998} and a sunspot \citep[][]{Wallace-2005}. In the same way as for the simulation of stellar oscillations and granulation presented in Sect. \ref{sect:2-0}, SOAP 2.0 uses real observations to avoid modeling the complex spectrum emerging from the quiet photosphere and active regions \citep[e.g.,][]{Balasubramaniam-2002}. The convective blueshift effect in the quiet photosphere and its inhibition in active regions is naturally included in the different average line profiles. The line profile in each cell is weighted according to the emerging flux: 1 for quiet-photosphere regions, smaller than 1 and constant for spot regions, and larger than 1 but varying with distance from the limb for plages to account for the limb-brightening effect \citep[][]{Akimov-1982,Frazier-1971}. Finally all the cells are summed, taking into account stellar rotation and limb darkening to obtain a full-disk average line profile from which it is possible to derive the photometry, RV, BIS SPAN, and FWHM. For more details we refer to  \citet{Dumusque-2014b}. 

We note that SOAP 2.0, as published, does not consider a wavelength dependency for the contrast of spots and plages\footnote{The contrast of an active region is measured from the difference in temperature of the region with respect to the photosphere and for the center of the visible bandpass.} and no limb-center variation in the bisector of quiet regions \citep[][]{Cavallini-1985b}, although this last effect seems negligible \citep[see Appendix B in][]{Dumusque-2014b}. In addition, the code is not capable of estimating the variation in calcium activity index \logrhk, which is an important observable that probes the presence of active regions. In the next section, we modify the SOAP 2.0 code to derive an estimate for the \logrhk\,variation in presence of active regions.

\subsubsection{Deriving the calcium activity index with SOAP 2.0} \label{sect:2-1-0}

The calcium activity index $S_{\mathrm{index}}$ is a measure of the emission seen in the core of the \ion{Ca}{ii} H and K lines \citep[e.g.,][]{Vaughan-1978}:
 \makeatletter 
 \def\@eqnnum{{\normalsize \normalcolor (\theequation)}} 
  \makeatother
{\small
\begin{eqnarray} \label{eq:sect:2-1-0-0}
S_{\mathrm{index}} &=& \frac{H+K}{C_{3900} + C_{4000}},\\
\mathrm{Log}(\mathrm{R'}_{HK}) &=& \mathrm{Log}\left[1.34.10^{-4}\,C_{cf}\,S_{\mathrm{index}}\right],\nonumber\\
C_{cf} &=& 1.13(B-V)^3 - 3.91 (B-V)^2 + 2.84 (B-V) - 0.47,\nonumber
\end{eqnarray}
}
where $H$ and $K$ are the total fluxes measured in the core of the \ion{Ca}{ii} H and K lines, and $C_{3900}$ and $C_{4000}$ are the total fluxes in the 20\,\AA \,pseudo-continuums centered on 3900 and 4000\,\AA. We also show here the conversion from $S_{\mathrm{index}}$ to \logrhk\,\citep[][]{Noyes-1984}. The calcium activity indicator \logrhk\,is corrected for the stellar energy distribution and therefore is used to compare the activity between stars of different spectral types. This calcium activity index is often used as a proxy for stellar activity when analyzing RV measurements. Therefore, we decided to model the variation of this observable using SOAP 2.0 for the purpose of the RV fitting challenge. 

It is possible in SOAP 2.0 to place a high-resolution spectrum rather than its CCF in each simulation cell. To estimate the calcium activity index, we included the part of the NSO solar spectral atlases that includes the \ion{Ca}{ii} H and K lines and the surrounding continuums used to calculate $S_{\mathrm{index}}$. Unfortunately, the NSO solar spectral atlases used only begin at 3920\,\AA, which prevents us from measuring the flux in the pseudo-continuum centered on 3900\,\AA. Considering that the $S_{\mathrm{index}}$ for the quiet Sun\footnote{\logrhk\,for the quiet Sun is -5.0. Using Eq. \ref{eq:sect:2-1-0-0} and a B-V of 0.656, \logrhk$=-5.0$ implies that $S_{\mathrm{index}}=0.16.$} is 0.16, we can estimate the total flux measured in the two 20\,\AA\,pseudo continuums, $C_{3900} + C_{4000} =  \frac{H_{\mathrm{quiet\,Sun}}+K_{\mathrm{quiet\,Sun}}}{0.16}$, and therefore we have
\begin{equation} \label{eq:sect:2-1-0-1}
S_{\mathrm{index}} = \frac{0.16\,(H+K)}{H_{\mathrm{quiet\,Sun}}+K_{\mathrm{quiet\,Sun}}}.
\end{equation}
\begin{figure}[t]
\begin{center}
\includegraphics[width=8.2cm]{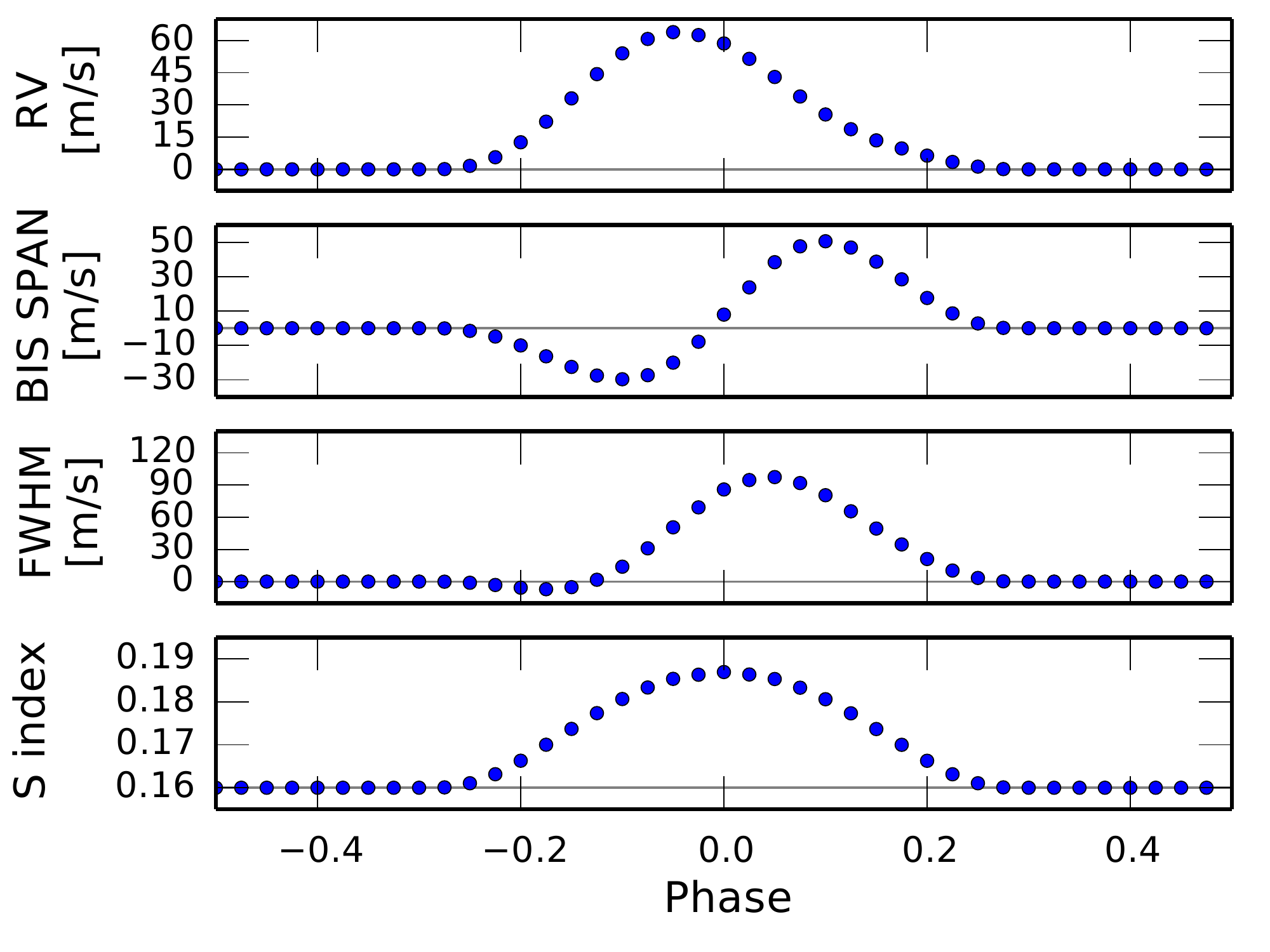}
\caption{Effect in RV, BIS SPAN, FWHM, and $S_{\mathrm{index}}$ induced by stellar rotation in presence of an equatorial plage. The plage is 5\% in area of the visible hemisphere, and the stellar rotation period is 25 days. The plage is on the limb at a rotation phase of -0.25 and 0.25 and at disk center for phase 0.}
\label{fig:2-1-0-0}
\end{center}
\end{figure}
Using this formula, we can estimate the variation of the calcium activity index with SOAP 2.0 as a visible active region rotates with the solar disk. Figure \ref{fig:2-1-0-0} shows the result of a SOAP 2.0 simulation for an equatorial plage covering 5\% in surface area of the visible hemisphere. In this example, the amplitude in $S_{\mathrm{index}}$ observed is 0.027, from 0.160 to 0.187, which corresponds to a variation in \logrhk\,of 0.15, from -5.0 to -4.85. Figure \ref{fig:2-1-0-1} illustrates the variation in emission seen in the core of the \ion{Ca}{ii} H and K lines between a case when no plage is present (quiet Sun) and when a plage that covers 5\% of the surface is at disk center (active Sun).
\begin{figure*}[t]
\begin{center}
\includegraphics[width=16.4cm]{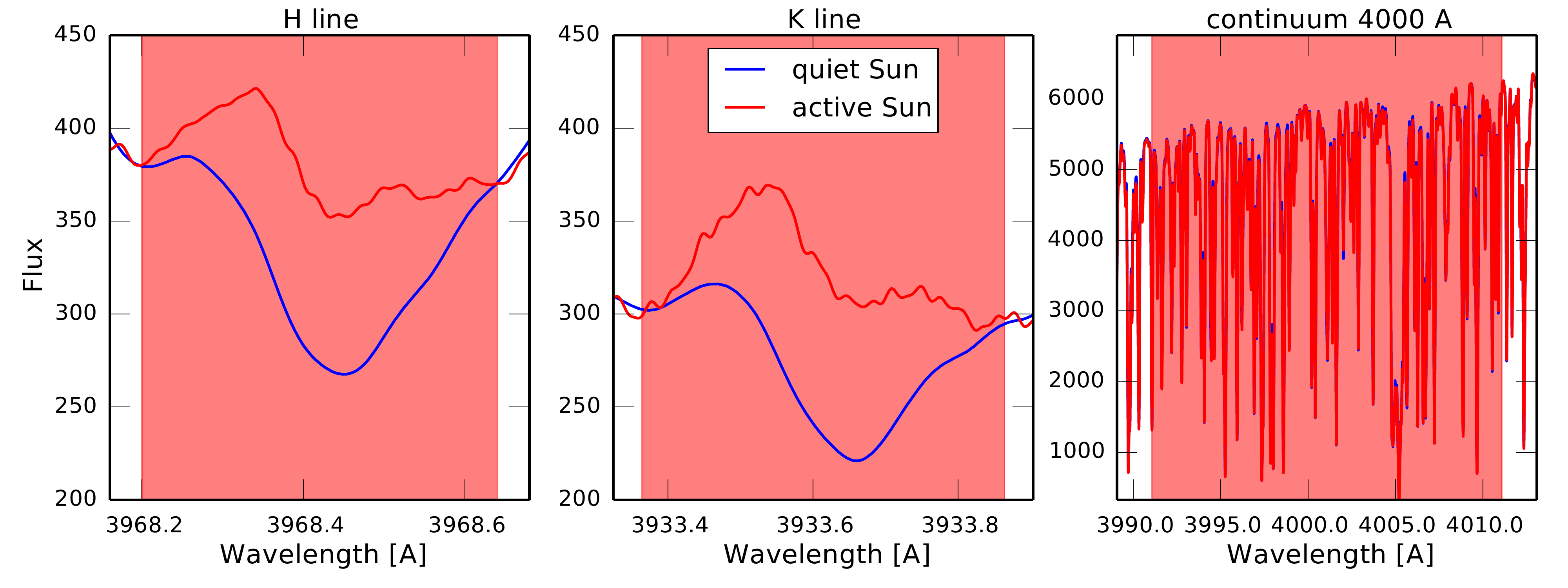}
\caption{Comparison between the core of the \ion{Ca}{ii} H and K lines and the 20\,\AA\,pseudo-continum centered on 4000\,\AA\,when the star is quiet (blue) and when an equatorial plage of 5\% in surface area of an hemisphere is at the disk center (red, corresponds to phase 0 in Fig. \ref{fig:2-1-0-0})}
\label{fig:2-1-0-1}
\end{center}
\end{figure*}

\begin{figure}[t]
\begin{center}
\includegraphics[width=8.2cm]{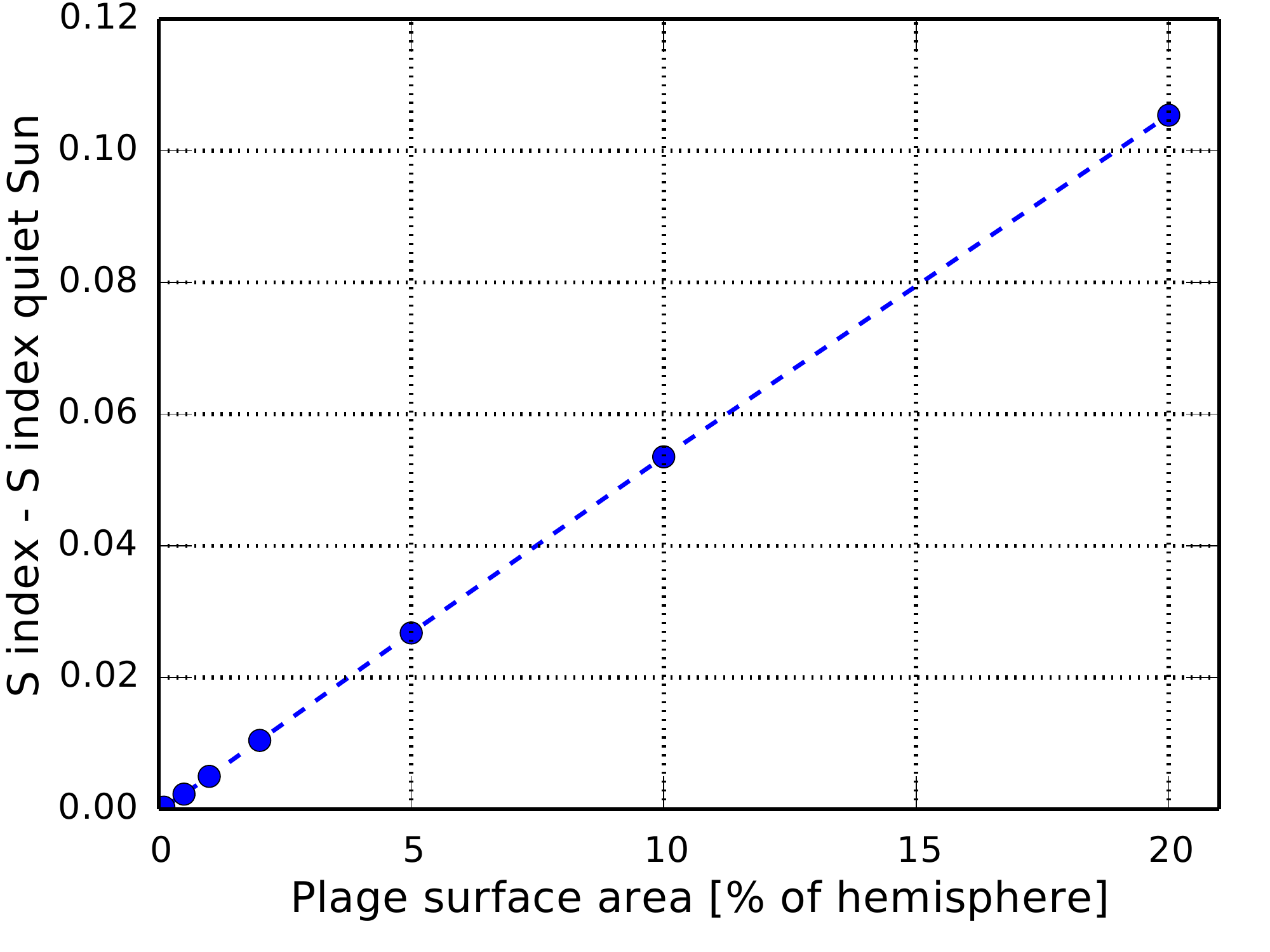}
\caption{
$S_{\mathrm{index}}$ as a function of the active region surface area. By removing the quiet-Sun $S_{\mathrm{index}}$ value, i.e., 0.16, from the measured $S_{\mathrm{index}}$, we observe a proportionality with the plage surface area.}
\label{fig:2-1-0-2}
\end{center}
\end{figure}

The effect of an active region on photometry, RV, BIS SPAN, and FWHM is proportional to the active region surface area \citep[see Fig. 8 in][]{Dumusque-2014}. This is expected because a large active region should induce the same effect on these observables as several smaller active regions of which it might consist. We show in Fig. \ref{fig:2-1-0-2} that the $S_{\mathrm{index}}$ is also proportional to the covered surface area. This proportionality is crucial for our simulation of stellar activity because it will allow us to measure the total effect of these regions by summing the individual contribution of each of them when we consider several active regions on the
stellar surface in Sect. \ref{sect:2-1-2} .

We note that the study presented here shows the effect of plages on the calcium activity index. A similar behavior can be seen for spots, but with a smaller $S_{\mathrm{index}}$ variation, as spots tend to be an order of magnitude smaller than plages \citep[][]{Chapman-2001} and much fainter than the quiet photosphere.

We are aware that this model does not correctly estimate the maximum variation amplitude in $S_{\mathrm{index}}$ for a given active region. This model can only give an approximate estimate because
\begin{itemize}
\item we were unable to use the pseudo-continuum at 3900\,\AA\,to measure the $S_{\mathrm{index}}$ because this pseudo-continuum is absent from the NSO solar spectral atlases we used,
\item we did not consider that the flux in the core of the \ion{Ca}{ii} H and K lines would need to be multiplied by a triangular function to match the Mount Wilson survey observations \citep[][]{Vaughan-1978},
\item we used in this simulation the NSO solar spectral atlas of a spot to model the \ion{Ca}{ii} H and K emission in a plage. These different active regions have different magnetic field strengths, therefore different emission in the chromosphere,
and\item the magnetic field strength varies from one active region to another, implying different emission in the core of the \ion{Ca}{ii} H and K lines.
\end{itemize}
However, important for the RV fitting challenge is not the maximum amplitude of the variation observed in $S_{\mathrm{index}}$, but its variation as a function of time. This time variation is assumed to be correctly taken into account because the value of $S_{\mathrm{index}}$ changes with the projection of the active region toward the line of sight, and its variation is symmetric with respect to the stellar disk center (see Fig. \ref{fig:2-1-0-0}).

\subsubsection{Simulating the RV short- and long-term activity signals of the Sun} \label{sect:2-1-2}

In the preceding section, we showed that we are able to estimate the effect of active regions on the different observables obtained from HARPS-like spectrographs using SOAP 2.0. It is not possible,
however, to consider the appearance and size evolution of active regions with SOAP 2.0, which is necessary to model the activity of Sun-like stars. To do so, we used a model developed in \citet{Dumusque-2011b}. In this previous work, the authors used empirical laws from solar observations to simulate the appearance of spot groups on the solar surface because spots tend to appear in groups. Here, we improve this model by also considering plages, which are generally found to surround spots. In the new model, a plage group will surround a spot group to form what we  refer to as an active region group (ARG).

Considering an ARG is possible because the amplitude of the variations observed in RV, BIS SPAN, FWHM, and \logrhk\,are proportional to the active region surface area \citep[see Fig. 8 in][and Sect. \ref{sect:2-1-0} here for the calcium activity index]{Dumusque-2014b}. This ensures that considering the individual effect of ten active regions covering an ARG and then summing their contribution, or simply considering the effect of the ARG, will give the same results. 

We present here a summary of the empirical laws derived from solar observations that we used to simulate the appearance of ARGs on the surface of a star. For more details we refer to \citet{Dumusque-2011b}. We note that the goal here is not to simulate the activity of the Sun, as this can be extremely complex and requires sophisticated magnetohydrodynamic models like CO5BOLD \citep[][]{Freytag-2012}, but to model in a simple way the observed phenomena that lead to RV variations.\\

\emph{Appearance and evolution of active region groups}\\

We followed \citet{Dumusque-2011b} to simulate the appearance and evolution of spot groups and started with an empirical law from solar observations that describes their lifetime distribution \citep[][]{Howard-2000}: 
\begin{itemize}
\item 50\% of the spot groups have lifetimes of between 1 h and 2 days,
\item 40\% of the spot groups have lifetimes of between 2 days and 11 days, and
\item 10\% of the spot groups have lifetimes of between 11 days and 2 months.
\end{itemize}
For simplicity, we assumed that spot groups and plage regions surrounding them, forming what we call ARGs, are tightly related. Plage groups appear and disappear at the same time as spot groups and evolve in size in a similar way. Plage groups cover a surface area that is ten times larger than the area coverd
by spot groups \citep[][]{Chapman-2001}. 

The lifetime of an ARG is correlated to its maximum surface area \citep[][]{Howard-2000}. For a surface area measured in filling factors, that is, the surface area ratio between the active region and the visible hemisphere, the largest size of an ARG is given by%
\begin{equation}\label{eq:2-1-2-0}
f_{ARG,max}=10^{-4}T,
\end{equation}
where the maximum filling factor of the ARG $f_{ARG,max}$ is expressed in hemisphere, and the lifetime $T$ in days. As in \citet{Dumusque-2011b}, we did not consider small ARGs that last
shorter than one day and affect the RV signal only negligibly.

The appearance of ARGs is ruled by a Poisson law:
\begin{equation}\label{eq:2-1-2-1}
P[(N(t+\tau)-N(t))=k]=\frac{e^{-\lambda\tau}(\lambda\tau)^k}{k!} \qquad k=0, 1, \cdots,
\end{equation}
where $P$ is the appearance probability of an ARG, $N$ is the number of ARGs, $t$ is the time, $\tau$ is the time step, and $\lambda$ is the average appearance rate of ARGs per unit of time. When an ARG appears on the stellar surface, it starts with a null size that increases during the first third of the ARG lifetime up to its maximum size as defined by Eq. \ref{eq:2-1-2-0}. The size then decreases during the remaining two thirds of its lifetime.

Active region groups appear around preferred active longitudes, with a Gaussian distribution centered on 0 and a dispersion of 15 degrees. Compared to the work of \citet{Dumusque-2011b}, where 4, 5, or 6 active longitudes could exist, we here only considered the five-case scenario that breaks the symmetry between the northern and southern hemisphere.\\

\emph{Solar magnetic cycle}\\

The Sun has an 11-year magnetic cycle, during which the so-called \emph{\textup{sunspot number}} on the visible hemisphere varies from zero to $\sim$150-200, as seen for cycles\footnote{see \url{http://solarscience.msfc.nasa.gov/SunspotCycle.shtml}} 21,
22,
and 23. We note that because of difficulties in counting small spots inside spot groups, the \textup{sunspot number }does not correspond to the precise number of spots at a given time, but is defined as ten times the number of spot groups plus the number of individual spots. The average number of spots inside spot groups is $\sim$10, therefore the sunspot number\emph{} represents an average number of spots at a given time.

A solar magnetic cycle is characterized by a progressive increase in the number of ARGs until the maximum of the cycle is reached, followed by a progressive decrease until nearly no ARGs remain on the solar surface. During a magnetic cycle, it is also observed that ARGs migrate from high latitude, $\pm 30$ degrees at the beginning of the cycle, toward the equator \citep[][]{Maunder-1904,Lockyer-1904}. To include both of these effects in our simulation, we used a time-dependent average appearance rate of ARGs ($\lambda$ in Eq. \ref{eq:2-1-2-1}) and a time-dependent latitude:
{\small
\begin{eqnarray}\label{eq:2-1-2-2}
\lambda(t) &=& (\lambda_{\mathrm{max\,activity}}-0.5).\left[-0.5\,cos\left(2\pi\left[\frac{t}{P_{\mathrm{cycle}}}+\phi_0\right]\right)+0.5\right] + 0.5,  \nonumber \\ 
\theta(t) &=& \theta_{\mathrm{min\,activity}}.\left[1-\frac{1}{2\pi}.\left(2\pi\left[\frac{t}{P_{\mathrm{cycle}}}+\phi_0\right]\right).\mathrm{modulo}\left(2\pi\right)\right],
\end{eqnarray}
}
where $t$ is the time, $\lambda_{\mathrm{max\,activity}}$ is the appearance rate of ARGs per unit of time for the highest activity level, $P_{\mathrm{cycle}}$ is the period of the magnetic cycle, $\phi_0$ is the phase of the magnetic cycle at time 0 (between 0 and 1), and $\theta_{\mathrm{min\,activity}}=\pm30\,\mathrm{degrees}$ is the latitude at which ARGs appear at the beginning of the magnetic cycle. We note that we added $0.5\,\mathrm{ARG}.\mathrm{day}^{-1}$ to $\lambda(t)$ here to consider a low but non-zero appearance
probability of ARGs at the lowest activity level.

To reproduce an average number of 150 to 200 spots, as observed on the Sun for the last maximum activity level of cycle 21, 22,
and 23, we needed to consider\footnote{We note that the average appearance rate of ARGs per unit of time is slightly higher than twice the value reported in \citet{Dumusque-2011b}. This discrepancy is explained by the fact that we here counted the average appearance rate of ARGs per unit of time over the entire surface, while \citet{Dumusque-2011b} considered the average appearance rate of ARGs per unit of time over the visible hemisphere. This new definition makes more sense.} $\lambda_{\mathrm{max\,activity}} = 10.5\,\mathrm{ARGs}.\mathrm{day}^{-1}$. This value was selected after several simulations modeling solar maximum, that is, fixing $\lambda(t)$ to $\lambda_{\mathrm{max\,activity}}$ and considering $\theta(t) = \pm15\,\mathrm{degrees}$. Our simulation only gives us information about the total number of ARG or spot groups. To derive the total number of spots, we used as in \citet{Dumusque-2011b} the proportionality between the number of spots per group and the spot group surface area\footnote{From Fig. 4 in \citet{Hathaway-2008}, we derive that $N_{\mathrm{spot\,group}} = 3.3\,10^{2}\,S_{\mathrm{spot\,group}}$, where $N_{\mathrm{spot\,group}}$ is the number of spots in the spot group and $S_{\mathrm{spot\,group}}$ is the surface area of the spot group in percent of hemisphere.}, which is one tenth of the ARG surface area by construction.

Because ARGs migrate from high toward low latitudes, it is important to consider the observed solar differential rotation. We considered here an empirical law derived from solar observations by \citet{Howard-1996}:
\begin{equation}\label{eq:2-1-2-3}
\omega=A+B\sin^2\theta,
\end{equation}
where $\omega$ is the angular speed in $\mathrm{degree}.\mathrm{day}^{-1}$, $\theta$ is the latitude of the ARG, $A = 14.476 \pm 0.006\,\mathrm{degree}.\mathrm{day}^{-1}$ is the angular speed of the solar equator, and $B=-2.875 \pm 0.058\,\mathrm{day}^{-1}$ is the amplitude of the differential rotation. ARGs at higher latitudes therefore rotate more slowly than their lower latitude analogs.

\subsubsection{Solar activity model and comparison with observations}  \label{sect:2-1-3}

\begin{figure*}[t]
\begin{center}
\includegraphics[width=8.2cm]{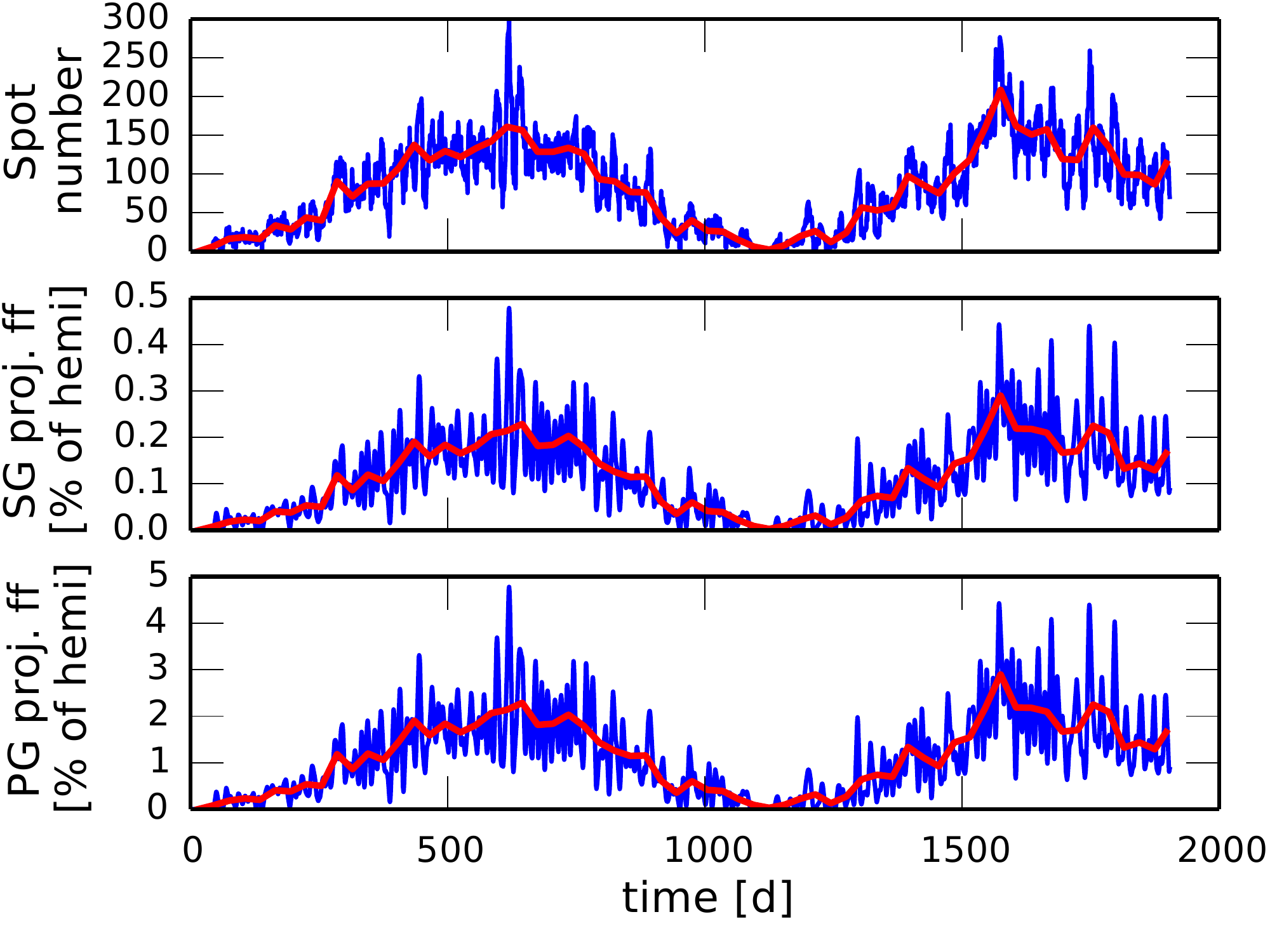}
\includegraphics[width=8.2cm]{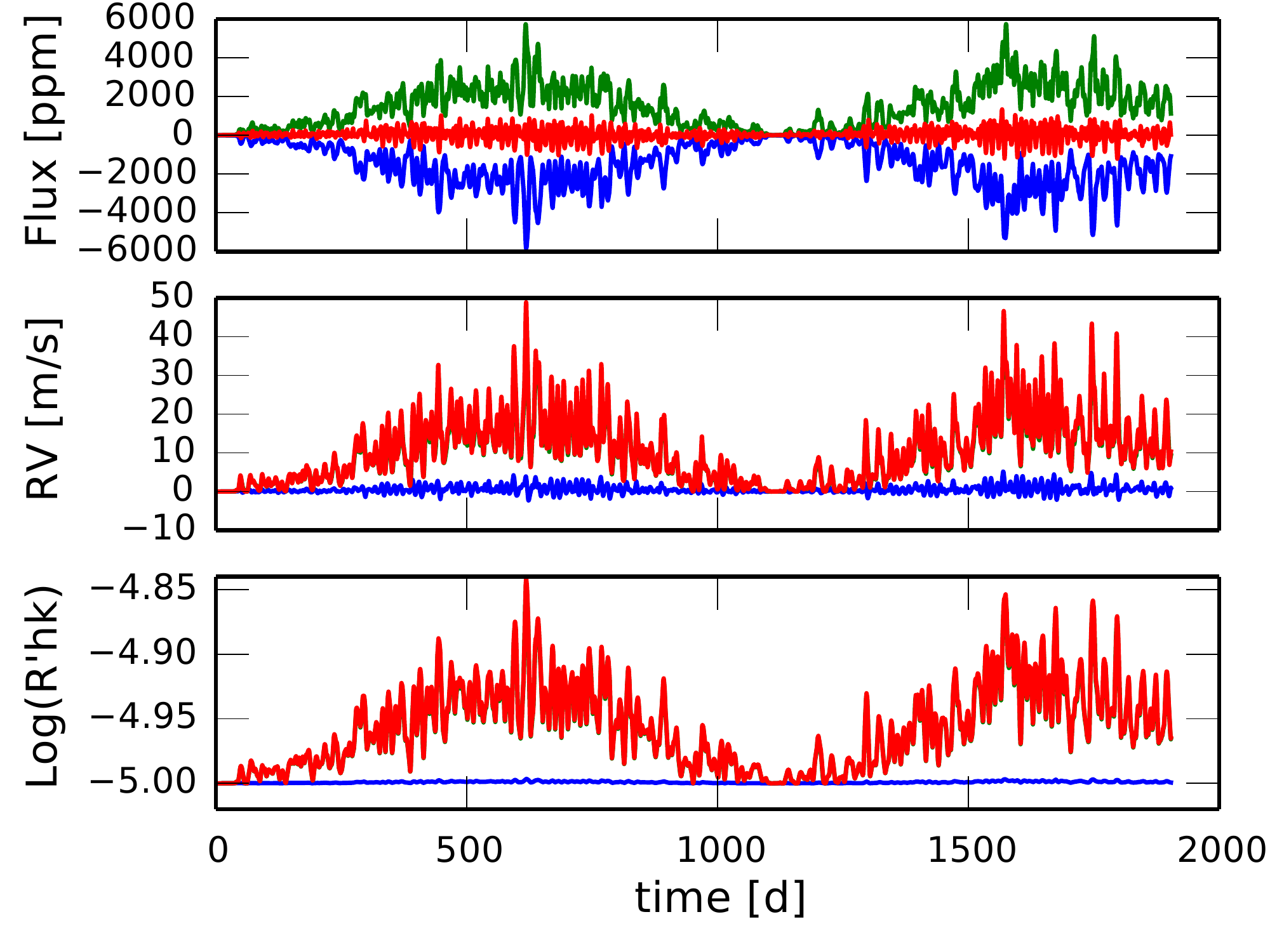}
\caption{Simulated solar-like magnetic cycle, with a period set to three years. {\emph Left:} Variation as a function of time of the total number of spots and of the projected filling factor of spot groups (SGs) and plage groups (PGs). The red line represents a monthly smoothing of the data. {\emph Right:} Flux, RV, and \logrhk\,variations as a function of time. Blue corresponds to the effect of spot groups, green to the effect of plage groups, and red to the combined effect. For the RV and the \logrhk, the red line overlaps the green one, implying that the variation of these observables is fully dominated by plage groups.}
\label{fig:2-1-2-0}
\end{center}
\end{figure*}

\begin{figure}[t]
\begin{center}
\includegraphics[width=8.2cm]{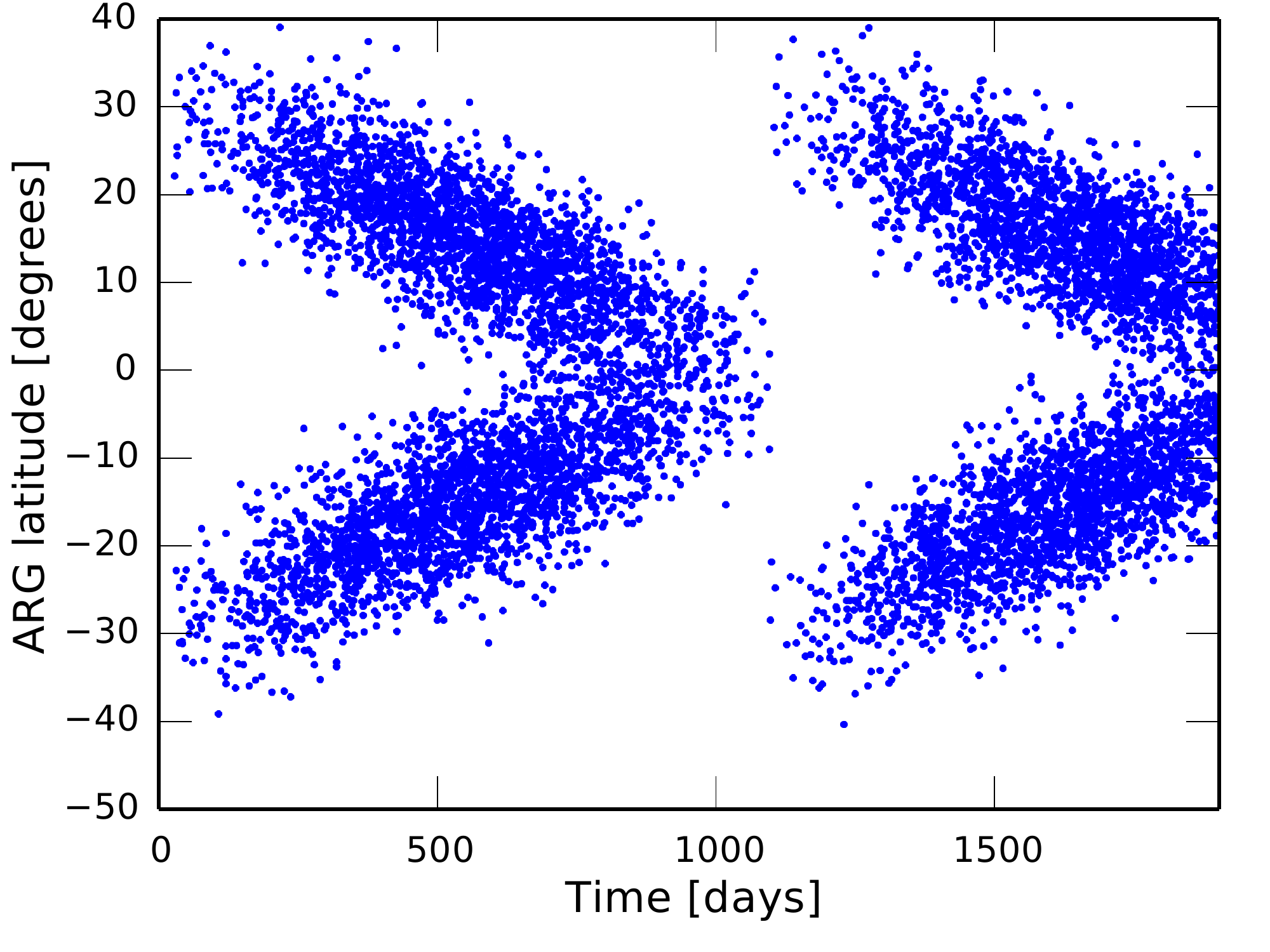}
\caption{Butterfly diagram showing the latitude of the ARGs as a function of time. During the three years of the magnetic cycle simulated here, the active regions migrate from a latitude of 30 degrees to the equator. The new cycle begins when ARGs again
appear at high latitude.}
\label{fig:2-1-2-1}
\end{center}
\end{figure}

\begin{figure}[t]
\begin{center}
\includegraphics[width=8.2cm]{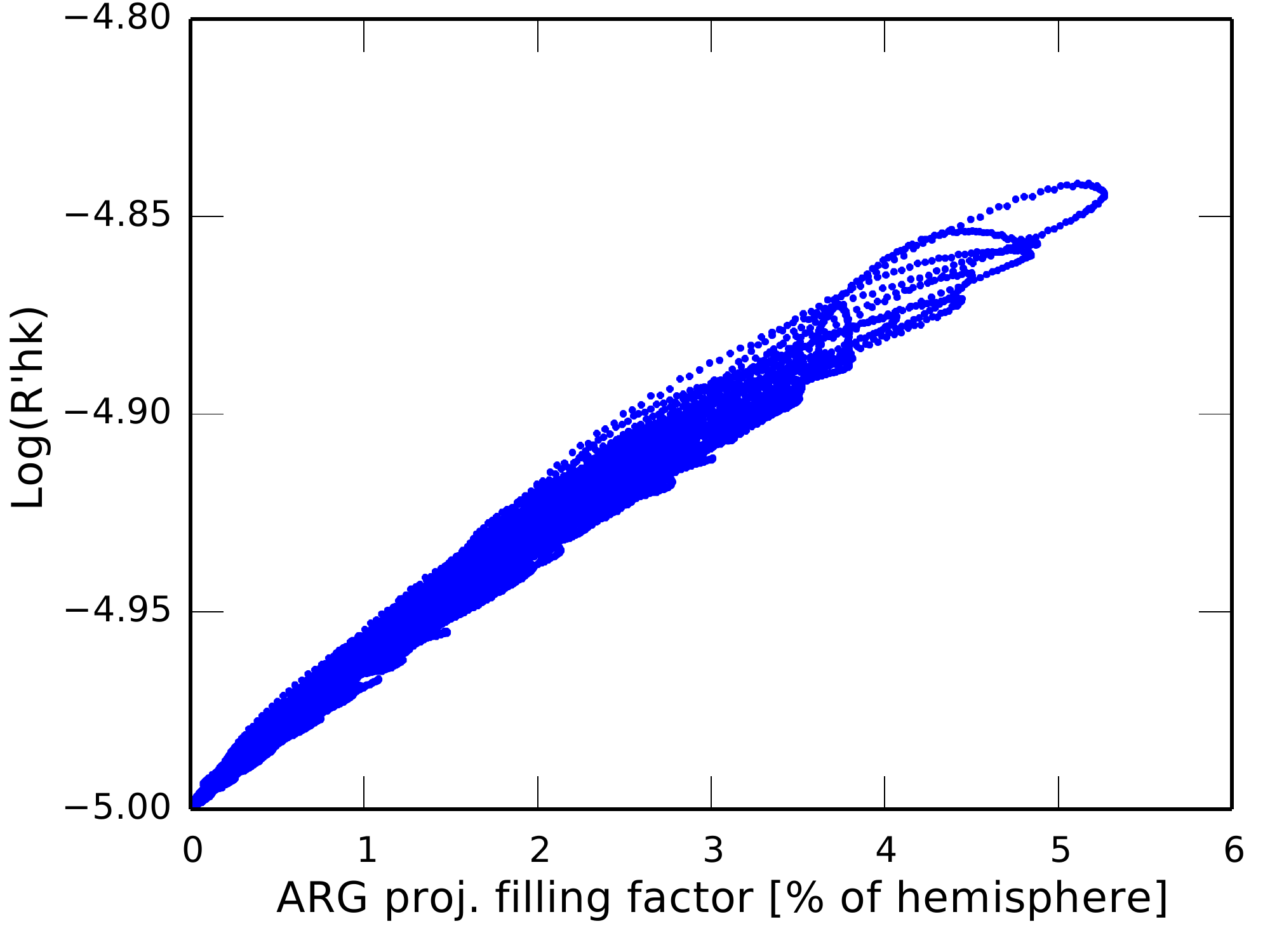}
\caption{\logrhk\,activity index as a function of the ARGs projected filling factor. Here we sum the contribution of all the ARGs including spots and plages.}
\label{fig:2-1-2-2}
\end{center}
\end{figure}

The physics included in our simulation of solar activity is rather simple, and we do not intend to model solar observations
precisely. For a more complex and realistic model, we refer to \citet{Borgniet-2015}. However, as the goal of this activity model is to simulate the activity of solar-type stars for the purpose of the RV fitting challenge, a first-order agreement with solar observations is sufficient. In Figs. \ref{fig:2-1-2-0}, \ref{fig:2-1-2-1}, and \ref{fig:2-1-2-2} we show the results of a five-year simulation of the Sun, with a magnetic cycle period set to three years.

We first compare the results of our simulation with the observed maximum number of spots, as well as the projected filling factor of spots and plages. In the left panel of Fig. \ref{fig:2-1-2-0} we show the evolution of the spot number and the variation in the projected filling factor for the spot and the plage groups. The number of spots on the visible hemisphere, averaged over a month, peaks at 160 and 208 for the first and second maxima of the magnetic cycle. For the same times, the projected filling factor of the spot groups is 0.23 and 0.29\%, and by construction, the projected filling factor of the plage groups is ten times larger.  \citet{Meunier-2010a} derived these projected filling factor values for cycle 23 using MDI/SOHO magnetograms, and \citet{Borgniet-2015} studied their monthly averages. The highest values found for cycle 23 are $\sim$0.2\% for the projected filling factor of spots and $\sim$3\% for the filling factor of plages. Considering these values, our simulation reproduces solar observations, although it may slightly underestimates the surface area covered by plages.

In the right panel of Fig. \ref{fig:2-1-2-0} we show the variations in flux, RV and \logrhk\,from our simulation\footnote{Our simulation is based on SOAP 2.0 and therefore return the $S_{\mathrm{index}}$ variation (see Sect. \ref{sect:2-1-0}). \logrhk\,is derived using Eq. \ref{eq:sect:2-1-0-0} \citep{Noyes-1984}, considering a solar B-V of 0.656. }. The flux effect induced by spots is nearly compensated for by plages, which can be explained by spots being fainter and plages being brighter than the solar quiet photosphere. For the RV, our simulation demonstrates that the observed variation is fully dominated by plages, as stated in several preceding studies of slow rotators like the Sun \citep[][]{Dumusque-2014,Meunier-2010a}. This is also expected for \logrhk\,because spots have a negligible surface area compared to plages, and they are much fainter. 

The peak-to-peak amplitude of the long-term RV variation in our simulated data set is about 20\ms, which is compatible with the amplitudes measured on a large sample of solar-like dwarfs \citep[][]{Lovis-2011b, Dumusque-2011}. We note, however, that the simulated \logrhk\,peak-to-peak variation is underestimated by a factor of two when compared to the Sun. It should be 0.2 dex, ranging between -5 and -4.8. As already discussed in Sect. \ref{sect:2-1-0}, our modeling of the $S_{\mathrm{index}}$ using SOAP 2.0, and therefore \logrhk, is missing some important information to correctly simulate the absolute variation observed in \logrhk. The different groups analyzing the data of the RV fitting challenge will use this observable to probe the presence of active regions,  to understand the RV red noise component that is induced by active regions, and to measure rotation periods by studying its periodicity (and not the estimated rotational period using the average level in \logrhk\,\citep[][]{Mamajek-2008, Noyes-1984}, which is incorrect in the simulated data). The correct variation amplitude is therefore not crucial, but the variation of \logrhk\,as a function of time should be modeled properly, which is the case.

Finally, we show in Fig. \ref{fig:2-1-2-1} the butterfly diagram, that is, the latitude of ARGs as a function of time. ARGs appear at a latitude of $\pm30$ degrees and migrate toward the equator during the course of a magnetic cycle. This simulated pattern agrees with the observations performed in cycle 23 \citep[see Fig. 3 in][]{Borgniet-2015}. We also show in Fig. \ref{fig:2-1-2-2} that \logrhk\,is proportional to the projected filling factor of ARGs, which is observed for the Sun \citep[][]{Meunier-2013}.

In conclusion, using simple empirical laws for the appearance and decay of ARGs on the solar surface, we are able to reproduce the activity observed in solar magnetic cycle 23 with good accuracy. Some more complex simulations are able to reproduce solar activity
more precisely \citep[][]{Borgniet-2015}, however, they use several additional parameters that can not be constrained for stars other
than the Sun. Because our goal is to simulate the activity of solar-type stars and then use the results to simulate RV observations for the RV fitting challenge, our simple activity simulation is sufficient.

\section{Data set for the RV fitting challenge} \label{sect:3}

The data set for the RV fitting challenge is composed of simulated and real planetary systems including instrumental noise and stellar signals. In this section we describe how these systems were generated and selected. The data of all these systems can be downloaded at CDS or on the \emph{wiki} of the RV fitting challenge at  \url{https://rv-challenge.wikispaces.com/Dataset+for+RV+challenge}.

\subsection{Simulated data} \label{sect:3-1}

To obtain realistic RV observations for the purpose of the RV fitting challenge, we first selected a realistic observational calendar and then generated RV measurements for these times, using the different models described above to simulate instrumental and stellar signals.

We selected the observational calendars of $\tau$\,Ceti (HD10700), HD192310, $\alpha$\,Cen\,B (HD128621), and Corot-7 from HARPS public available data\footnote{See the ESO archive: \url{http://archive.eso.org/cms.html}}. Except for Corot-7, these are the stars that have been the most
frequently followed using this instrument \citep[][]{Pepe-2011}. Given that those stars are observed on every possible night, over $\sim$200 nights allocated per year, it is difficult to obtain a better observational sampling using ground-based instruments \citep[however, see the CHIRON, the automatic planet finder, and the MINERVA RV facilities,][]{Tokovinin-2013,Vogt-2014a,Swift-2015b}.

To generate one set of data, we first selected an observational calendar from the available ones and simulated the expected signals induced by instrumental noise, stellar oscillations, granulation, and supergranulation for the same times, as explained in Sect. \ref{sect:2-0}. We then selected an 11-year period for the magnetic cycle of the star, the phase of the magnetic cycle for the first measurement in the calendar, and the rotation period of the star, and we simulated the stellar activity based on our solar model described in Sect. \ref{sect:2-1}. This model for solar activity will estimate the effect of active regions every two hours for the entire time span of the observations. We finally selected the points in the activity simulation that are closest in time to the calendar and added their contribution to the data that
were generated to include instrumental noise, stellar oscillations, granulation, and supergranulation. Because activity varies on the timescale of stellar rotation periods, longer than 20 days in all our cases, the maximum one-hour discrepancy in time that we can have between the activity model and the calendar is not relevant. In Fig. \ref{fig:2-0-2} we do not compare the rms of $\alpha$\,Cen\,B and Corot-7 to the simulated data because these stars present significant activity that will induce a higher rms for all observables in our ten day chunks. However, these stars probably have similar signals induced by granulation, stellar oscillations, and instrumental noise because they are close to $\tau$\,Ceti in spectral type.

In total, the data set for the RV fitting challenge includes ten simulated planetary systems: 1 to 5, 7, 8, 12, 13, and 15. In addition, before generating the data set for the RV fitting challenge, a test system was simulated so that the different teams ready to analyze the data of the RV fitting challenge could prepare themselves and test their method for recovering planetary signals despite stellar signals. This test data set included the RV, BIS SPAN, FWHM, and \logrhk\,variations induced by a planet with a 16-day period and stellar signals, like for the other systems of the RV fitting challenge. However, in addition, this test data set also included the RV, BIS SPAN, FWHM, and \logrhk\,variations of each individual signal component, that
is, the planetary signal, the oscillation, granulation, and supergranulation signals, the instrumental noise, and the activity signal. This test data can be downloaded at \url{https://rv-challenge.wikispaces.com/Test+dataset}.

\subsection{Real data}  \label{sect:3-2}

To test the realism of the simulated data, we included real RV measurements obtained by HARPS in five systems of the RV fitting challenge. We show in Table \ref{data} that the HARPS publicly available RV measurements of HD192310 (system 6), $\alpha$\,Cen\,B (HD128621, systems 9,10, and11) and Corot-7 (system 14) are included. System 6 includes the real HARPS measurements of HD192310, including therefore planets b and c \citep[][]{Howard-2011, Pepe-2011}. We note that this system was rejected because of problems when generating the times series (see Sect. \ref{sect:3-3}). System 9 includes the real data set of $\alpha$\,Cen\,B that was used to detect the Earth-mass planet orbiting the star \citep[][]{Dumusque-2012}. We note, however, that the 50 \cms signal claimed as a planet at the time is probably an artifact that is due to sampling and the model used to mitigate stellar signals \citep[][]{Rajpaul-2016}. To prevent the different teams involved in the RV fitting challenge from recognizing the real data of $\alpha$\,Cen\,B, we shifted the starting date to JDB=2455000, inverted the time, added a Gaussian noise of 5 \cms, and changed the gamma velocity of the star. The data therefore looked different, but have the same stellar and planetary signals (if existing), only with a different phase. Systems 10 and 11 also used the real RV measurements of $\alpha$\,Cen\,B, but this time we added simulated planetary signals (see Sect. \ref{sect:3-3} and Table \ref{data}). Adding these extra signals changed the original data significantly, and we therefore did not modify the original data like for system 9. Finally, system 14 includes the real RV data of Corot-7 taken with HARPS, including thus two to three planetary signals \citep[the third planet might be due to activity, see][]{Haywood-2014,Queloz-2009}. As in system 9, we disguised the data so that the different teams analyzing the times series could not recognize them.

\subsection{Injected planetary systems}   \label{sect:3-3}

The goal of the RV fitting challenge is to assess at which precision planetary signals can be detected in RV measurements that are affected by stellar signals. This can only be done if we precisely know the planetary signals present in the data. We therefore injected planetary systems that were detected by
\emph{Kepler} \citep[][]{Mullally-2015} and by HARPS \citep[][]{Mayor-2011}
in our simulated RV measurements and also in a few real RV time-series. We summarize  the different planetary signals present in the
data set for the RV fitting challenge in Table \ref{data}. We note that no planetary signals were injected in systems 4, 8, 9, 13, and 14. We also note that for the real planetary systems i) we did not include all the detected planets, ii) we changed the published mass and eccentricity by less than 10\%, and iii) we chose a random time at periastron passage to prevent the participant of the RV fitting challenge from recognizing the planetary systems. Finally, in addition to detected planets, we also injected simulated planetary signals. We decided to inject them to test
\begin{itemize}
\item when techniques relying on Gaussian process regression or other red-noise models were suppressing planetary signals when the orbital period was close to stellar rotation or an harmonic,
\item the ability of any correction techniques to detect Earth twins, that is, Earth-mass planets in the habitable zone,
\item the effect of long-term trends when searching for close-in planets.
\end{itemize}

We rejected system 6 because we discovered a bug in the way we generated the RVs. The idea of this data set was to use the HARPS measurements of HD192310, to have real stellar and instrumental signals, and to add some addition planets to the already present signals of HD192310b and HD192310c. This data set was supposed to be similar to the simulated data set 7 for an additional comparison between real and simulated data. However, in the process of adding extra planets, we also added a simulated signal for HD192310b and c. The RV time series of system 6 therefore includes the real planetary signals of planet b and c, plus simulated signals for these two planets, with the same period, amplitude, and eccentricity as the real signals, but different argument of periastron and time of periastron passage. As the sum of two Keplerians with same period, amplitude and eccentricity, but different argument of periastron and time of periastron passage cannot be represented as one Keplerian\footnote{this is the case if the eccentricity is null, however}, we decided to reject this system from the data set of the RV fitting challenge. To discover the real and simulated planetary signals present in the data, the different teams were supposed to fit two Keplerians at the period of planet b and two at the period of planet c, which is possible but normally excluded for planetary system stability reasons. Rejecting system 6, we are left with fourteen data sets, containing a total of 45 injected planetary signals. In addition to those 45 planets, we have a few others, probably those present in the real data sets 9, 10, and 11 ($\alpha$\,Cen\,Bb) and those that are present in system 14 (Corot-7b, c, and perhaps d).

\section{Comparison between simulated and real data} \label{sect:4}

The goal of the RV fitting challenge is to test the efficiency of different methods in recovering planetary signals despite stellar signals. Testing the efficiency of different techniques on simulated data can be useless when those simulated data are not realistic.

To be able to test the realism of our simulated data, we simulated some systems as closely as possible to the real systems selected for the data set of the RV fitting challenge. Thus systems 9 and 13, 11 and 12, and 14 and 15, including real and simulated data, can be compared. In Figs. \ref{fig:annex0}, \ref{fig:annex1}, \ref{fig:annex2}, \ref{fig:annex3}. \ref{fig:annex4}, \ref{fig:annex5}, and \ref{fig:annex6}, we plot the RV as a function of time and the correlation between all the time series of each RV fitting challenge system. We can perform a first comparison between real and simulated systems by considering the correlation between the different time series.

Comparing real system 9 and simulated system 13 shown in Figs. \ref{fig:annex3} and \ref{fig:annex5}, we see that the peak-to-peak amplitude of the raw RV variation is 19 \ms\,for the real data when the RV drift induced by $\alpha$\,Cen\,A is removed, using the parameters found in \citet{Dumusque-2012}, while it is twice that for the simulated data, 41 \ms. When we remove the effect of the magnetic cycle in the RVs of both systems using a linear correlation with \logrhk\,\citep[][]{Meunier-2013}, the peak amplitude of the RV variation in the real and simulated data becomes very similar, 11 and 13 \ms, respectively. This implies that the amplitude of the simulated stellar signal induced by oscillations, granulation, supergranulation, and short-term activity is similar to what is observed in the real data. However, the simulated magnetic cycle induces a RV variation that is twice larger than what is observed. The real measurements used for system 9 are the data gathered by HARPS on $\alpha$\,Cen\,B, a K0 dwarf. It has been shown that for a similar amplitude magnetic cycle in \logrhk, the induced RV variation has a larger amplitude for G dwarfs than for K dwarfs because of the lower convective velocities in stars of later spectral type \citep[][]{Lovis-2011b, Dumusque-2011}. The simulation of the magnetic cycle presented in Sect. \ref{sect:2-1-2} is based on the Sun, a G2 dwarf, therefore it is not surprising that our simulation overestimates the RV effect of the magnetic cycle when simulating $\alpha$\,Cen\,B data. Equation 13 in \citet{Lovis-2011b} shows that the slope of the linear correlation between RV and \logrhk\,is 30.3 for the Sun ($T_{eff}=5778\,K$) and 15.1 for $\alpha$\,Cen\,B ($T_{eff}=5214\,K$ and considering solar metallicity). This implies that for the same magnetic cycle in \logrhk, the RV peak-to-peak amplitude will be twice as high for the Sun as for $\alpha$\,Cen\,B, exactly the factor of two observed here. A similar behavior can explain why the peak-to-peak amplitude in BIS SPAN and FWHM is larger for the simulated than the real measurements. The magnetic cycle effects on the RVs, BIS SPAN and FWHM can be modeled and corrected for using \logrhk, therefore this overestimation should not complicate the detection of planets in systems 13 compared to system 9. As already stated in Sect. \ref{sect:2-1-3}, this difference in amplitude is a problem if the different teams that analyze the data aim to estimate the rotational period using the average level in \logrhk\,\citep[][]{Mamajek-2008, Noyes-1984}. Comparing the correlations between all the parameters, we see very similar behaviors, with tighter correlations for the simulated data than for the real ones, however, which should not be the case because the simulated data have twice the dispersion
in this case. This can probably be explained by our simplistic modeling of short- and long-term stellar activity that considered that all spots and plages induce the same variation\footnote{as estimate from SOAP 2.0} that is just scaled to account for the active region size.

Comparing real system 11 and simulated system 12 shown in Figs. \ref{fig:annex4} and \ref{fig:annex5}, we arrive at similar conclusions as for system 9 and 13. This is expected because the only difference between system 11 from 9 and system 12 from 13 is the addition of planetary signals. These injected planetary signals all have a smaller semi-amplitude than the effects of stellar signals (see Table \ref{data}) and therefore do not contribute significantly to the correlations present in the data.

Comparing real system 14 and simulated system 15 shown in Fig. \ref{fig:annex6}, we directly see that the simulated \logrhk\,has values far below the measured \logrhk. The Corot-7 data were used to generate system 14. Corot-7 is a G9 dwarf \citep[][]{Leger-2009} far more active than the Sun. Our simulation based on the Sun is therefore not able to reproduce similar activity levels in \logrhk. When we simulated system 15, we artificially increased the level of stellar activity so that the peak-to-peak variation in the simulated data would be similar to the real data. It is therefore not possible to simply compare these two systems at this stage, but a trend is seen in the RVs of system 15 and a strong correlation can be seen between the RVs and the \logrhk, which is not the case in system 14. This is because we generated the data for system 15 as if it were a quiet solar-like star and then multiplied the RV, BIS SPAN, and FWHM time series to account for the higher dispersion seen in the active star Corot-7 compared to the quiet Sun. However, in this process, we also multiplied the magnetic cycle effect, which should not be done as the amplitude of the magnetic cycle is proportional to the velocity of convection and its inhibition in active regions, which is not related to the higher activity of Corot-7 that is due to its younger age.

A first comparison between real and simulated data shows that the simulation of stellar signals presented in this paper, although simple, manages to reproduce the RV, BIS SPAN, FWHM, and \logrhk\,variations observed in real measurements fairly well, mainly in terms of the observed correlation between the different time
series. A more detailed discussion is left for a forthcoming paper, which will present the result of the different teams that analyzed the data set of the RV fitting challenge \citep[][]{Dumusque-2016b}.

\section{Conclusion}  \label{sect:5}

We presented here the simulation of instrumental and stellar signals that led to the data set of the RV fitting challenge. This simulation, based on solar and stellar observations, allows generating realistic instrumental signals from HARPS and stellar signals induced by oscillations, granulation, and supergranulation, and magnetic activity on the rotation period and magnetic cycle timescales. The simulation is able to reproduce the variations seen in RV, BIS SPAN, FWHM, and \logrhk.

The simulation of instrumental and stellar signals was used to generate two-thirds of the 15 systems included in the data set
of the RV fitting challenge. The remaining third of the data set corresponds to real measurements obtained with HARPS. These real measurements are important for checking the realism of simulated systems, which is critical if we wish to draw valid conclusions from the results of the RV fitting challenge \citep[][]{Dumusque-2016b}. A first comparison between simulated and real data shows that the simulation of stellar signals presented in this paper manages to reproduce the correlation between the different observables given in the data set of the RV fitting challenge fairly well, that is, RV, BIS SPAN, FWHM, and \logrhk. A more detailed comparison is left for a forthcoming analysis \citep[][]{Dumusque-2016b}.

Two versions of the data set of the RV fitting challenge exist and can be downloaded at the CDS or on the RV fitting challenge \emph{wiki}. The first version is the blind-test data set given to the different teams to recover planetary signals embedded in stellar signals; it can be downloaded at \url{https://rv-challenge.wikispaces.com/Dataset+for+RV+challenge}. In this version, only the RV, BIS SPAN, FWHM, and \logrhk\,variations are given without extra information. The second version includes the variation of the same observables, but separately for each signal component present in the data, in addition to stellar and planet properties; it can be downloaded at \url{https://rv-challenge.wikispaces.com/Details+about+the+dataset}. It is therefore possible to use this more detailed version to check \emph{\textup{a posteriori}} that planetary signals recovered in the data correspond to true signals, that models to fit activity are indeed adjusting the correct activity component, and so on. This more detailed version of the data set of the RV fitting challenge was given to the different teams after they completed
their analysis of the blind-test data set.

We hope that in addition to the different teams that participated in the analysis of the data set of the RV fitting challenge, other teams will use these data to search for new methods for recovering planetary signals embedded in stellar signals. Determining the best methods is critical for the immediate future, when TESS \citep[][]{Ricker-2014} and PLATO \citep[][]{Rauer-2014} will deliver hundreds of good candidates for RV follow-up with ultra-precise spectrographs such as ESPRESSO \citep[][]{Pepe-2014}, G-CLEF \citep[][]{Fzresz-2014}, and EXPRES (PI: D. A. Fischer and C. Jurgenson).

\begin{acknowledgements}
 X. Dumusque is extremely grateful to S. Aigrain, G. Anglada-Escud\'e, G. Bou\'e, A. Correia, R. D\'iaz, D. A. Fischer, E. B. Ford, P. C. Gregory, A. Hatzes, J. S. Jenkins, A. Santerne, S. H. Saar, G. Scandariato, and M. Tuomi for interesting discussions about the design of the data set of the RV fitting challenge and the best way to organize it. X. Dumusque also acknowledges the Society in Science$-$The Branco Weiss Fellowship for its financial support.

\end{acknowledgements}

\bibliographystyle{aa}
\bibliography{dumusque_bibliography}

\begin{thebibliography}{118}
\expandafter\ifx\csname natexlab\endcsname\relax\def\natexlab#1{#1}\fi

\bibitem[{{Akimov} {et~al.}(1982){Akimov}, {Belkina}, \&
  {Dyatel}}]{Akimov-1982}
{Akimov}, L.~A., {Belkina}, I.~L., \& {Dyatel}, N.~P. 1982, \sovast, 26, 334

\bibitem[{{Allende Prieto} {et~al.}(2013){Allende Prieto}, {Koesterke},
  {Ludwig}, {Freytag}, \& {Caffau}}]{Allende-Prieto-2013}
{Allende Prieto}, C., {Koesterke}, L., {Ludwig}, H.-G., {Freytag}, B., \&
  {Caffau}, E. 2013, \aap, 550, A103

\bibitem[{{Andersen} {et~al.}(1994){Andersen}, {Leifsen}, \&
  {Toutain}}]{Andersen-1994}
{Andersen}, B.~N., {Leifsen}, T.~E., \& {Toutain}, T. 1994, \solphys, 152, 247

\bibitem[{{Anglada-Escud{\'e}} {et~al.}(2012){Anglada-Escud{\'e}}, {Arriagada},
  {Vogt}, {Rivera}, {Butler}, {Crane}, {Shectman}, {Thompson}, {Minniti},
  {Haghighipour}, {Carter}, {Tinney}, {Wittenmyer}, {Bailey}, {O'Toole},
  {Jones}, \& {Jenkins}}]{Anglada-Escude-2012a}
{Anglada-Escud{\'e}}, G., {Arriagada}, P., {Vogt}, S.~S., {et~al.} 2012, \apjl,
  751, L16

\bibitem[{{Anglada-Escud{\'e}} \& {Tuomi}(2015)}]{Anglada-Escude-2015}
{Anglada-Escud{\'e}}, G. \& {Tuomi}, M. 2015, ArXiv e-prints
  [\eprint[arXiv]{1503.01976}]

\bibitem[{{Arentoft} {et~al.}(2008){Arentoft}, {Kjeldsen}, {Bedding}, {Bazot},
  {Christensen-Dalsgaard}, {Dall}, {Karoff}, {Carrier}, {Eggenberger},
  {Sosnowska}, {Wittenmyer}, {Endl}, {Metcalfe}, {Hekker}, {Reffert}, {Butler},
  {Bruntt}, {Kiss}, {O'Toole}, {Kambe}, {Ando}, {Izumiura}, {Sato}, {Hartmann},
  {Hatzes}, {Bouchy}, {Mosser}, {Appourchaux}, {Barban}, {Berthomieu},
  {Garcia}, {Michel}, {Provost}, {Turck-Chi{\`e}ze}, {Marti{\'c}}, {Lebrun},
  {Schmitt}, {Bertaux}, {Bonanno}, {Benatti}, {Claudi}, {Cosentino}, {Leccia},
  {Frandsen}, {Brogaard}, {Glowienka}, {Grundahl}, \&
  {Stempels}}]{Arentoft-2008}
{Arentoft}, T., {Kjeldsen}, H., {Bedding}, T.~R., {et~al.} 2008, \apj, 687,
  1180

\bibitem[{{Asplund} {et~al.}(2000){Asplund}, {Nordlund}, {Trampedach}, {Allende
  Prieto}, \& {Stein}}]{Asplund-2000}
{Asplund}, M., {Nordlund}, {\AA}., {Trampedach}, R., {Allende Prieto}, C., \&
  {Stein}, R.~F. 2000, \aap, 359, 729

\bibitem[{{Balasubramaniam}(2002)}]{Balasubramaniam-2002}
{Balasubramaniam}, K.~S. 2002, \apj, 575, 553

\bibitem[{{Baliunas} {et~al.}(1995){Baliunas}, {Donahue}, {Soon}, {Horne},
  {Frazer}, {Woodard-Eklund}, {Bradford}, {Rao}, {Wilson}, {Zhang}, {Bennett},
  {Briggs}, {Carroll}, {Duncan}, {Figueroa}, {Lanning}, {Misch}, {Mueller},
  {Noyes}, {Poppe}, {Porter}, {Robinson}, {Russell}, {Shelton}, {Soyumer},
  {Vaughan}, \& {Whitney}}]{Baliunas-1995}
{Baliunas}, S.~L., {Donahue}, R.~A., {Soon}, W.~H., {et~al.} 1995, \apj, 438,
  269

\bibitem[{{Baluev}(2013)}]{Baluev-2013}
{Baluev}, R.~V. 2013, \mnras, 429, 2052

\bibitem[{{Baranne} {et~al.}(1996){Baranne}, {Queloz}, {Mayor}, {Adrianzyk},
  {Knispel}, {Kohler}, {Lacroix}, {Meunier}, {Rimbaud}, \&
  {Vin}}]{Baranne-1996}
{Baranne}, A., {Queloz}, D., {Mayor}, M., {et~al.} 1996, \aaps, 119, 373

\bibitem[{{Bazot} {et~al.}(2007){Bazot}, {Bouchy}, {Kjeldsen}, {Charpinet},
  {Laymand}, \& {Vauclair}}]{Bazot-2007}
{Bazot}, M., {Bouchy}, F., {Kjeldsen}, H., {et~al.} 2007, \aap, 470, 295

\bibitem[{{Bedding} {et~al.}(2001){Bedding}, {Butler}, {Kjeldsen}, {Baldry},
  {O'Toole}, {Tinney}, {Marcy}, {Kienzle}, \& {Carrier}}]{Bedding-2001}
{Bedding}, T.~R., {Butler}, R.~P., {Kjeldsen}, H., {et~al.} 2001, \apjl, 549,
  L105

\bibitem[{{Beeck} {et~al.}(2013{\natexlab{a}}){Beeck}, {Cameron}, {Reiners}, \&
  {Sch{\"u}ssler}}]{Beeck-2013a}
{Beeck}, B., {Cameron}, R.~H., {Reiners}, A., \& {Sch{\"u}ssler}, M.
  2013{\natexlab{a}}, \aap, 558, A48

\bibitem[{{Beeck} {et~al.}(2013{\natexlab{b}}){Beeck}, {Cameron}, {Reiners}, \&
  {Sch{\"u}ssler}}]{Beeck-2013b}
{Beeck}, B., {Cameron}, R.~H., {Reiners}, A., \& {Sch{\"u}ssler}, M.
  2013{\natexlab{b}}, \aap, 558, A49

\bibitem[{{Borgniet} {et~al.}(2015){Borgniet}, {Meunier}, \&
  {Lagrange}}]{Borgniet-2015}
{Borgniet}, S., {Meunier}, N., \& {Lagrange}, A.-M. 2015, \aap, 581, A133

\bibitem[{{Bouchy} {et~al.}(2005){Bouchy}, {Bazot}, {Santos}, {Vauclair}, \&
  {Sosnowska}}]{Bouchy-2005b}
{Bouchy}, F., {Bazot}, M., {Santos}, N.~C., {Vauclair}, S., \& {Sosnowska}, D.
  2005, \aap, 440, 609

\bibitem[{{Bouchy} \& {Carrier}(2001)}]{Bouchy-2001}
{Bouchy}, F. \& {Carrier}, F. 2001, \aap, 374, L5

\bibitem[{{Bouchy} {et~al.}(2001){Bouchy}, {Pepe}, \& {Queloz}}]{Bouchy-2001b}
{Bouchy}, F., {Pepe}, F., \& {Queloz}, D. 2001, \aap, 374, 733

\bibitem[{{Brandt} \& {Solanki}(1990)}]{Brandt-1990}
{Brandt}, P.~N. \& {Solanki}, S.~K. 1990, \aap, 231, 221

\bibitem[{{Campbell} {et~al.}(1988){Campbell}, {Walker}, \&
  {Yang}}]{Campbell-1988}
{Campbell}, B., {Walker}, G.~A.~H., \& {Yang}, S. 1988, \apj, 331, 902

\bibitem[{Cavallini(1985{\natexlab{a}})}]{Cavallini-1985a}
Cavallini, F.;~Ceppatelli, G. R.~A. 1985{\natexlab{a}}, \aap, 143, 116

\bibitem[{Cavallini(1985{\natexlab{b}})}]{Cavallini-1985b}
Cavallini, F.;~Ceppatelli, G. R.~A. 1985{\natexlab{b}}, \aap, 150, 256

\bibitem[{{Cegla} {et~al.}(2013){Cegla}, {Shelyag}, {Watson}, \&
  {Mathioudakis}}]{Cegla-2013}
{Cegla}, H.~M., {Shelyag}, S., {Watson}, C.~A., \& {Mathioudakis}, M. 2013,
  \apj, 763, 95

\bibitem[{{Chapman} {et~al.}(2001){Chapman}, {Cookson}, {Dobias}, \&
  {Walton}}]{Chapman-2001}
{Chapman}, G.~A., {Cookson}, A.~M., {Dobias}, J.~J., \& {Walton}, S.~R. 2001,
  \apj, 555, 462

\bibitem[{{de la Cruz Rodr{\'{\i}}guez} {et~al.}(2011){de la Cruz
  Rodr{\'{\i}}guez}, {Kiselman}, \& {Carlsson}}]{de-la-Cruz-Rodriguez-2011}
{de la Cruz Rodr{\'{\i}}guez}, J., {Kiselman}, D., \& {Carlsson}, M. 2011,
  \aap, 528, A113

\bibitem[{{Deming} \& {Plymate}(1994)}]{Deming-1994}
{Deming}, D. \& {Plymate}, C. 1994, \apj, 426, 382

\bibitem[{{Desort} {et~al.}(2007){Desort}, {Lagrange}, {Galland}, {Udry}, \&
  {Mayor}}]{Desort-2007}
{Desort}, M., {Lagrange}, A.-M., {Galland}, F., {Udry}, S., \& {Mayor}, M.
  2007, \aap, 473, 983

\bibitem[{{D{\'{\i}}az} {et~al.}(2016){D{\'{\i}}az}, {S{\'e}gransan}, {Udry},
  {Lovis}, {Pepe}, {Dumusque}, {Marmier}, {Alonso}, {Benz}, {Bouchy},
  {Coffinet}, {Collier Cameron}, {Deleuil}, {Figueira}, {Gillon}, {Lo Curto},
  {Mayor}, {Mordasini}, {Motalebi}, {Moutou}, {Pollacco}, {Pompei}, {Queloz},
  {Santos}, \& {Wyttenbach}}]{Diaz-2016}
{D{\'{\i}}az}, R.~F., {S{\'e}gransan}, D., {Udry}, S., {et~al.} 2016, \aap,
  585, A134

\bibitem[{{Dravins}(1982)}]{Dravins-1982}
{Dravins}, D. 1982, \araa, 20, 61

\bibitem[{{Dravins}(1985)}]{Dravins-1985}
{Dravins}, D. 1985, in Stellar Radial Velocities, ed. {A.~G.~D.~Philip \&
  D.~W.~Latham}, 311--320

\bibitem[{Dravins(1981)}]{Dravins-1981}
Dravins, D.;~Lindegren, L. N.~A. 1981, \aap, 96, 345

\bibitem[{{Dressing} {et~al.}(2015){Dressing}, {Charbonneau}, {Dumusque},
  {Gettel}, {Pepe}, {Collier Cameron}, {Latham}, {Molinari}, {Udry}, {Affer},
  {Bonomo}, {Buchhave}, {Cosentino}, {Figueira}, {Fiorenzano}, {Harutyunyan},
  {Haywood}, {Johnson}, {Lopez-Morales}, {Lovis}, {Malavolta}, {Mayor},
  {Micela}, {Motalebi}, {Nascimbeni}, {Phillips}, {Piotto}, {Pollacco},
  {Queloz}, {Rice}, {Sasselov}, {S{\'e}gransan}, {Sozzetti}, {Szentgyorgyi}, \&
  {Watson}}]{Dressing-2015}
{Dressing}, C.~D., {Charbonneau}, D., {Dumusque}, X., {et~al.} 2015, \apj, 800,
  135

\bibitem[{{Dumusque}(2014)}]{Dumusque-2014}
{Dumusque}, X. 2014, in American Astronomical Society Meeting Abstracts, Vol.
  224, American Astronomical Society Meeting Abstracts 224, 301.06

\bibitem[{{Dumusque} {et~al.}(2014){Dumusque}, {Boisse}, \&
  {Santos}}]{Dumusque-2014b}
{Dumusque}, X., {Boisse}, I., \& {Santos}, N.~C. 2014, \apj, 796, 132

\bibitem[{Dumusque {et~al.}(2016)Dumusque, Borsa, Damasso, Diaz, Gregory, Hara,
  Hatzes, Rajpaul, \& Tuomi}]{Dumusque-2016b}
Dumusque, X., Borsa, F., Damasso, M., {et~al.} 2016, submitted to A\&A

\bibitem[{{Dumusque} {et~al.}(2015){Dumusque}, {Glenday}, {Phillips},
  {Buchschacher}, {Collier Cameron}, {Cecconi}, {Charbonneau}, {Cosentino},
  {Ghedina}, {Latham}, {Li}, {Lodi}, {Lovis}, {Molinari}, {Pepe}, {Udry},
  {Sasselov}, {Szentgyorgyi}, \& {Walsworth}}]{Dumusque-2015b}
{Dumusque}, X., {Glenday}, A., {Phillips}, D.~F., {et~al.} 2015, \apjl, 814,
  L21

\bibitem[{{Dumusque} {et~al.}(2011{\natexlab{a}}){Dumusque}, {Lovis},
  {S{\'e}gransan}, {Mayor}, {Udry}, {Benz}, {Bouchy}, {Lo Curto}, {Mordasini},
  {Pepe}, {Queloz}, {Santos}, \& {Naef}}]{Dumusque-2011c}
{Dumusque}, X., {Lovis}, C., {S{\'e}gransan}, D., {et~al.} 2011{\natexlab{a}},
  \aap, 535, A55

\bibitem[{{Dumusque} {et~al.}(2011{\natexlab{b}}){Dumusque}, {Lovis}, {Udry},
  \& {Santos}}]{Dumusque-2011}
{Dumusque}, X., {Lovis}, C., {Udry}, S., \& {Santos}, N.~C. 2011{\natexlab{b}},
  in IAU Symposium, Vol. 276, IAU Symposium, ed. A.~{Sozzetti}, M.~G.
  {Lattanzi}, \& A.~P. {Boss}, 530--532

\bibitem[{{Dumusque} {et~al.}(2012){Dumusque}, {Pepe}, {Lovis}, {Segransan},
  {Sahlmann}, {Benz}, {Bouchy}, {Mayor}, {Queloz}, {Santos}, \&
  {Udry}}]{Dumusque-2012}
{Dumusque}, X., {Pepe}, F., {Lovis}, C., {et~al.} 2012, \nat, 491, 207

\bibitem[{{Dumusque} {et~al.}(2011{\natexlab{c}}){Dumusque}, {Santos}, {Udry},
  {Lovis}, \& {Bonfils}}]{Dumusque-2011b}
{Dumusque}, X., {Santos}, N.~C., {Udry}, S., {Lovis}, C., \& {Bonfils}, X.
  2011{\natexlab{c}}, \aap, 527, A82

\bibitem[{{Dumusque} {et~al.}(2011{\natexlab{d}}){Dumusque}, {Udry}, {Lovis},
  {Santos}, \& {Monteiro}}]{Dumusque-2011a}
{Dumusque}, X., {Udry}, S., {Lovis}, C., {Santos}, N.~C., \& {Monteiro},
  M.~J.~P.~F.~G. 2011{\natexlab{d}}, \aap, 525, A140

\bibitem[{{Evans} \& {Michard}(1962)}]{Evans-1962}
{Evans}, J.~W. \& {Michard}, R. 1962, \apj, 136, 493

\bibitem[{{Feroz} \& {Hobson}(2014)}]{Feroz-2014}
{Feroz}, F. \& {Hobson}, M.~P. 2014, \mnras, 437, 3540

\bibitem[{{Frazier}(1971)}]{Frazier-1971}
{Frazier}, E.~N. 1971, \solphys, 21, 42

\bibitem[{{Freytag} {et~al.}(2012){Freytag}, {Steffen}, {Ludwig},
  {Wedemeyer-B{\"o}hm}, {Schaffenberger}, \& {Steiner}}]{Freytag-2012}
{Freytag}, B., {Steffen}, M., {Ludwig}, H.-G., {et~al.} 2012, Journal of
  Computational Physics, 231, 919

\bibitem[{{F{\.z}r{\'e}sz} {et~al.}(2014){F{\.z}r{\'e}sz}, {Epps}, {Barnes},
  {Podgorski}, {Szentgyorgyi}, {Mueller}, {Baldwin}, {Bean}, {Bergner}, {Chun},
  {Crane}, {Evans}, {Evans}, {Foster}, {Gauron}, {Guzman}, {Hertz},
  {Jord{\'a}n}, {Kim}, {McCracken}, {Norton}, {Ordway}, {Park}, {Park},
  {Plummer}, {Uomoto}, \& {Yuk}}]{Fzresz-2014}
{F{\.z}r{\'e}sz}, G., {Epps}, H., {Barnes}, S., {et~al.} 2014, in \procspie,
  Vol. 9147, Ground-based and Airborne Instrumentation for Astronomy V, 91479G

\bibitem[{{Gregory}(2011)}]{Gregory-2011}
{Gregory}, P.~C. 2011, \mnras, 415, 2523

\bibitem[{{Gregory}(2012)}]{Gregory-2012}
{Gregory}, P.~C. 2012, ArXiv e-prints [\eprint[arXiv]{1212.4058}]

\bibitem[{{Hall} {et~al.}(2007){Hall}, {Lockwood}, \& {Skiff}}]{Hall-2007}
{Hall}, J.~C., {Lockwood}, G.~W., \& {Skiff}, B.~A. 2007, \aj, 133, 862

\bibitem[{{Harvey}(1984)}]{Harvey-1984}
{Harvey}, J.~W. 1984, {In probing the depths of star : the study of Solar
  oscillation from space} (ed. R.W. Noyes \& E.J. Rhodes Jr. (Pasadena,
  JPL/NASA)), 327

\bibitem[{{Hathaway} \& {Choudhary}(2008)}]{Hathaway-2008}
{Hathaway}, D.~H. \& {Choudhary}, D.~P. 2008, \solphys, 250, 269

\bibitem[{{Hatzes}(2016)}]{Hatzes-2016}
{Hatzes}, A.~P. 2016, \aap, 585, A144

\bibitem[{{Haywood} {et~al.}(2014){Haywood}, {Collier Cameron}, {Queloz},
  {Barros}, {Deleuil}, {Fares}, {Gillon}, {Lanza}, {Lovis}, {Moutou}, {Pepe},
  {Pollacco}, {Santerne}, {S{\'e}gransan}, \& {Unruh}}]{Haywood-2014}
{Haywood}, R.~D., {Collier Cameron}, A., {Queloz}, D., {et~al.} 2014, \mnras,
  443, 2517

\bibitem[{{Holzreuter} \& {Solanki}(2013)}]{Holzreuter-2013}
{Holzreuter}, R. \& {Solanki}, S.~K. 2013, \aap, 558, A20

\bibitem[{{Howard} {et~al.}(2011){Howard}, {Johnson}, {Marcy}, {Fischer},
  {Wright}, {Henry}, {Isaacson}, {Valenti}, {Anderson}, \&
  {Piskunov}}]{Howard-2011}
{Howard}, A.~W., {Johnson}, J.~A., {Marcy}, G.~W., {et~al.} 2011, \apj, 730, 10

\bibitem[{{Howard} {et~al.}(2013){Howard}, {Sanchis-Ojeda}, {Marcy}, {Johnson},
  {Winn}, {Isaacson}, {Fischer}, {Fulton}, {Sinukoff}, \&
  {Fortney}}]{Howard-2013b}
{Howard}, A.~W., {Sanchis-Ojeda}, R., {Marcy}, G.~W., {et~al.} 2013, Nature,
  503, 381

\bibitem[{{Howard}(2000)}]{Howard-2000}
{Howard}, R. 2000, {Sunspot Evolution}, ed. P.~{Murdin}

\bibitem[{{Howard}(1996)}]{Howard-1996}
{Howard}, R.~F. 1996, \araa, 34, 75

\bibitem[{{Isaacson} \& {Fischer}(2010)}]{Isaacson-2010}
{Isaacson}, H. \& {Fischer}, D. 2010, \apj, 725, 875

\bibitem[{{Kjeldsen} \& {Bedding}(1995)}]{Kjeldsen-1995}
{Kjeldsen}, H. \& {Bedding}, T.~R. 1995, \aap, 293, 87

\bibitem[{{Lefebvre} {et~al.}(2008){Lefebvre}, {Garc{\'{\i}}a},
  {Jim{\'e}nez-Reyes}, {Turck-Chi{\`e}ze}, \& {Mathur}}]{Lefebvre-2008}
{Lefebvre}, S., {Garc{\'{\i}}a}, R.~A., {Jim{\'e}nez-Reyes}, S.~J.,
  {Turck-Chi{\`e}ze}, S., \& {Mathur}, S. 2008, \aap, 490, 1143

\bibitem[{L{\'e}ger {et~al.}(2009)L{\'e}ger, Rouan, Schneider, Barge, Fridlund,
  Samuel, Ollivier, Guenther, Deleuil, Deeg, Auvergne, Alonso, Aigrain, \&
  Alapini}]{Leger-2009}
L{\'e}ger, A., Rouan, D., Schneider, J., {et~al.} 2009, \aap, 506, 287

\bibitem[{{Leighton} {et~al.}(1962){Leighton}, {Noyes}, \&
  {Simon}}]{Leighton-1962}
{Leighton}, R.~B., {Noyes}, R.~W., \& {Simon}, G.~W. 1962, \apj, 135, 474

\bibitem[{{Lemke} \& {Reiners}(2016)}]{Lemke-2016}
{Lemke}, U. \& {Reiners}, A. 2016, ArXiv e-prints [\eprint[arXiv]{1603.00470}]

\bibitem[{{Lindegren} \& {Dravins}(2003)}]{Lindegren-2003}
{Lindegren}, L. \& {Dravins}, D. 2003, \aap, 401, 1185

\bibitem[{{Livingston}(1982)}]{Livingston-1982}
{Livingston}, W.~C. 1982, \nat, 297, 208

\bibitem[{{Lockyer}(1904)}]{Lockyer-1904}
{Lockyer}, W.~J.~S. 1904, Proceedings of the Royal Society of London Series I,
  73, 142

\bibitem[{Lomb(1976)}]{Lomb-1976a}
Lomb, N.~R. 1976, Astrophysics and Space Science, 39, 447

\bibitem[{{Lovis} {et~al.}(2011){Lovis}, {Dumusque}, {Santos}, {Bouchy},
  {Mayor}, {Pepe}, {Queloz}, {S{\'e}gransan}, \& {Udry}}]{Lovis-2011b}
{Lovis}, C., {Dumusque}, X., {Santos}, N.~C., {et~al.} 2011, ArXiv e-prints
  [\eprint[arXiv]{1107.5325}]

\bibitem[{{Makarov}(2010)}]{Makarov-2010}
{Makarov}, V.~V. 2010, \apj, 715, 500

\bibitem[{{Mamajek} \& {Hillenbrand}(2008)}]{Mamajek-2008}
{Mamajek}, E.~E. \& {Hillenbrand}, L.~A. 2008, \apj, 687, 1264

\bibitem[{{Marti{\'c}} {et~al.}(1999){Marti{\'c}}, {Schmitt}, {Lebrun},
  {Barban}, {Connes}, {Bouchy}, {Michel}, {Baglin}, {Appourchaux}, \&
  {Bertaux}}]{Martic-1999}
{Marti{\'c}}, M., {Schmitt}, J., {Lebrun}, J.-C., {et~al.} 1999, \aap, 351, 993

\bibitem[{{Maunder}(1904)}]{Maunder-1904}
{Maunder}, E.~W. 1904, Popular Astronomy, 12, 616

\bibitem[{{Mayor} {et~al.}(2009){Mayor}, {Bonfils}, {Forveille}, {Delfosse},
  {Udry}, {Bertaux}, {Beust}, {Bouchy}, {Lovis}, {Pepe}, {Perrier}, {Queloz},
  \& {Santos}}]{Mayor-2009b}
{Mayor}, M., {Bonfils}, X., {Forveille}, T., {et~al.} 2009, \aap, 507, 487

\bibitem[{{Mayor} {et~al.}(2011){Mayor}, {Marmier}, {Lovis}, {Udry},
  {S{\'e}gransan}, {Pepe}, {Benz}, {Bertaux}, {Bouchy}, {Dumusque}, {Lo Curto},
  {Mordasini}, {Queloz}, \& {Santos}}]{Mayor-2011}
{Mayor}, M., {Marmier}, M., {Lovis}, C., {et~al.} 2011, ArXiv e-prints
  [\eprint[arXiv]{1109.2497}]

\bibitem[{{McMillan} {et~al.}(1993){McMillan}, {Moore}, {Perry}, \&
  {Smith}}]{McMillan-1993}
{McMillan}, R.~S., {Moore}, T.~L., {Perry}, M.~L., \& {Smith}, P.~H. 1993,
  \apj, 403, 801

\bibitem[{{Meunier} {et~al.}(2010){Meunier}, {Desort}, \&
  {Lagrange}}]{Meunier-2010a}
{Meunier}, N., {Desort}, M., \& {Lagrange}, A.-M. 2010, \aap, 512, A39

\bibitem[{{Meunier} \& {Lagrange}(2013)}]{Meunier-2013}
{Meunier}, N. \& {Lagrange}, A.-M. 2013, \aap, 551, A101

\bibitem[{{Meunier} {et~al.}(2015){Meunier}, {Lagrange}, {Borgniet}, \&
  {Rieutord}}]{Meunier-2015}
{Meunier}, N., {Lagrange}, A.-M., {Borgniet}, S., \& {Rieutord}, M. 2015, \aap,
  583, A118

\bibitem[{{Moutou} {et~al.}(2005){Moutou}, {Pont}, {Barge}, {Aigrain},
  {Auvergne}, {Blouin}, {Cautain}, {Erikson}, {Guis}, {Guterman}, {Irwin},
  {Lanza}, {Queloz}, {Rauer}, {Voss}, \& {Zucker}}]{Moutou-2005b}
{Moutou}, C., {Pont}, F., {Barge}, P., {et~al.} 2005, \aap, 437, 355

\bibitem[{{Mullally} {et~al.}(2015){Mullally}, {Coughlin}, {Thompson}, {Rowe},
  {Burke}, {Latham}, {Batalha}, {Bryson}, {Christiansen}, {Henze}, {Ofir},
  {Quarles}, {Shporer}, {Van Eylen}, {Van Laerhoven}, {Shah}, {Wolfgang},
  {Chaplin}, {Xie}, {Akeson}, {Argabright}, {Bachtell}, {Barclay}, {Borucki},
  {Caldwell}, {Campbell}, {Catanzarite}, {Cochran}, {Duren}, {Fleming},
  {Fraquelli}, {Girouard}, {Haas}, {He{\l}miniak}, {Howell}, {Huber}, {Larson},
  {Gautier}, {Jenkins}, {Li}, {Lissauer}, {McArthur}, {Miller}, {Morris},
  {Patil-Sabale}, {Plavchan}, {Putnam}, {Quintana}, {Ramirez}, {Silva Aguirre},
  {Seader}, {Smith}, {Steffen}, {Stewart}, {Stober}, {Still}, {Tenenbaum},
  {Troeltzsch}, {Twicken}, \& {Zamudio}}]{Mullally-2015}
{Mullally}, F., {Coughlin}, J.~L., {Thompson}, S.~E., {et~al.} 2015, \apjs,
  217, 31

\bibitem[{{Noyes} {et~al.}(1984){Noyes}, {Hartmann}, {Baliunas}, {Duncan}, \&
  {Vaughan}}]{Noyes-1984}
{Noyes}, R.~W., {Hartmann}, L.~W., {Baliunas}, S.~L., {Duncan}, D.~K., \&
  {Vaughan}, A.~H. 1984, \apj, 279, 763

\bibitem[{{Pall{\'e}} {et~al.}(2013){Pall{\'e}}, {Grundahl}, {Trivi{\~n}o
  Hage}, {Christensen-Dalsgaard}, {Frandsen}, {Garc{\'{\i}}a}, {Uytterhoeven},
  {Andersen}, {Rasmussen}, {S{\o}rensen}, {Kjeldsen}, {Spano}, {Nilsson},
  {Hartman}, {J{\o}rgensen}, {Skottfelt}, {Harps{\o}e}, \&
  {Andersen}}]{Palle-2013}
{Pall{\'e}}, P.~L., {Grundahl}, F., {Trivi{\~n}o Hage}, A., {et~al.} 2013,
  Journal of Physics Conference Series, 440, 012051

\bibitem[{{Palle} {et~al.}(1995){Palle}, {Jimenez}, {Perez Hernandez},
  {Regulo}, {Roca Cortes}, \& {Sanchez}}]{Palle-1995}
{Palle}, P.~L., {Jimenez}, A., {Perez Hernandez}, F., {et~al.} 1995, \apj, 441,
  952

\bibitem[{{Paulson} {et~al.}(2002){Paulson}, {Saar}, {Cochran}, \&
  {Hatzes}}]{Paulson-2002}
{Paulson}, D.~B., {Saar}, S.~H., {Cochran}, W.~D., \& {Hatzes}, A.~P. 2002,
  \aj, 124, 572

\bibitem[{{Pepe} {et~al.}(2013){Pepe}, {Cameron}, {Latham}, {Molinari}, {Udry},
  {Bonomo}, {Buchhave}, {Charbonneau}, {Cosentino}, {Dressing}, {Dumusque},
  {Figueira}, {Fiorenzano}, {Gettel}, {Harutyunyan}, {Haywood}, {Horne},
  {Lopez-Morales}, {Lovis}, {Malavolta}, {Mayor}, {Micela}, {Motalebi},
  {Nascimbeni}, {Phillips}, {Piotto}, {Pollacco}, {Queloz}, {Rice}, {Sasselov},
  {S{\'e}gransan}, {Sozzetti}, {Szentgyorgyi}, \& {Watson}}]{Pepe-2013}
{Pepe}, F., {Cameron}, A.~C., {Latham}, D.~W., {et~al.} 2013, \nat, 503, 377

\bibitem[{{Pepe} {et~al.}(2011){Pepe}, {Lovis}, {S{\'e}gransan}, {Benz},
  {Bouchy}, {Dumusque}, {Mayor}, {Queloz}, {Santos}, \& {Udry}}]{Pepe-2011}
{Pepe}, F., {Lovis}, C., {S{\'e}gransan}, D., {et~al.} 2011, \aap, 534, A58

\bibitem[{Pepe {et~al.}(2002)Pepe, Mayor, Galland, Naef, Queloz, Santos, Udry,
  \& Burnet}]{Pepe-2002a}
Pepe, F., Mayor, M., Galland, F., {et~al.} 2002, \aap, 388, 632

\bibitem[{{Pepe} {et~al.}(2014){Pepe}, {Molaro}, {Cristiani}, {Rebolo},
  {Santos}, {Dekker}, {M{\'e}gevand}, {Zerbi}, {Cabral}, {Di Marcantonio},
  {Abreu}, {Affolter}, {Aliverti}, {Allende Prieto}, {Amate}, {Avila},
  {Baldini}, {Bristow}, {Broeg}, {Cirami}, {Coelho}, {Conconi}, {Coretti},
  {Cupani}, {D'Odorico}, {De Caprio}, {Delabre}, {Dorn}, {Figueira}, {Fragoso},
  {Galeotta}, {Genolet}, {Gomes}, {Gonz{\'a}lez Hern{\'a}ndez}, {Hughes},
  {Iwert}, {Kerber}, {Landoni}, {Lizon}, {Lovis}, {Maire}, {Mannetta},
  {Martins}, {Monteiro}, {Oliveira}, {Poretti}, {Rasilla}, {Riva}, {Santana
  Tschudi}, {Santos}, {Sosnowska}, {Sousa}, {Span{\'o}}, {Tenegi}, {Toso},
  {Vanzella}, {Viel}, \& {Zapatero Osorio}}]{Pepe-2014}
{Pepe}, F., {Molaro}, P., {Cristiani}, S., {et~al.} 2014, ArXiv e-prints
  [\eprint[arXiv]{1401.5918}]

\bibitem[{{Queloz} {et~al.}(2009){Queloz}, {Bouchy}, {Moutou}, {Hatzes},
  {H{\'e}brard}, {Alonso}, {Auvergne}, {Baglin}, {Barbieri}, {Barge}, {Benz},
  {Bord{\'e}}, {Deeg}, {Deleuil}, {Dvorak}, {Erikson}, {Ferraz Mello},
  {Fridlund}, {Gandolfi}, {Gillon}, {Guenther}, {Guillot}, {Jorda}, {Hartmann},
  {Lammer}, {L{\'e}ger}, {Llebaria}, {Lovis}, {Magain}, {Mayor}, {Mazeh},
  {Ollivier}, {P{\"a}tzold}, {Pepe}, {Rauer}, {Rouan}, {Schneider},
  {Segransan}, {Udry}, \& {Wuchterl}}]{Queloz-2009}
{Queloz}, D., {Bouchy}, F., {Moutou}, C., {et~al.} 2009, \aap, 506, 303

\bibitem[{{Queloz} {et~al.}(2001){Queloz}, {Henry}, {Sivan}, {Baliunas},
  {Beuzit}, {Donahue}, {Mayor}, {Naef}, {Perrier}, \& {Udry}}]{Queloz-2001}
{Queloz}, D., {Henry}, G.~W., {Sivan}, J.~P., {et~al.} 2001, \aap, 379, 279

\bibitem[{{Rajpaul} {et~al.}(2016){Rajpaul}, {Aigrain}, \&
  {Roberts}}]{Rajpaul-2016}
{Rajpaul}, V., {Aigrain}, S., \& {Roberts}, S. 2016, \mnras, 456, L6

\bibitem[{{Rauer} {et~al.}(2014){Rauer}, {Catala}, {Aerts}, {Appourchaux},
  {Benz}, {Brandeker}, {Christensen-Dalsgaard}, {Deleuil}, {Gizon}, {Goupil},
  {G{\"u}del}, {Janot-Pacheco}, {Mas-Hesse}, {Pagano}, {Piotto}, {Pollacco},
  {Santos}, {Smith}, {Su{\'a}rez}, {Szab{\'o}}, {Udry}, {Adibekyan}, {Alibert},
  {Almenara}, {Amaro-Seoane}, {Ammer-von Eiff}, {Asplund}, {Antonello},
  {Barnes}, {Baudin}, {Belkacem}, {Bergemann}, {Bihain}, {Birch}, {Bonfils},
  {Boisse}, {Bonomo}, {Borsa}, {Brand{\~a}o}, {Brocato}, {Brun}, {Burleigh},
  {Burston}, {Cabrera}, {Cassisi}, {Chaplin}, {Charpinet}, {Chiappini},
  {Church}, {Csizmadia}, {Cunha}, {Damasso}, {Davies}, {Deeg}, {D{\'{\i}}az},
  {Dreizler}, {Dreyer}, {Eggenberger}, {Ehrenreich}, {Eigm{\"u}ller},
  {Erikson}, {Farmer}, {Feltzing}, {Oliveira Fialho}, {Figueira}, {Forveille},
  {Fridlund}, {Garc{\'{\i}}a}, {Giommi}, {Giuffrida}, {Godolt}, {Gomes da
  Silva}, {Granzer}, {Grenfell}, {Grotsch-Noels}, {G{\"u}nther}, {Haswell},
  {Hatzes}, {H{\'e}brard}, {Hekker}, {Helled}, {Heng}, {Jenkins}, {Johansen},
  {Khodachenko}, {Kislyakova}, {Kley}, {Kolb}, {Krivova}, {Kupka}, {Lammer},
  {Lanza}, {Lebreton}, {Magrin}, {Marcos-Arenal}, {Marrese}, {Marques},
  {Martins}, {Mathis}, {Mathur}, {Messina}, {Miglio}, {Montalban}, {Montalto},
  {Monteiro}, {Moradi}, {Moravveji}, {Mordasini}, {Morel}, {Mortier},
  {Nascimbeni}, {Nelson}, {Nielsen}, {Noack}, {Norton}, {Ofir}, {Oshagh},
  {Ouazzani}, {P{\'a}pics}, {Parro}, {Petit}, {Plez}, {Poretti}, {Quirrenbach},
  {Ragazzoni}, {Raimondo}, {Rainer}, {Reese}, {Redmer}, {Reffert},
  {Rojas-Ayala}, {Roxburgh}, {Salmon}, {Santerne}, {Schneider}, {Schou},
  {Schuh}, {Schunker}, {Silva-Valio}, {Silvotti}, {Skillen}, {Snellen}, {Sohl},
  {Sousa}, {Sozzetti}, {Stello}, {Strassmeier}, {{\v S}vanda}, {Szab{\'o}},
  {Tkachenko}, {Valencia}, {Van Grootel}, {Vauclair}, {Ventura}, {Wagner},
  {Walton}, {Weingrill}, {Werner}, {Wheatley}, \& {Zwintz}}]{Rauer-2014}
{Rauer}, H., {Catala}, C., {Aerts}, C., {et~al.} 2014, Experimental Astronomy
  [\eprint[arXiv]{1310.0696}]

\bibitem[{{Ricker} {et~al.}(2014){Ricker}, {Winn}, {Vanderspek}, {Latham},
  {Bakos}, {Bean}, {Berta-Thompson}, {Brown}, {Buchhave}, {Butler}, {Butler},
  {Chaplin}, {Charbonneau}, {Christensen-Dalsgaard}, {Clampin}, {Deming},
  {Doty}, {De Lee}, {Dressing}, {Dunham}, {Endl}, {Fressin}, {Ge}, {Henning},
  {Holman}, {Howard}, {Ida}, {Jenkins}, {Jernigan}, {Johnson}, {Kaltenegger},
  {Kawai}, {Kjeldsen}, {Laughlin}, {Levine}, {Lin}, {Lissauer}, {MacQueen},
  {Marcy}, {McCullough}, {Morton}, {Narita}, {Paegert}, {Palle}, {Pepe},
  {Pepper}, {Quirrenbach}, {Rinehart}, {Sasselov}, {Sato}, {Seager},
  {Sozzetti}, {Stassun}, {Sullivan}, {Szentgyorgyi}, {Torres}, {Udry}, \&
  {Villasenor}}]{Ricker-2014}
{Ricker}, G.~R., {Winn}, J.~N., {Vanderspek}, R., {et~al.} 2014, in Society of
  Photo-Optical Instrumentation Engineers (SPIE) Conference Series, Vol. 9143,
  Society of Photo-Optical Instrumentation Engineers (SPIE) Conference Series,
  20

\bibitem[{{Robertson} {et~al.}(2014){Robertson}, {Mahadevan}, {Endl}, \&
  {Roy}}]{Robertson-2014}
{Robertson}, P., {Mahadevan}, S., {Endl}, M., \& {Roy}, A. 2014, Science, 345,
  440

\bibitem[{{Saar} \& {Donahue}(1997)}]{Saar-1997b}
{Saar}, S.~H. \& {Donahue}, R.~A. 1997, \apj, 485, 319

\bibitem[{{Sanchis-Ojeda} {et~al.}(2013){Sanchis-Ojeda}, {Rappaport}, {Winn},
  {Levine}, {Kotson}, {Latham}, \& {Buchhave}}]{Sanchis-Ojeda-2013}
{Sanchis-Ojeda}, R., {Rappaport}, S., {Winn}, J.~N., {et~al.} 2013, \apj, 774,
  54

\bibitem[{{Santos} {et~al.}(2010){Santos}, {Gomes da Silva}, {Lovis}, \&
  {Melo}}]{Santos-2010a}
{Santos}, N.~C., {Gomes da Silva}, J., {Lovis}, C., \& {Melo}, C. 2010, \aap,
  511, A54+

\bibitem[{{Scargle}(1982)}]{Scargle-1982}
{Scargle}, J.~D. 1982, \apj, 263, 835

\bibitem[{{Schrijver} {et~al.}(1989){Schrijver}, {Cote}, {Zwaan}, \&
  {Saar}}]{Schrijver-1989}
{Schrijver}, C.~J., {Cote}, J., {Zwaan}, C., \& {Saar}, S.~H. 1989, \apj, 337,
  964

\bibitem[{{Strassmeier} {et~al.}(2015){Strassmeier}, {Ilyin}, {J{\"a}rvinen},
  {Weber}, {Woche}, {Barnes}, {Bauer}, {Beckert}, {Bittner}, {Bredthauer},
  {Carroll}, {Denker}, {Dionies}, {DiVarano}, {D{\"o}scher}, {Fechner},
  {Feuerstein}, {Granzer}, {Hahn}, {Harnisch}, {Hofmann}, {Lesser}, {Paschke},
  {Pankratow}, {Plank}, {Pl{\"u}schke}, {Popow}, {Sablowski}, \&
  {Storm}}]{Strassmeier-2015}
{Strassmeier}, K.~G., {Ilyin}, I., {J{\"a}rvinen}, A., {et~al.} 2015, ArXiv
  e-prints [\eprint[arXiv]{1505.06492}]

\bibitem[{{Swift} {et~al.}(2015){Swift}, {Bottom}, {Johnson}, {Wright},
  {McCrady}, {Wittenmyer}, {Plavchan}, {Riddle}, {Muirhead}, {Herzig}, {Myles},
  {Blake}, {Eastman}, {Beatty}, {Barnes}, {Gibson}, {Lin}, {Zhao}, {Gardner},
  {Falco}, {Criswell}, {Nava}, {Robinson}, {Sliski}, {Hedrick}, {Ivarsen},
  {Hjelstrom}, {de Vera}, \& {Szentgyorgyi}}]{Swift-2015b}
{Swift}, J.~J., {Bottom}, M., {Johnson}, J.~A., {et~al.} 2015, Journal of
  Astronomical Telescopes, Instruments, and Systems, 1, 027002

\bibitem[{{Teixeira} {et~al.}(2009){Teixeira}, {Kjeldsen}, {Bedding}, {Bouchy},
  {Christensen-Dalsgaard}, {Cunha}, {Dall}, {Frandsen}, {Karoff}, {Monteiro},
  \& {Pijpers}}]{Teixeira-2009}
{Teixeira}, T.~C., {Kjeldsen}, H., {Bedding}, T.~R., {et~al.} 2009, \aap, 494,
  237

\bibitem[{{Tokovinin} {et~al.}(2013){Tokovinin}, {Fischer}, {Bonati},
  {Giguere}, {Moore}, {Schwab}, {Spronck}, \& {Szymkowiak}}]{Tokovinin-2013}
{Tokovinin}, A., {Fischer}, D.~A., {Bonati}, M., {et~al.} 2013, \pasp, 125,
  1336

\bibitem[{{Tuomi} {et~al.}(2013){Tuomi}, {Anglada-Escud{\'e}}, {Gerlach},
  {Jones}, {Reiners}, {Rivera}, {Vogt}, \& {Butler}}]{Tuomi-2013a}
{Tuomi}, M., {Anglada-Escud{\'e}}, G., {Gerlach}, E., {et~al.} 2013, \aap, 549,
  A48

\bibitem[{{Ulrich}(1970)}]{Ulrich-1970}
{Ulrich}, R.~K. 1970, \apj, 162, 993

\bibitem[{{Unruh} {et~al.}(1999){Unruh}, {Solanki}, \& {Fligge}}]{Unruh-1999}
{Unruh}, Y.~C., {Solanki}, S.~K., \& {Fligge}, M. 1999, \aap, 345, 635

\bibitem[{{Vaughan} {et~al.}(1978){Vaughan}, {Preston}, \&
  {Wilson}}]{Vaughan-1978}
{Vaughan}, A.~H., {Preston}, G.~W., \& {Wilson}, O.~C. 1978, \pasp, 90, 267

\bibitem[{{Vogt} {et~al.}(2012){Vogt}, {Butler}, \& {Haghighipour}}]{Vogt-2012}
{Vogt}, S.~S., {Butler}, R.~P., \& {Haghighipour}, N. 2012, Astronomische
  Nachrichten, 333, 561

\bibitem[{{Vogt} {et~al.}(2010){Vogt}, {Butler}, {Rivera}, {Haghighipour},
  {Henry}, \& {Williamson}}]{Vogt-2010b}
{Vogt}, S.~S., {Butler}, R.~P., {Rivera}, E.~J., {et~al.} 2010, \apj, 723, 954

\bibitem[{{Vogt} {et~al.}(2014){Vogt}, {Radovan}, {Kibrick}, {Butler},
  {Alcott}, {Allen}, {Arriagada}, {Bolte}, {Burt}, {Cabak}, {Chloros},
  {Cowley}, {Deich}, {Dupraw}, {Earthman}, {Epps}, {Faber}, {Fischer}, {Gates},
  {Hilyard}, {Holden}, {Johnston}, {Keiser}, {Kanto}, {Katsuki}, {Laiterman},
  {Lanclos}, {Laughlin}, {Lewis}, {Lockwood}, {Lynam}, {Marcy}, {McLean},
  {Miller}, {Misch}, {Peck}, {Pfister}, {Phillips}, {Rivera}, {Sandford},
  {Saylor}, {Stover}, {Thompson}, {Walp}, {Ward}, {Wareham}, {Wei}, \&
  {Wright}}]{Vogt-2014a}
{Vogt}, S.~S., {Radovan}, M., {Kibrick}, R., {et~al.} 2014, ArXiv e-prints
  [\eprint[arXiv]{1402.6684}]

\bibitem[{{Wallace} {et~al.}(1998){Wallace}, {Hinkle}, \&
  {Livingston}}]{Wallace-1998}
{Wallace}, L., {Hinkle}, K., \& {Livingston}, W. 1998, {An atlas of the
  spectrum of the solar photosphere from 13,500 to 28,000 cm-1 (3570 to 7405
  A)}

\bibitem[{{Wallace} {et~al.}(2005){Wallace}, {Hinkle}, \&
  {Livingston}}]{Wallace-2005}
{Wallace}, L., {Hinkle}, K., \& {Livingston}, W.~C. 2005, {An atlas of sunspot
  umbral spectra in the visible from 15,000 to 25,500 cm-1 (3920 to 6664
  {\AA})}

\bibitem[{{Wilson}(1963)}]{Wilson-1963}
{Wilson}, O.~C. 1963, \apj, 138, 832

\bibitem[{{Wilson}(1978)}]{Wilson-1978}
{Wilson}, O.~C. 1978, \apj, 226, 379

\bibitem[{{Wright}(2005)}]{Wright-2005}
{Wright}, J.~T. 2005, \pasp, 117, 657

\bibitem[{{Zechmeister} \& {K{\"u}rster}(2009)}]{Zechmeister-2009}
{Zechmeister}, M. \& {K{\"u}rster}, M. 2009, \aap, 496, 577

\end{thebibliography}

\newpage
\onecolumn

\begin{landscape}

\begin{longtab}
\begin{longtable}{ccccccccccccc}
\caption{\label{data} Properties of the data set of the  RV fitting challenge.}\\
\hline\hline
System & Star & Data & M star & Rotation & Magn. cycle & \multicolumn{7}{c}{Planet properties} \\
 & (calendar) & type & [M$_{\odot}$] & period [d] & phase & Planet & Period [d] & Mass [M$_{\oplus}$] & Ecc. & T transit [d] & Semi-amplitude [m/s] & $\omega$ [rad]  \\
\hline
\endfirsthead
\caption{continued.}\\
\hline\hline
System  & Star & Data & M star & Rotation & Magn. cycle & \multicolumn{7}{c}{Planet properties} \\
 & (calendar) & type & [M$_{\odot}$] & period [d] & phase & Planet & Period [d] & Mass [M$_{\oplus}$] & Ecc. & T0 [d] & Semi-amplitude [m/s] & $\omega$ [rad]  \\
\hline
\endhead
\hline
\endfoot
test & HD10700 & simulated & 0.783 & 25.05 & 0.0 & Dummy & 16.00 & 5.02 & 0.00 & 8.000000 & 1.5 & 0.00\\
\hline
1 & HD10700 & simulated & 0.783 & 25.05 & 0.0 & Kepler-11b & 9.89 & 4.13 & 0.10 & 55494.86566 & 1.45 & 3.73 \\
 & & & &  & & Kepler-11d & 23.37 & 6.28 & 0.12 & 55490.59677 & 1.67 & 2.55 \\
 & & & &  & & Kepler-11e & 33.28 & 8.74 & 0.08 & 55473.28821 & 2.05 & 0.23 \\
 & & & &  & & Kepler-11g & 112.46 & 2.38 & 0.21 & 55457.43153 & 0.38 & 4.36 \\
 & & & &  & & Dummy & 273.20 & 1.90 & 0.16 & 55293.88276 & 0.22 & 5.94 \\
\hline
2 & HD10700 & simulated & 0.783 & 25.05 & 0.0 & Kepler-20b & 3.77 & 5.68 & 0.05 & 55499.70529 & 2.75 & 5.51 \\
 & & & &  & & Kepler-20e & 5.79 & 0.63 & 0.11 & 55499.58933 & 0.27 & 3.68 \\
 & & & &  & & Kepler-20c & 10.64 & 8.24 & 0.14 & 55489.92296 & 2.85 & 0.53 \\
 & & & &  & & Kepler-20f & 20.16 & 1.23 & 0.08 & 55480.55466 & 0.34 & 1.78 \\
 & & & &  & & Kepler-20d & 75.28 & 7.41 & 0.19 & 55430.69600 & 1.35 & 1.06 \\
\hline
3 & HD10700 & simulated & 0.783 & 25.05 & 0.5 & HD10180b & 1.12 & 1.32 & 0.00 & 55498.92368 & 0.96 & 0.00 \\
 & & & &  & & HD10180d & 17.01 & 12.42 & 0.15 & 55488.48886 & 3.68 & 5.92 \\
 & & & &  & & Dummy & 26.30 & 1.50 & 0.08 & 55484.03573 & 0.38 & 4.95 \\
 & & & &  & & HD10180e & 48.75 & 24.89 & 0.06 & 55484.21874 & 5.14 & 5.92 \\
 & & & &  & & Dummy & 201.50 & 3.20 & 0.20 & 55423.40832 & 0.42 & 1.48 \\
 & & & &  & & HD10180g & 595.98 & 21.19 & 0.13 & 55122.54156 & 1.91 & 2.01 \\
 & & & &  & & HD10180h & 2315.44 & 67.26 & 0.16 & 54859.73755 & 3.87 & 5.55 \\
\hline
4 & HD10700 & simulated & 0.783 & 25.05 & 0.5 & - & - & - & - & - & - & -\\
\hline
5 & HD192310 & simulated & 0.800 & 40.00 & 0.5 & HD10700b & 14.66 & 2.10 & 0.17 & 55486.81922 & 0.65 & 2.92 \\
 & & & &  & & Dummy & 26.20 & 1.70 & 0.25 & 55481.40963 & 0.44 & 5.45 \\
 & & & &  & & HD10700c & 34.65 & 3.04 & 0.03 & 55467.37320 & 0.69 & 5.52 \\
 & & & &  & & HD10700e & 173.16 & 4.43 & 0.05 & 55421.08234 & 0.59 & 4.43 \\
 & & & &  & & Dummy & 283.10 & 3.50 & 0.30 & 55462.39275 & 0.41 & 4.10 \\
 & & & &  & & HD10700f & 616.32 & 6.34 & 0.03 & 55414.69085 & 0.55 & 1.45 \\
\hline
\bf{6} & \bf{HD192310} & \bf{real} & \bf{0.800} & \multicolumn{3}{c}{{\bf REJECTED}} & & & & & & \\
\hline
7 & HD192310 & simulated & 0.800 & 40.00 & 0.0 & Dummy & 38.32 & 1.54 & 0.02 & 55464.01402 & 0.34 & 2.46 \\
 & & & &  & & HD192310b & 72.48 & 16.39 & 0.13 & 55467.62326 & 2.94 & 4.14 \\
 & & & &  & & Dummy & 100.99 & 1.92 & 0.24 & 55412.94504 & 0.32 & 5.07 \\
 & & & &  & & Dummy & 303.80 & 1.47 & 0.12 & 55263.18577 & 0.16 & 4.54 \\
 & & & &  & & HD192310c & 541.57 & 24.72 & 0.33 & 55251.45429 & 2.38 & 3.49 \\
\hline
8 & HD192310 & simulated & 0.800 & 40.00 & 0.5 & - & - & - & - & - & - & -\\
\hline
\bf{9} & \bf{HD128621} & \bf{\bf{real}} & \bf{0.934} & \bf{36-40} & - & - & - & - & - & - & - & -\\
\hline
\bf{10} & \bf{HD128621} & \bf{\bf{real}} & \bf{0.934} & \bf{36-40} & - & Dummy & 0.82 & 0.93 & 0.05 & 55499.88682 & 0.67 & 0.34 \\
 & & & & & & HD85512b & 56.68 & 3.49 & 0.11 & 55498.78199 & 0.61 & 5.67 \\
 & & & & & & Dummy & 296.30 & 1.62 & 0.05 & 55212.65748 & 0.16 & 3.30 \\
\hline
\bf{11} & \bf{HD128621} & \bf{\bf{real}} & \bf{0.934} & \bf{36-40} & - & HD10700b & 14.66 & 2.10 & 0.17 & 55486.23930 & 0.58 & 3.73 \\
 & & & & & & HD10700c & 34.65 & 3.04 & 0.03 & 55483.21240 & 0.62 & 2.38 \\
 & & & & & & HD10700d & 96.93 & 3.71 & 0.08 & 55488.09514 & 0.54 & 4.23 \\
 & & & & & & Dummy & 283.10 & 3.50 & 0.30 & 55492.94069 & 0.37 & 0.52 \\
 & & & & & & Dummy & 3245.20 & 27.20 & 0.60 & 53636.69331 & 1.54 & 3.81 \\
 & & & &  & & & & & & & &\\
 & & & &  & & & & & & & &\\
12 & HD128621 & simulated & 0.934 & 40.00 & 0.0 & HD128621b & 3.08 & 1.04 & 0.00 & 55498.05350 & 0.48 & 0.00 \\
 & & & &  & & HD10700b & 14.66 & 2.10 & 0.17 & 55489.46771 & 0.58 & 4.50 \\
 & & & &  & & HD10700c & 34.65 & 3.04 & 0.03 & 55477.87032 & 0.62 & 5.36 \\
 & & & &  & & HD10700d & 96.93 & 3.71 & 0.08 & 55488.62457 & 0.54 & 1.82 \\
 & & & &  & & Dummy & 268.94 & 3.33 & 0.28 & 55421.38699 & 0.36 & 3.47 \\
 & & & &  & & Dummy & 3407.46 & 28.56 & 0.63 & 54870.04638 & 1.64 & 1.74 \\
\hline
13 & HD128621 & simulated & 0.934 & 40.00 & 0.5 & - & - & - & - & - & - & -\\
\hline
\bf{14} & \bf{Corot-7} & \bf{\bf{real}} & \bf{0.930} & \bf{22.32} & - & - & - & - & - & - & - & -\\
\hline
15 & Corot-7 & simulated & 0.930 & 22.32 & 0.5 & Corot-7b & 0.88 & 4.87 & 0.12 & 55499.85310 & 3.44 & 0.04 \\
 & & & &  & & Corot-7c & 3.63 & 13.29 & 0.12 & 55499.32164 & 5.85 & 5.80 \\
 & & & &  & & Corot-7d & 9.47 & 17.54 & 0.00 & 55496.58304 & 5.56 & 0.00 \\
\hline
\end{longtable}
\end{longtab}

\end{landscape}

\begin{appendix}

\section{Correlation plots between all time series for each system}

\begin{figure*}
\begin{center}
\includegraphics[width=15cm]{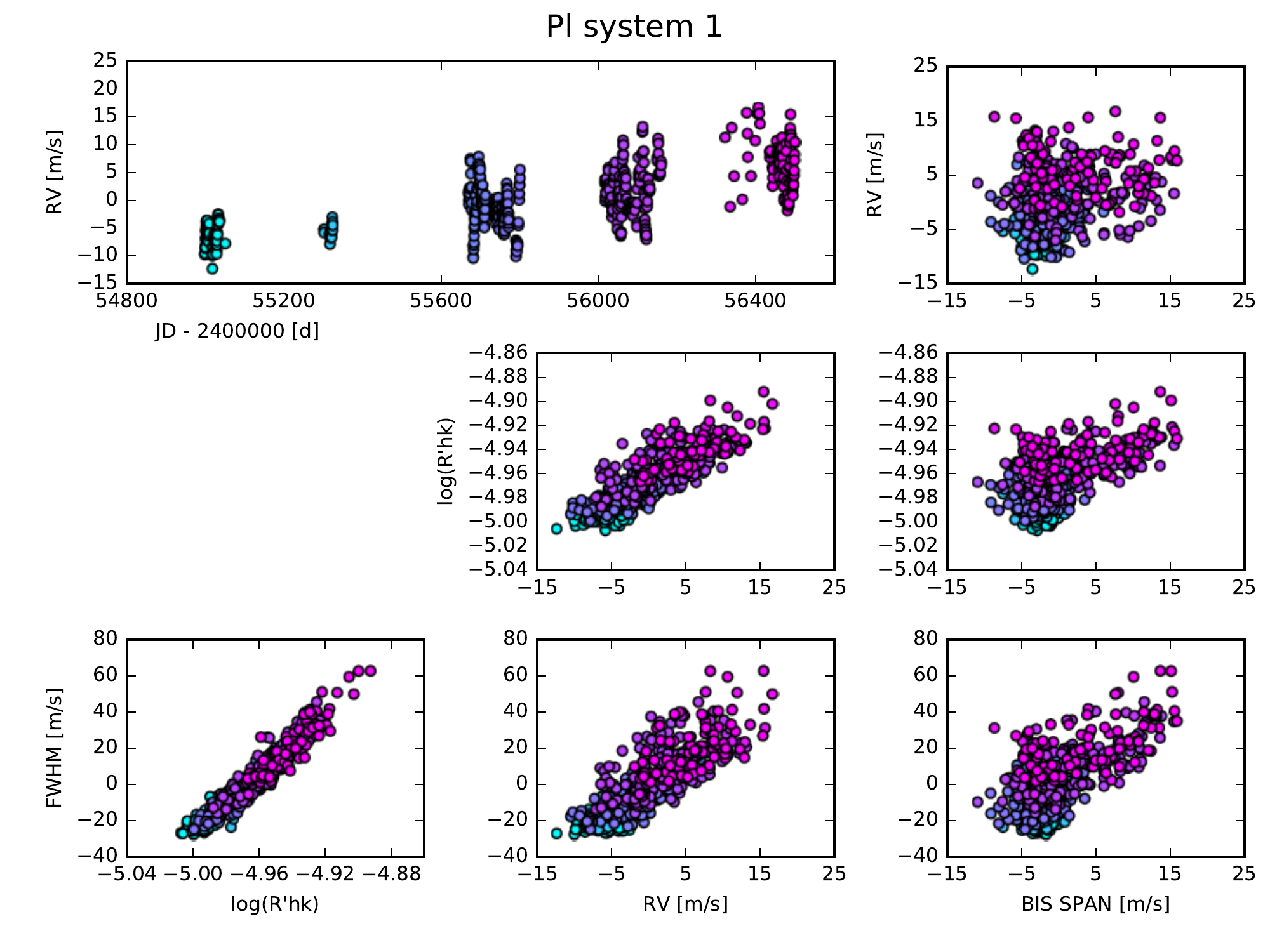}
\includegraphics[width=15cm]{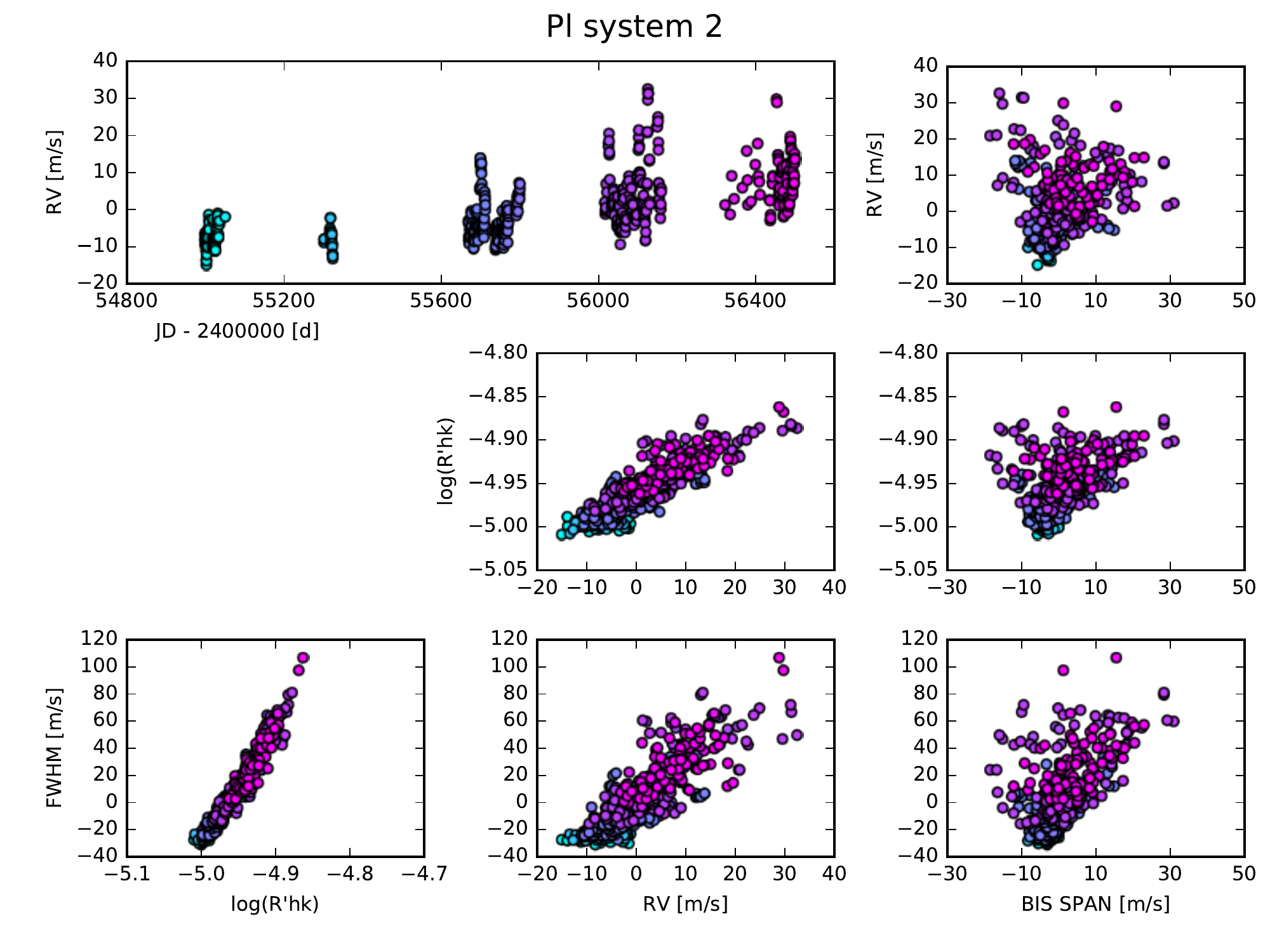}
\caption{RV as a function of time and the correlation between all the parameters for systems 1 and 2.}
\label{fig:annex0}
\end{center}
\end{figure*}

\begin{figure*}
\begin{center}
\includegraphics[width=15cm]{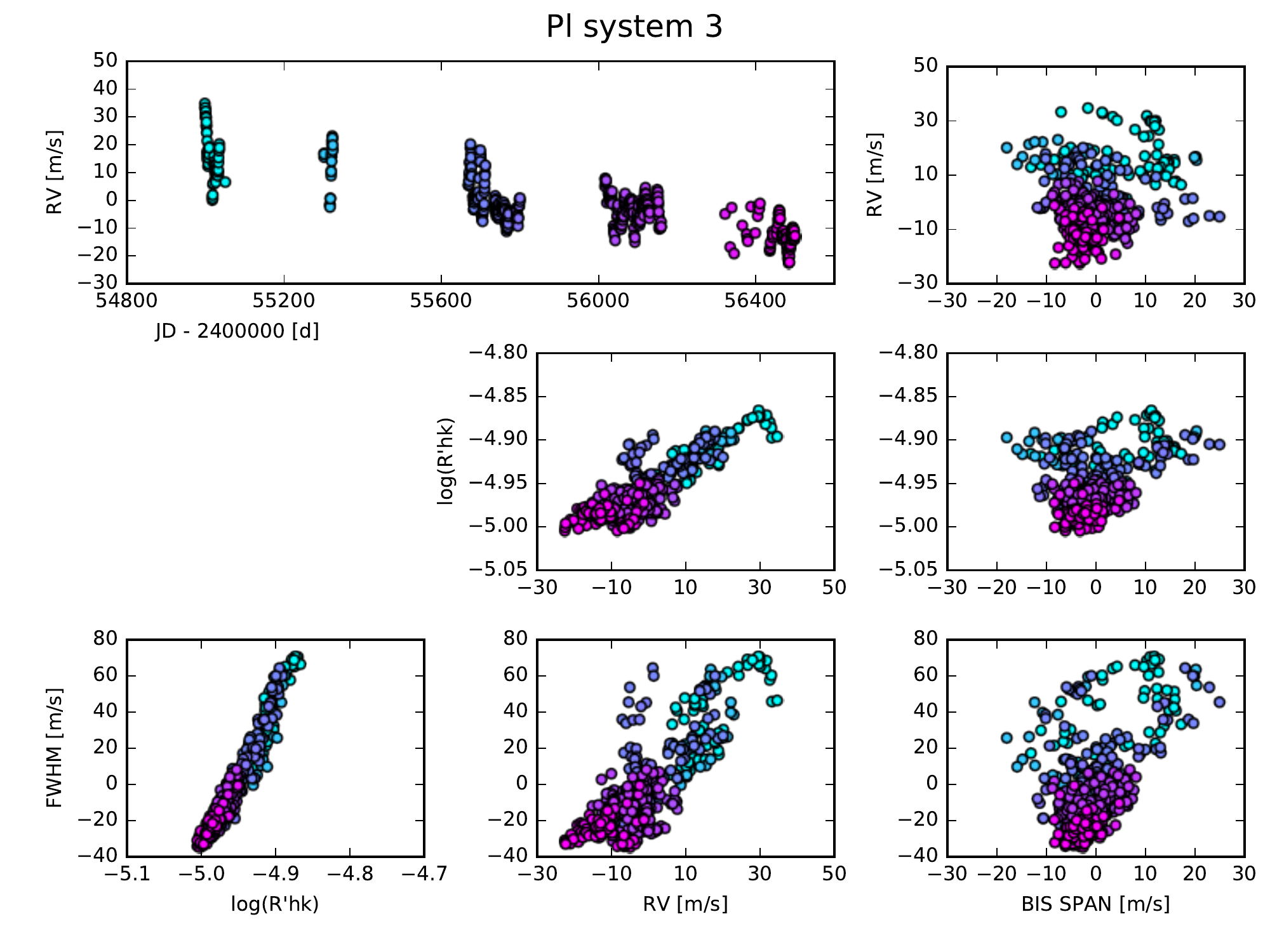}
\includegraphics[width=15cm]{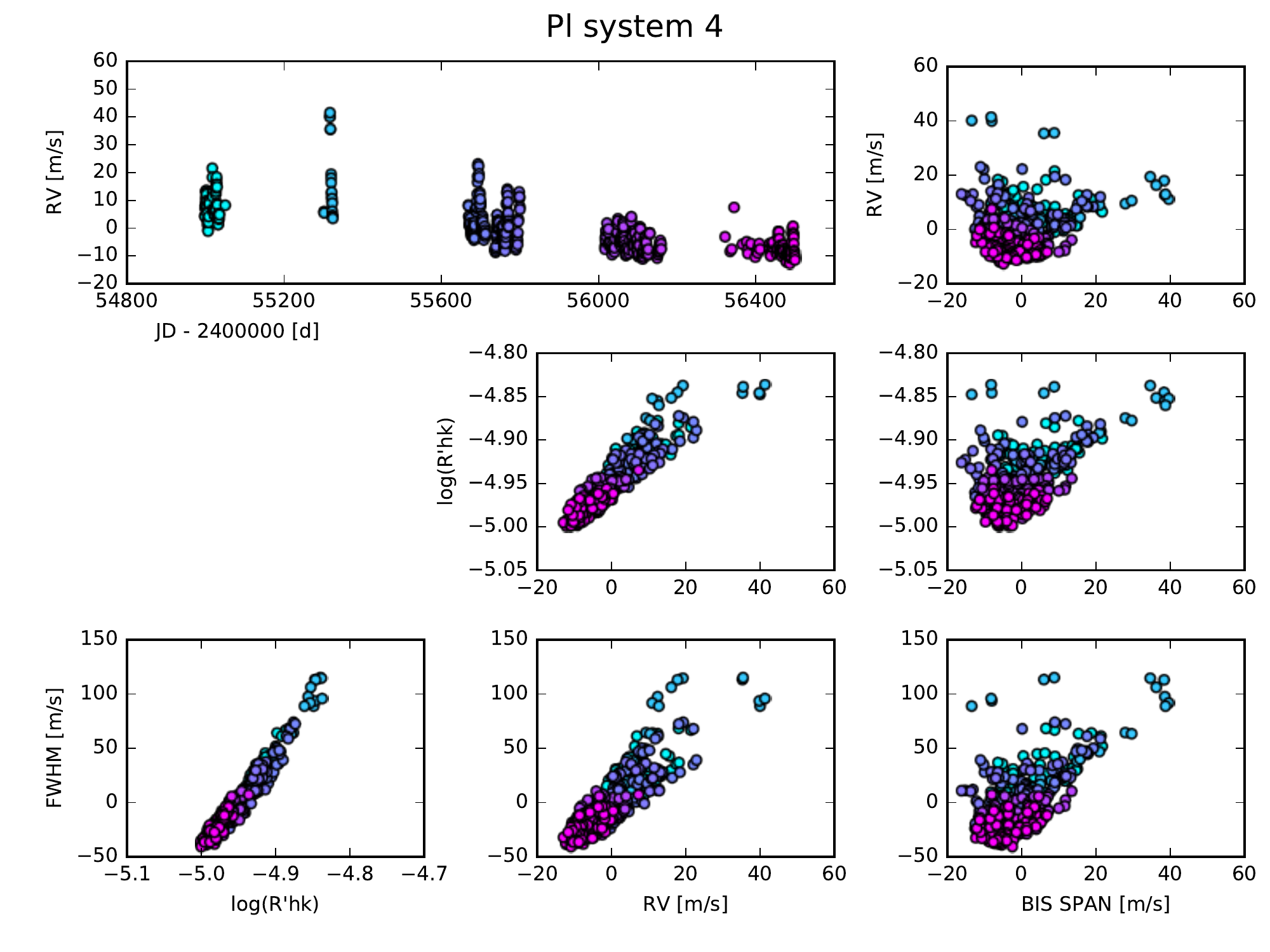}
\caption{Same as Fig. \ref{fig:annex0} for systems 3 and 4.}
\label{fig:annex1}
\end{center}
\end{figure*}

\begin{figure*}
\begin{center}
\includegraphics[width=15cm]{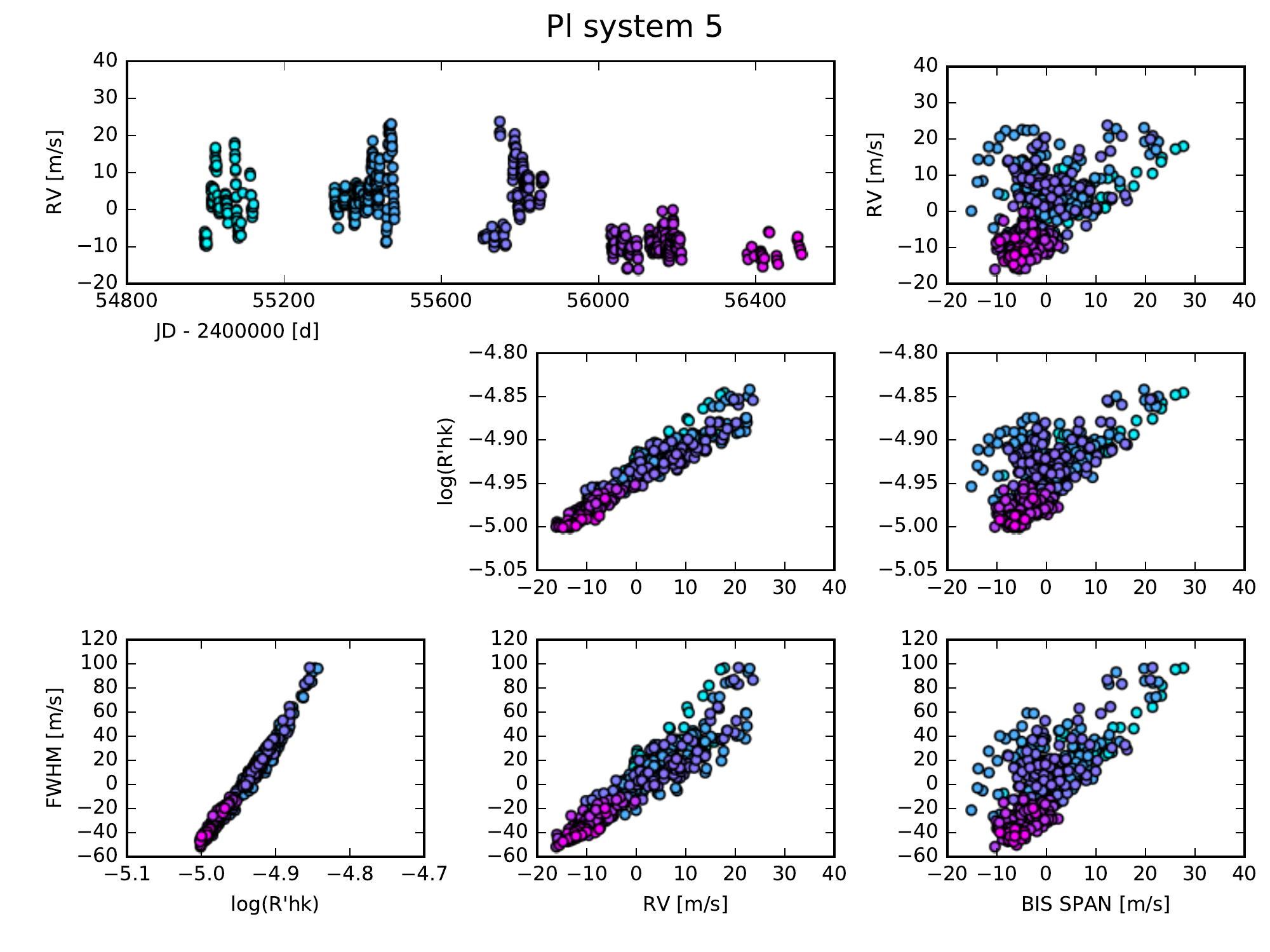}
\includegraphics[width=15cm]{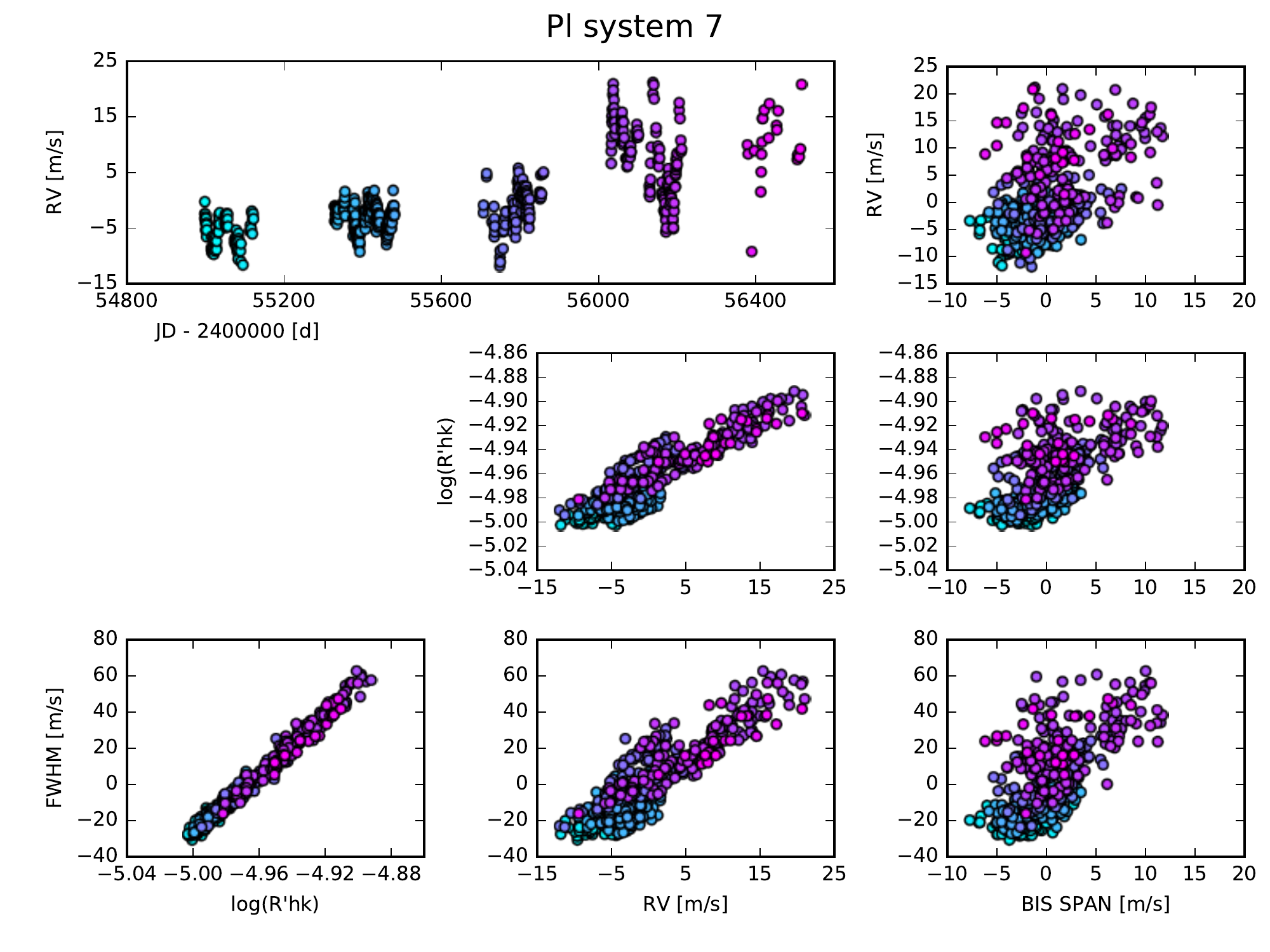}
\caption{Same as Fig. \ref{fig:annex0} for systems 5 and 7.}
\label{fig:annex2}
\end{center}
\end{figure*}

\begin{figure*}
\begin{center}
\includegraphics[width=15cm]{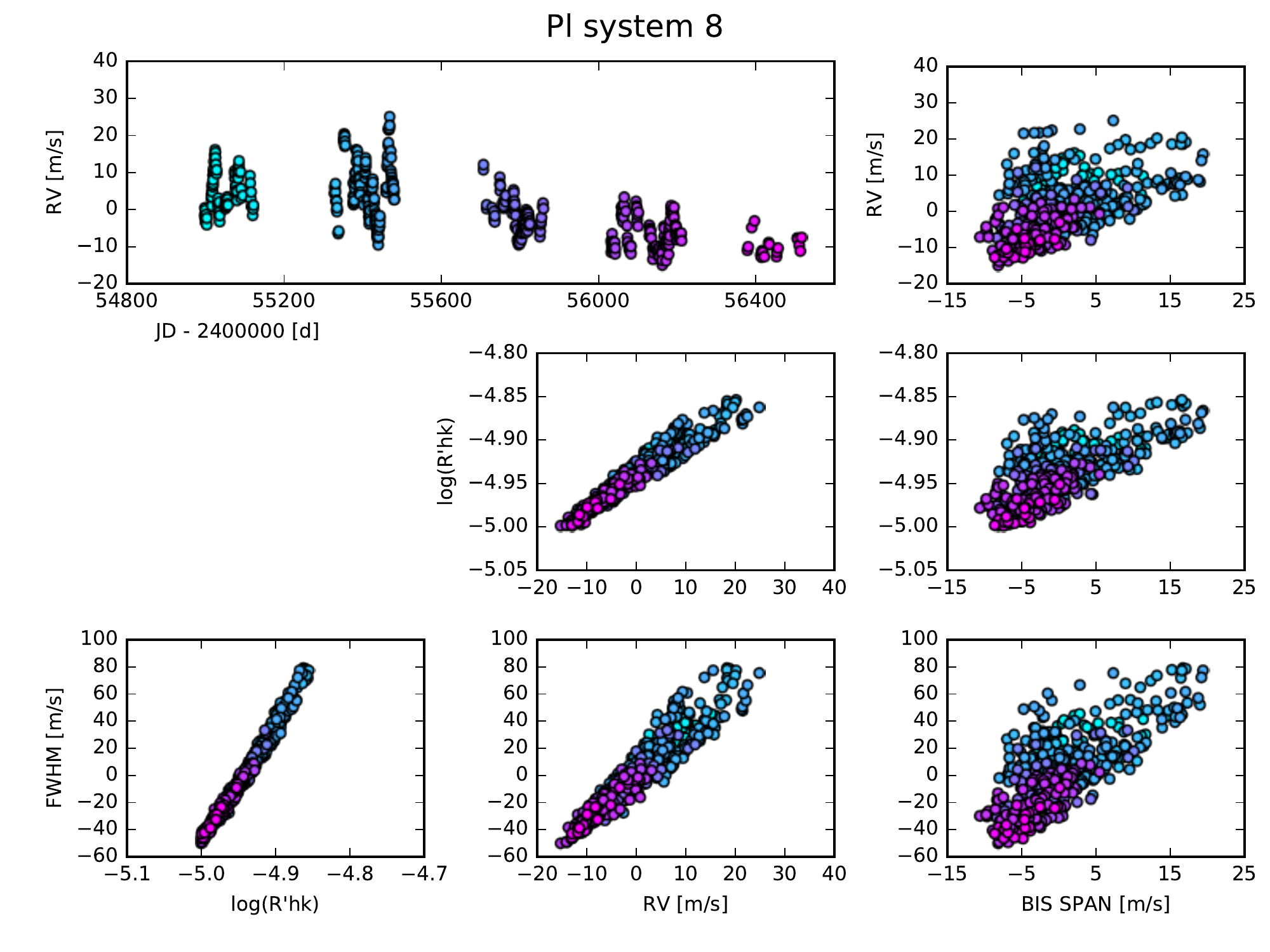}
\includegraphics[width=15cm]{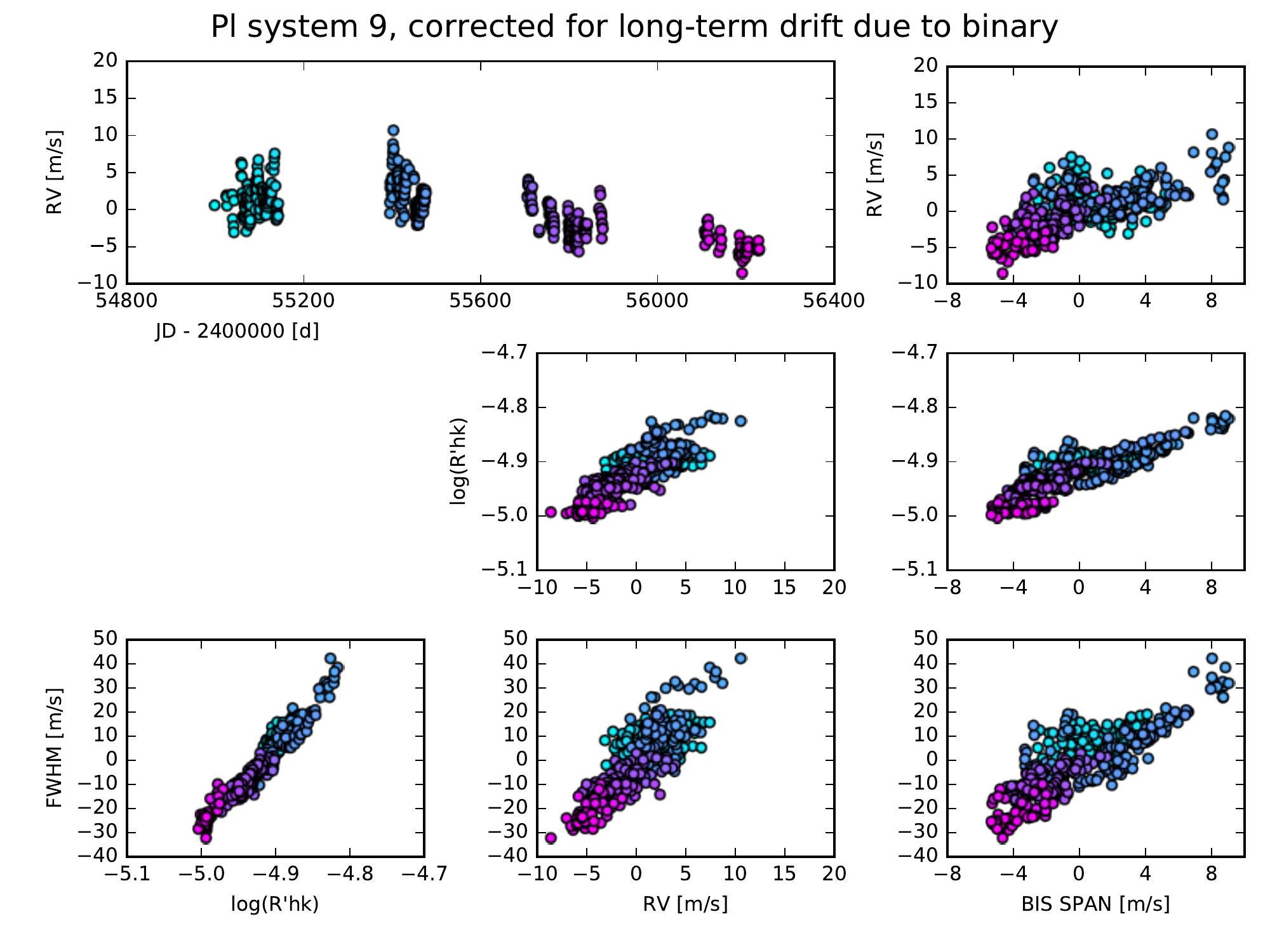}
\caption{Same as Fig. \ref{fig:annex0} for systems 8 and 9. We removed the drift due to $\alpha$\,Cen\,A in the RVs of system 9 using the parameters calculated in \citet{Dumusque-2012} to show for the correlations induced by stellar signals for the different observables.}
\label{fig:annex3}
\end{center}
\end{figure*}

\begin{figure*}
\begin{center}
\includegraphics[width=15cm]{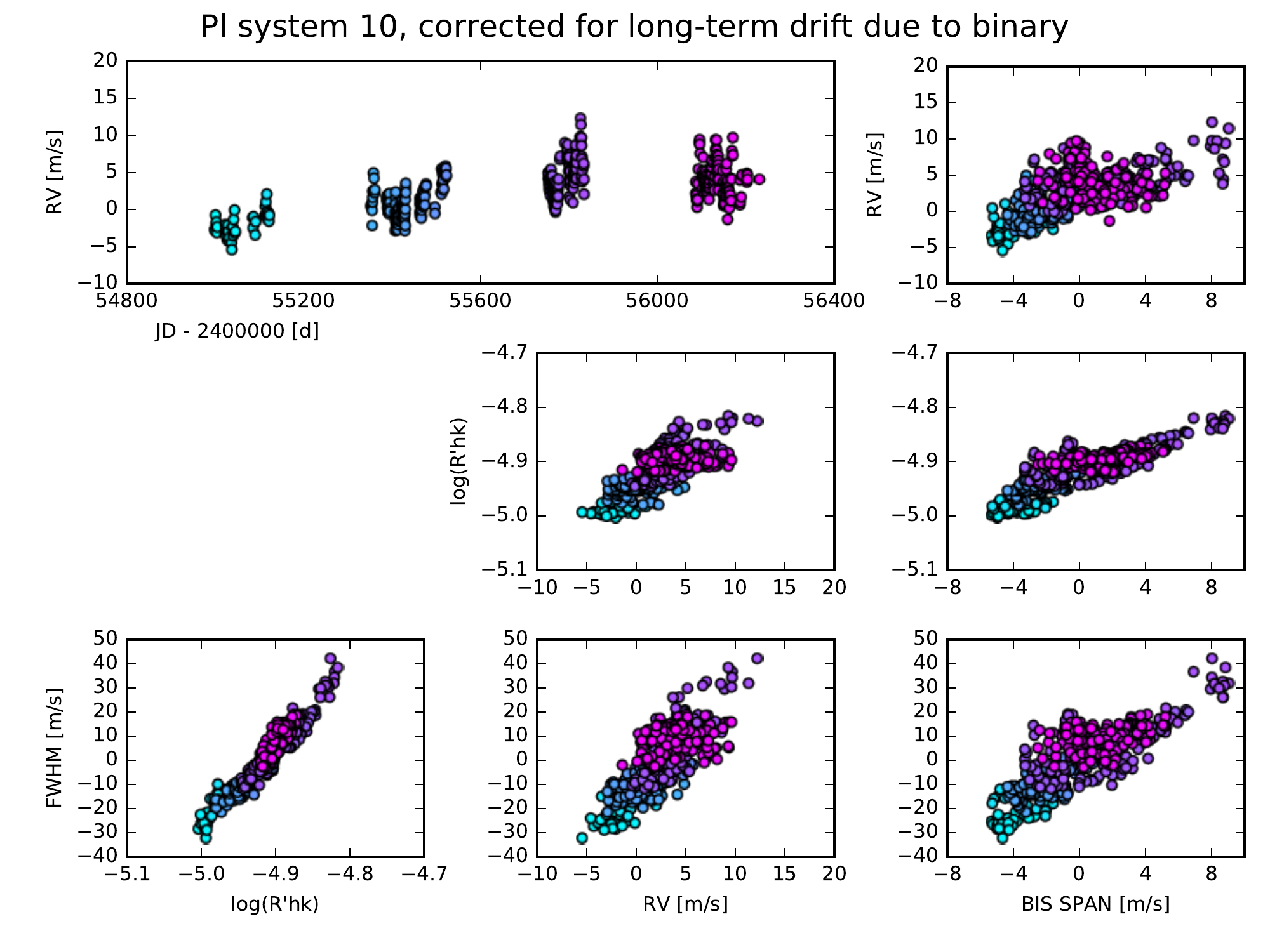}
\includegraphics[width=15cm]{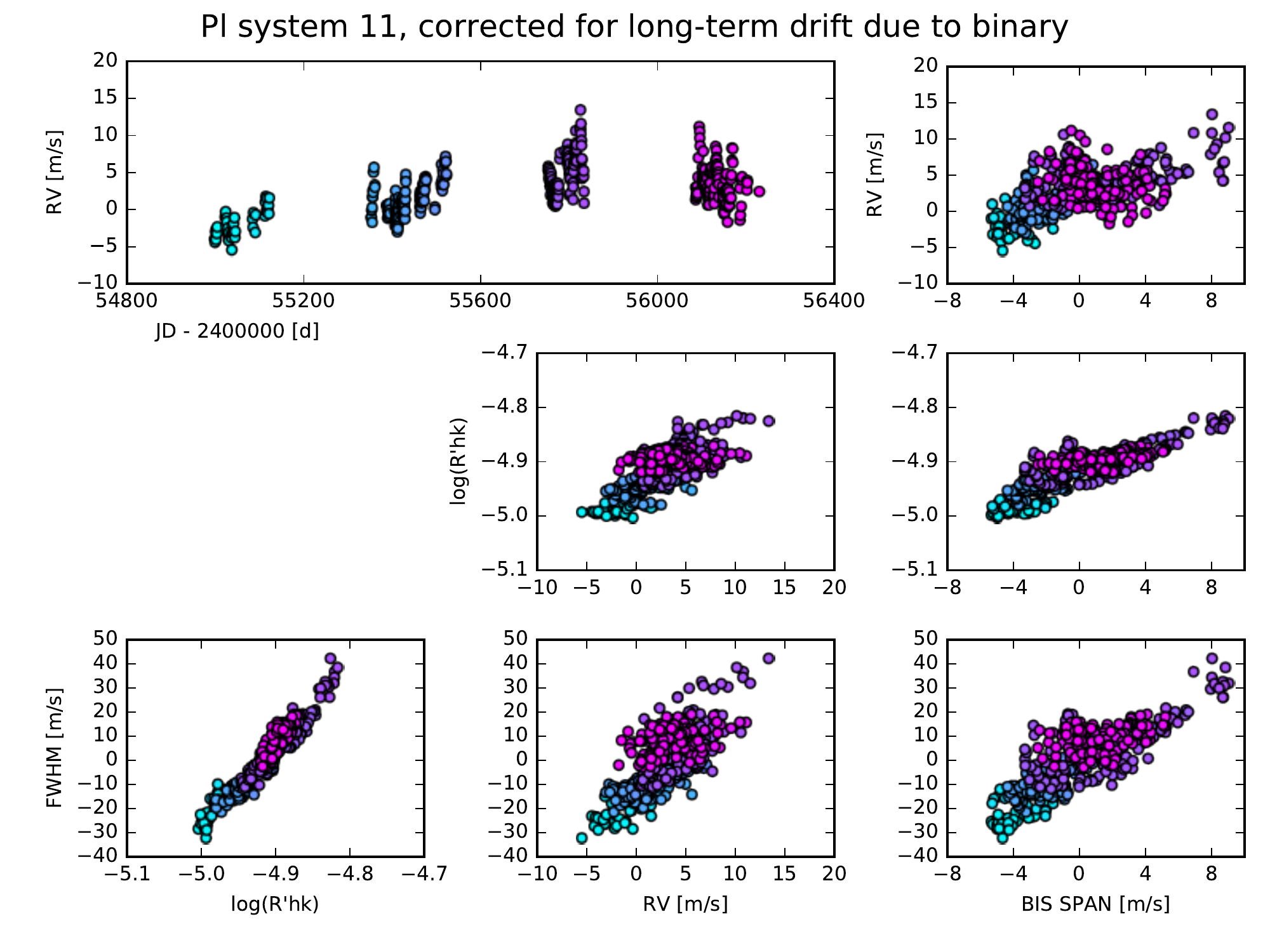}
\caption{Same as Fig. \ref{fig:annex0} for systems 10 and 11. We removed the drift due to $\alpha$\,Cen\,A in the RVs of system 10 and 11 using the parameters calculated in \citet{Dumusque-2012} to show the correlations induced by stellar signals for the different observables.}
\label{fig:annex4}
\end{center}
\end{figure*}

\begin{figure*}
\begin{center}
\includegraphics[width=15cm]{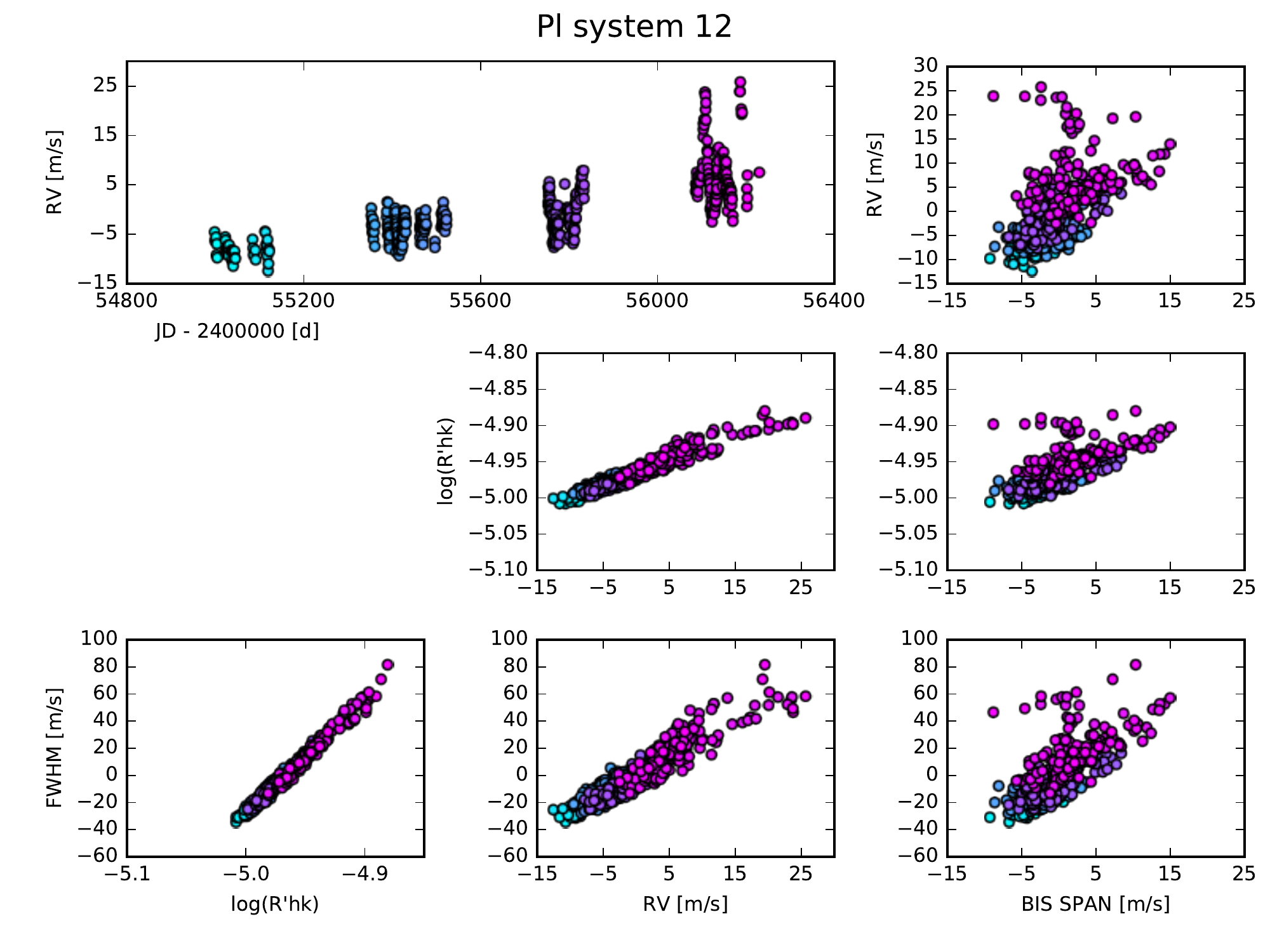}
\includegraphics[width=15cm]{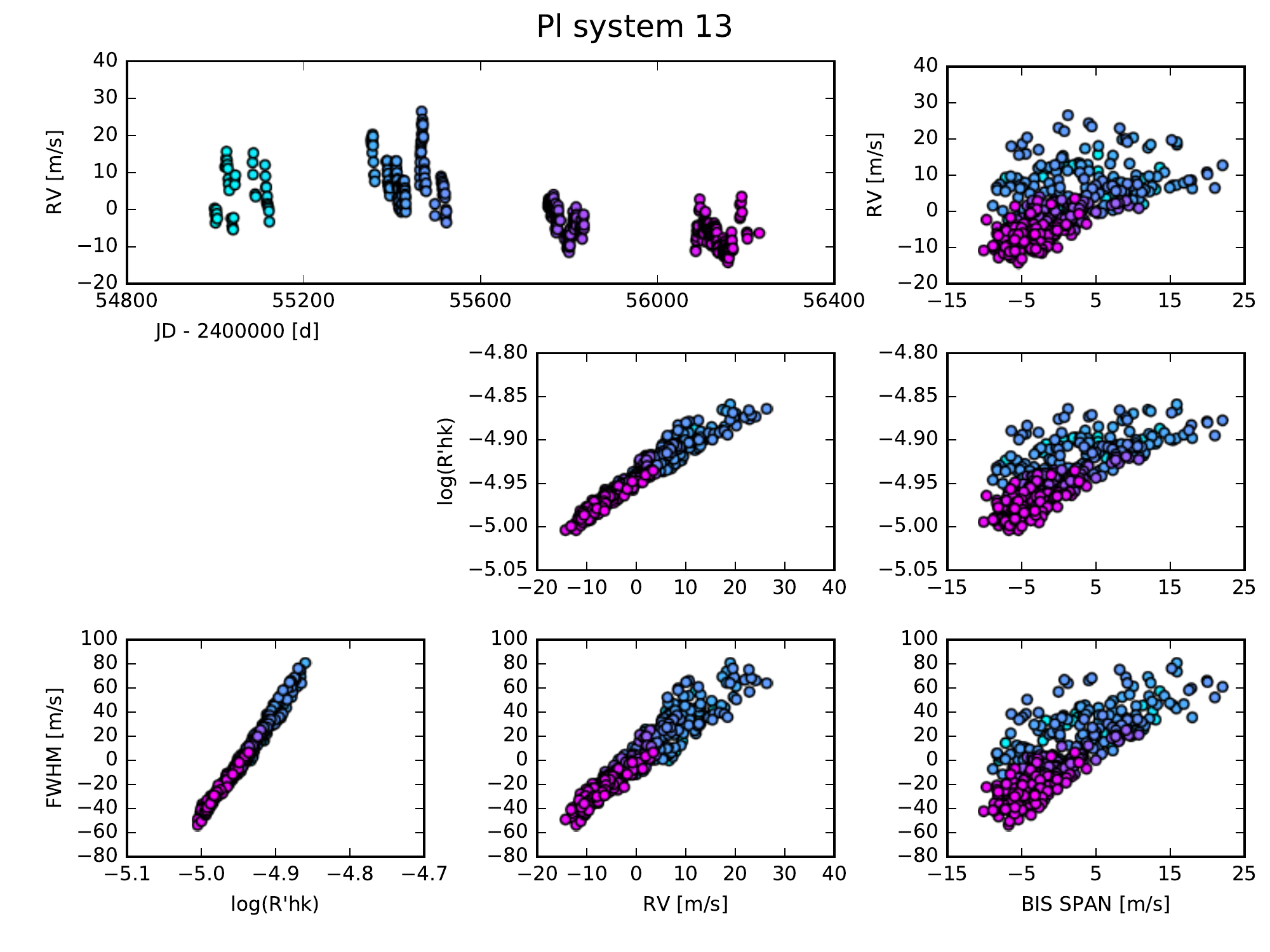}
\caption{Same as Fig. \ref{fig:annex0} for systems 12 and 13.}
\label{fig:annex5}
\end{center}
\end{figure*}

\begin{figure*}
\begin{center}
\includegraphics[width=15cm]{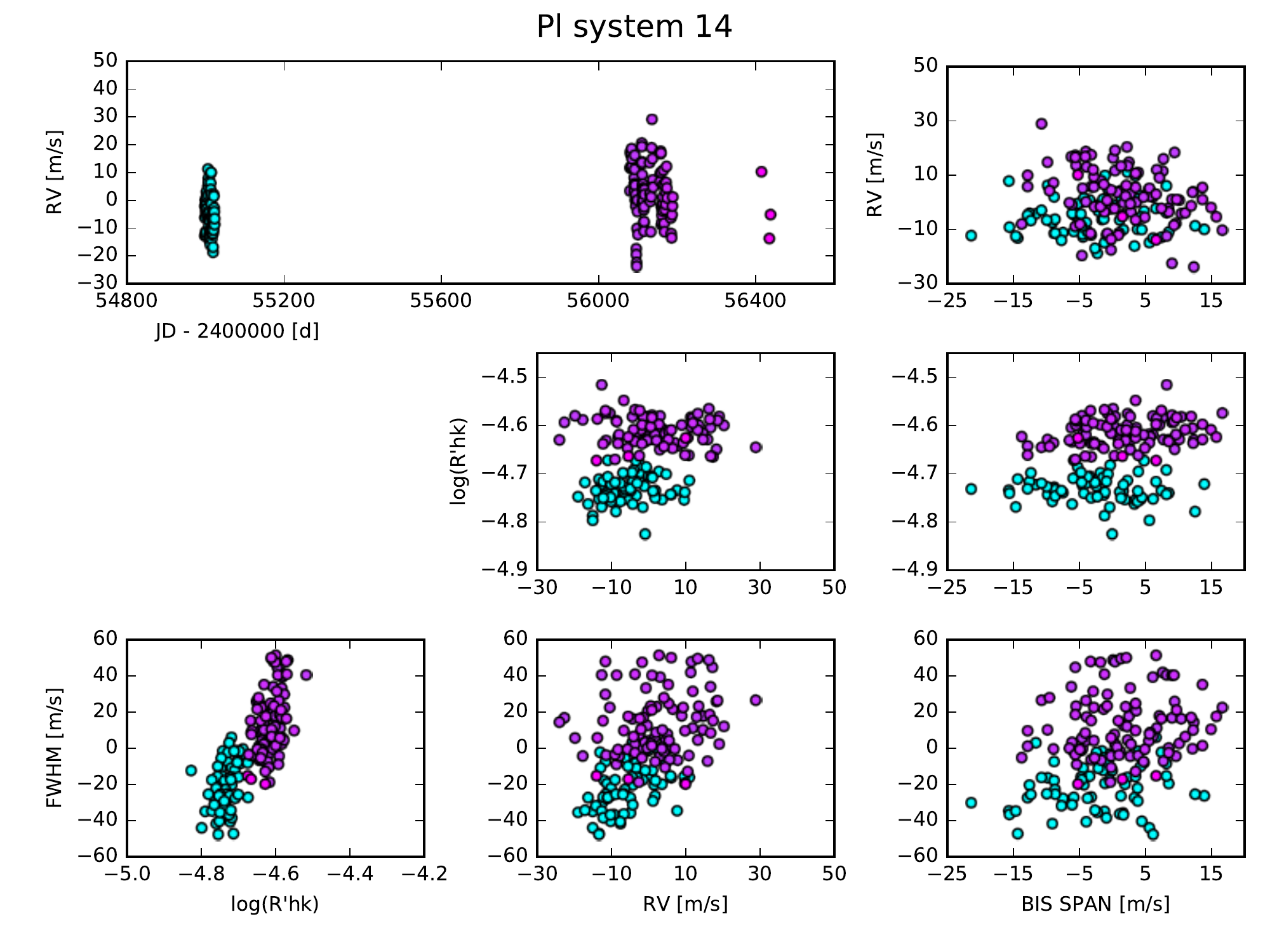}
\includegraphics[width=15cm]{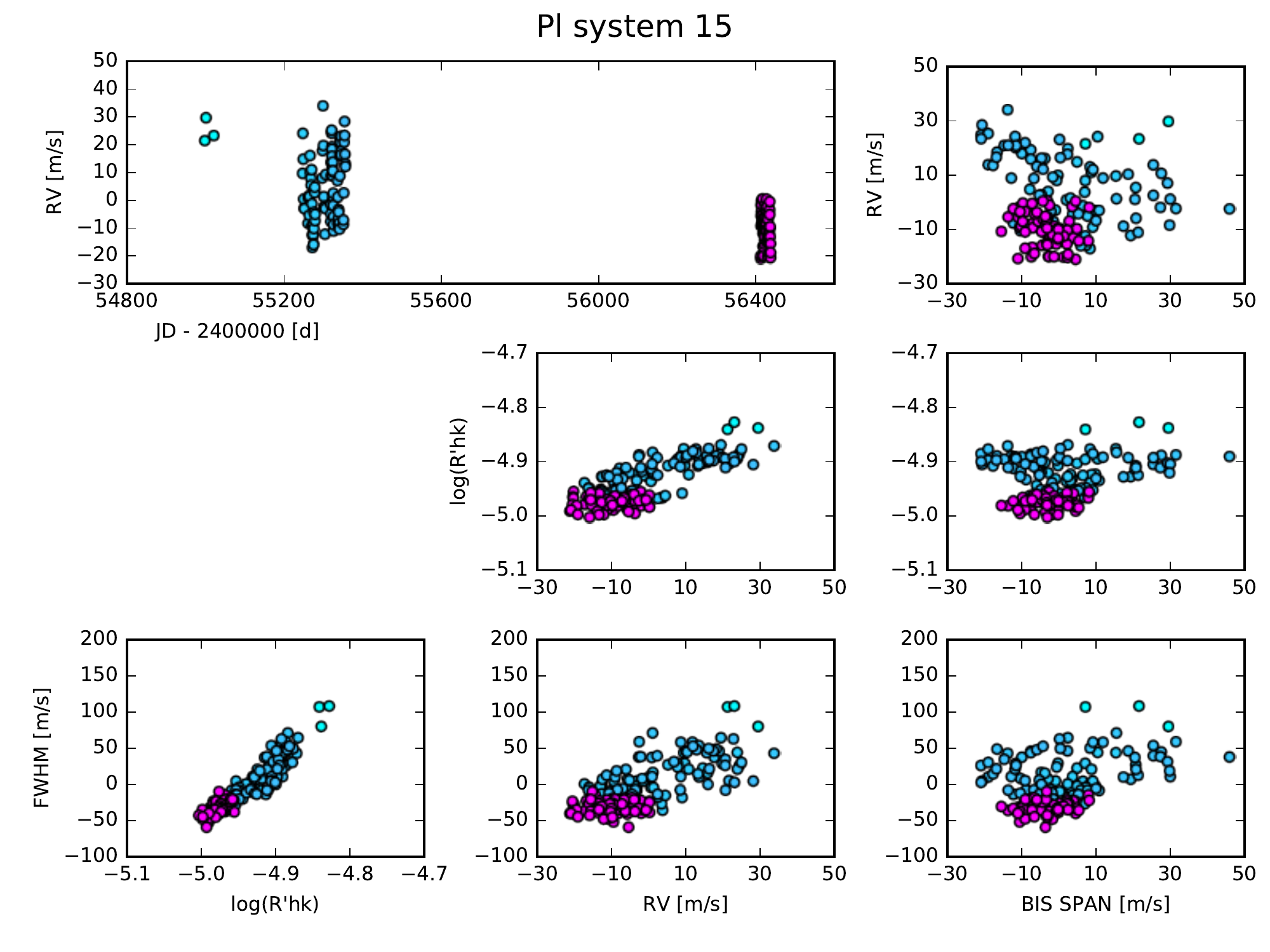}
\caption{Same as Fig. \ref{fig:annex0} for systems 14 and 15.}
\label{fig:annex6}
\end{center}
\end{figure*}

\end{appendix}


\end{document}